\documentclass[english]{article}
\usepackage[T1]{fontenc}
\usepackage[latin9]{inputenc}
\usepackage{amsbsy}
\usepackage{amssymb}
\usepackage{graphicx}
\usepackage{caption}
\usepackage{subcaption}
\usepackage{fancybox}
\usepackage{fullpage}
\usepackage{hyperref}
\usepackage{color}
\usepackage{url,textcomp}
\usepackage[framed,numbered,autolinebreaks,useliterate]{mcode}
\usepackage{amsmath}
\usepackage{amsfonts}
\usepackage{amsthm}
\usepackage{babel}
\usepackage{enumerate}
\usepackage[section]{placeins}
\usepackage{bbm}

\definecolor{bugs}{rgb}{0.5,0.1,0.1}
\definecolor{matlab}{rgb}{0.1,0.1,0.5}
\definecolor{grc}{rgb}{0.0,0.55,0.0}
\definecolor{olive}{rgb}{0.4,0.7,0.2}
  \newcommand{\ttmat}[1]{{\small \color{matlab}{\tt #1}}}
  
  \newcommand{\ttgrc}[1]{{\small \color{grc}{\tt #1}}}



\title{Least Squares Wavelet-based Estimation for Additive Regression Models using Non Equally-Spaced Designs}

\author{German A. Schnaidt Grez  \\ email \href{mailto:gschnaidt@gatech.edu}{gschnaidt@gatech.edu} \and Brani Vidakovic \\ email
\href{mailto:brani@gatech.edu}{brani@gatech.edu}}

\date{Georgia Institute of Technology}

\begin{document}
\maketitle
\begin{abstract}
Additive regression models are actively researched in the statistical field because of their usefulness in the analysis of responses determined by non-linear relationships with multivariate predictors. In this kind of statistical models, the response depends linearly on unknown functions of predictor variables and typically, the goal of the analysis is to make inference about these functions.

\medskip

In this paper, we study the problem of additive regression using a least squares approach based on periodic orthogonal wavelets on the interval [0,1]. For this estimator, we obtain strong consistency (with respect to the $\mathbb{L}_{2}$ norm) characterized by optimal convergence rates up to a logarithmic factor, independent of the dimensionality of the problem. This is achieved by truncating the model estimates by a properly chosen parameter, and selecting the multiresolution level $J$ used for the wavelet expansion, as a function of the sample size. In this approach, we obtain these results without the assumption of an equispaced design, a condition that is typically assumed in most wavelet-based procedures.

\medskip

Finally, we show practical results obtained from a simulation study and a real life application, demonstrating the applicability of the proposed methods for the problem of non-linear robust additive regression models.

\end{abstract}


\newpage

\section{Introduction}\label{Intro}

Additive regression models are popular in the statistical field because of their usefulness in the analysis of responses determined by non-linear relationships involving multivariate predictors. In this kind of statistical models, the response depends linearly on unknown functions of the predictors and typically, the goal of the analysis is to make inferences about these functions. This model has been extensively studied through the application of piecewise polynomial approximations, splines, marginal integration, as well as back-fitting or functional principal components. Chapter 15 of \cite{Ramsay2005}, Chapter 22 of \cite{Gyorfi2002} and \cite{Mammen2003}, \cite{Buja1989} and \cite{Hastie1990} feature thorough discussions of the issues related to fitting such models and provide a comprehensive overview and analysis of various estimation techniques for this problem.

\medskip

In general, the additive regression model relates a univariate response $Y$ to predictor variables $\textbf{X}\in \mathbb{R}^{p}\,,\,p\geq1$, via a set of unknown non-linear functions $\left\{f_{l}\,|\,f_{l}:\mathbb{R}\rightarrow R\,,\,l=1,...,p\right\}$. The functions $f_{l}$ may be assumed to have a specified parametric form (e.g. polynomial) or may be specified non-parametrically, simply as "smooth functions" that satisfy a set of constraints (e.g. belong to a certain functional space such as a Besov or Sobolev, Lipschitz continuity, spaces of functions with bounded derivatives, etc.). Though the parametric estimates may seem more attractive from the modeling perspective, they can have a major drawback: a parametric model automatically restricts the space of functions that is used to approximate the unknown regression function, regardless of the available data. As a result, when the elicited parametric family is not "close" to the assumed functional form the results obtained through the parametric approach can be misleading. For this reason, the non-parametric approach has gained more popularity in statistical research, providing a more general, flexible and robust approach in tasks of functional inference.

\medskip

In this paper we study the problem of additive regression with random designs using a least squares methodology based on a periodic orthogonal wavelet basis on the interval [0,1]. We show that it is possible to choose the detail level $J=J(n)$ of the multiresolution space $V_{J}$ in order to prevent an ill-conditioned design matrix and then, to obtain a strongly consistent estimators (with respect to the $\mathbb{L}_{2}$ norm) by truncating the estimated regression function using a suitable threshold parameter that depends on the sample size $n$. In this setting, we show that it is possible to achieve optimal convergence rates up to a logarithmic factor, independent of the dimensionality of the problem. Moreover, we obtain these results without the assumption of an equispaced design for the application of the wavelet procedures.

\medskip
Our choice of wavelets as an orthonormal basis is motivated by the fact that wavelets could be well localized in both time and scale (frequency), and possess superb approximation properties for signals with rapid local changes such as discontinuities, cusps, sharp spikes, etc.. Moreover, the representation of these signals in the form of wavelet decompositions can be accurately done using only a few wavelet coefficients, enabling sparsity and dimensionality reduction. This adaptivity does not, in general, hold for other standard orthonormal bases (e.g. Fourier basis) which may require many compensating coefficients to describe signal discontinuities or local bursts. 
\medskip

In addition, we show the potential of the proposed methodology via a simulation study and evaluate its performance using different exemplary functions and random designs, under different sample sizes. Here, we demonstrate that the proposed method is suitable for the problem of non-linear additive regression models and behave in coherence with the obtained theoretical results. Finally, we compare the results obtained through our proposed methodology against a previously published study, using a real life data set.

As it was mentioned, additive regression models have been studied by many authors using a wide variety of approaches. The approaches include marginal integration, back-fitting, least squares (including penalized least squares), orthogonal series approximations, and local polynomials. Short descriptions of the most commonly used techniques are provided next:

\begin{enumerate}[(i)]
\item {\textbf{Marginal Integration}.} This method was proposed by Tjostheim and Auestad (1994)\cite{Tjostheim1994} and Linton and Nielsen (1995)\cite{Nielsen1995} and later generalized by Chen et al. (1996)\cite{Chen1996}. The marginal integration idea is based on the estimation of the effects of each function in the model using sample averages of kernel functions by keeping a variable of interest fixed at each observed sample point, while changing the remaining ones. This method has been shown to produce good results in simulation studies (Sperlich et al., 1999)\cite{Sperlich1999}. However, the marginal integration performance over finite samples tends to be inadequate when the dimension of the predictors is large. In particular, the bias-variance trade-off of the estimator in this case is challenging: for a given bandwidth there may be too few data points $\textbf{x}_{i}$ for any given $\textbf{x}$, which inflates the estimator variance and reduces its numerical stability. On the other hand, choosing larger bandwidth may reduce the variability but also enlarge the bias.

\item {\textbf{Back-fitting}.} This approach was first introduced by Buja et al. (1989)\cite{Buja1989a} and further developed by Hastie and Tibshirani (1990)\cite{Hastie1990a}. This technique uses nonparametric regression to estimate each additive component, and then updates the preliminary estimates. This process continues in an iterative fashion until convergence. One of the drawbacks of this method is that it has been proven to be theoretically challenging to analize. In this context, Opsomer and Ruppert (1997)\cite{Opsomer1997} investigated the properties of a version of back-fitting, and found that the estimator was not oracle efficient\footnote{An oracle efficient estimator is such that each component of the model can be estimated with the same convergence rate as if the rest of the model components were known.}. Later on, Mammen et al. (1999)\cite{Mammen1999} and Mammen and Park (2006)\cite{Mammen2006} proposed ways to modify the backfitting approach to produce estimators with better statistical properties such as oracle efficiency and asymptotic normality, and also free of the curse of dimensionality. Even though this is a popular method, it has been shown that its efficiency decreases when the unknown functions are observed at non-equispaced locations.

\item {\textbf{Series based methods using wavelets}.} One important benefit of wavelets is that they are able to adapt to unknown smoothness of functions (Donoho et al. (1995)\cite{Donoho1995}). Most of the work using wavelets is based on the requirement of equally spaced measurements (e.g. at equal time intervals or a certain response observed on a regularly spaced grid). Antoniadis et al. (1997)\cite{Antoniadis1997a} propose a method using interpolations and averaging; based on the observed sample, the function is approximated at equally spaced dyadic points. In this context, most of the methods that use this kind of approach lead to wavelet coefficients that can be computed via a matrix transformation of the original data and are formulated in terms of a continuous wavelet transformation applied to a constant piecewise interpolation of the observed samples. Pensky and Vidakovic (2001)\cite{Pensky2001} propose a method that uses a probabilistic model on the design of the independent variables and can be applied to non-equally spaced designs (NESD). Their approach is based on a linear wavelet-based estimator that is similar to the wavelet modification of the Nadaraja-Watson estimator (Antoniadis et al. (1994)). In the same context, Amato and Antoniadis (2001)\cite{Amato2001} propose a wavelet series estimator based on tensor wavelet series and a regularization rule that guarantees an adaptive solution to the estimation problem in the presence of NESD.

\item {\textbf{Other methods based on wavelets}.} Different approaches from the previously described that are wavelet-based have been also investigated. Donoho et al. (1992)\cite{Donoho1992} proposed an estimator that is the solution of a penalized Least squares optimization problem preventing the problem of ill-conditioned design matrices. Zhang and Wong (2003) proposed a two-stage wavelet thresholding procedure using local polynomial fitting and marginal integration for the estimation of the additive components. Their method is adaptive to different degrees of smoothness of the components and has good asymptotic properties. Later on Sardy and Tseng (2004)\cite{Sardy2004} proposed a non-linear smoother and non-linear back-fitting algorithm that is based on \verb"WaveShrink", modeling each function in the model as a parsimonious expansion on a wavelet basis that is further subjected to variable selection (i.e. which wavelets to use in the expansion) via non-linear shrinkage.
\end{enumerate}

As was discussed before in the context of the application of wavelets to the problem of additive models in NESD, another possibility is just simply ignore the non-equispaced condition on the predictors and apply the wavelet methods directly to the observed sample. Even though this might seem a somewhat crude approach, we will show that it is possible to implement this procedure via a relatively simple algorithm, obtaining good statistical properties and estimation results.

\subsection{About Periodic Wavelets}\label{wavelets}

For the implementation of the functional estimator, we choose periodic wavelets as an orthonormal basis. Even though this kind of wavelets exhibit poor behaviour near the boundaries (when the analyzed function is not periodic, high amplitude wavelet coefficients are generated in the neighborhood of the boundaries) they are typically used due to the relatively simple numerical implementation and compact support. Also, as was suggested by Johnstone (1994), this simplification affects only a small number of wavelet coefficients at each resolution level.

\medskip

Periodic wavelets in $[0,1]$ are defined by a modification of the standard scaling and wavelet functions:
\begin{eqnarray}
 & \phi^{per}_{j,k}(x)=\sum_{l \in \mathbb{Z}}\phi_{j,k}(x-l) \,,\\
 & \psi^{per}_{j,k}(x)=\sum_{l \in \mathbb{Z}}\psi_{j,k}(x-l)\,.
\end{eqnarray}

It is possible to show, as in \cite{Restrepo1996}, that $\left\{ \phi^{per}_{j,k}(x), 0\leq k \leq 2^{j}-1 , j\geq 0\right\}$ constitutes an orthonormal basis
for $\mathbb{L}_{2}[0,1]$. Consequently, $\cup_{j=0}^{\infty} V_{j}^{per}=\mathbb{L}_{2}[0,1]$, where $V_{j}^{per}$ is the space spanned by $\left\{ \phi^{per}_{j,k}(x), 0\leq k \leq 2^{j}-1 \right\}$. This allows to represent an $\mathbb{L}_{2}[0,1]$ function $f$ as:

\begin{equation}\label{eq:1b}
f(x)=\langle f(x),\phi^{per}_{0,0}(x) \rangle \phi^{per}_{0,0}(x) + \sum_{j\geq 0}\sum_{k=0}^{2^{j}-1}\langle f(x),\psi^{per}_{j,k}(x) \rangle \psi^{per}_{j,k}(x)\,.
\end{equation}
\medskip
Also, for a fixed $j=J$, we can obtain an orthogonal projection of $f(x)$ onto $V_{J}$ denoted as $\textbf{P}_{J}(f(x))$, and given by:
\begin{equation}\label{eq:1c}
\textbf{P}_{J}(f(x))=\sum_{k=0}^{2^{J}-1}\langle f(x),\phi^{per}_{J,k}(x) \rangle \phi^{per}_{J,k}(x)\,.
\end{equation}
Since periodized wavelets provide a basis for $\mathbb{L}^{2}([0,1])$, we have that $\parallel f(x) - \textbf{P}_{J}(f(x)) \parallel_{2} \rightarrow 0$ as $ J \rightarrow  \infty $. Also, it can be shown that $\parallel f(x) - \textbf{P}_{J}(f(x)) \parallel_{\infty} \rightarrow 0$ as $ J \rightarrow  \infty $. Therefore, we can see that $\textbf{P}_{J}(f(x))$ uniformly converges to $f$ as $J \rightarrow \infty$.
\medskip
Similarly, as discussed in \cite{Daubechies1992} it is possible to assess the approximation error for a certain density of interest $f$ using a truncated projection (i.e. for a certain chosen detail space $J$). For example, using the $s$-th Sobolev norm of a function defined as:

\begin{equation}
\parallel f(x) \parallel_{H^{s}}=\sqrt{\int(1+|x|^{2})^{s}|f(x)|^{2}dx}\,,
\end{equation}
one defines the $H^{s}$ sobolev space, as the space that consists of all functions $f$ whose s-Sobolev norm exists and is finite. As it is shown in \cite{Daubechies1992}:

\begin{equation}\label{eq:1d}
\parallel f(x) - \textbf{P}_{J}(f(x)) \parallel_{2} \leq 2^{-J\cdot s}\cdot \parallel f \parallel _{H^{s}[0,1]}\,.
\end{equation}
From (\ref{eq:1d}), for a pre-specified $\epsilon>0$ one can choose $J$ such that $\parallel f(x) - \textbf{P}_{J}(f(x)) \parallel_{2} \leq \epsilon$. In fact, a possible choice of $J$ could be:

\begin{equation}\label{eq:1e}
J \geq -\lceil \frac{1}{s} \log_{2}\left(\frac{\epsilon}{\parallel f \parallel _{H^{s}[0,1]}}\right) \rceil\,.
\end{equation}
Therefore, it is possible to approximate a desired function to arbitrary precision using the MRA generated by a wavelet basis.

\newpage
\section{Wavelet-based Estimation in Additive Regression Models}\label{ARM}

Suppose that instead of the typical linear regression model $y=\sum_{j=1}^{p}\beta_{j}x_{j}+\beta_{0}+\epsilon$ which assumes linearity
in the predictors $\textbf{x}=\left(x_{1},...,x_{p}\right)$, we have the following:

\begin{eqnarray}
\nonumber
f(\textbf{x})&=&\beta_{0}+f_{A}(\textbf{x})+\sigma\cdot\epsilon \\
 &=&\beta_{0}+\sum_{j=1}^{p}f_{j}(x_{j})+\sigma\cdot\epsilon\,, \label{eq:3.1}
\end{eqnarray}
where $\epsilon$, independent of $\textbf{x}$, $\mathbb{E}[\epsilon]=0$, $\mathbb{E}[\epsilon^{2}]=1$, $\sigma>0$, $\sigma<\infty$. Similarly, $\textbf{x}_{i} \mathop{\sim}\limits^{\text{iid}} h(\textbf{x})$, an unknown design density of observations and $\left\{f_{1}(),...,f_{p}()\right\}$ are unknown functions to be estimated.
\medskip
\medskip

Suppose that we are able to observe a sample $\{y_{i}=f(\textbf{x}_{i}),\textbf{x}_{i} \}_{i=1}^{n}$ where $\textbf{x}_{1},...,\textbf{x}_{n} \mathop{\sim}\limits^{iid} h(\textbf{x})$. We are interested in
estimating $\beta_{0}$ and $\{f_{1}(),...,f_{p}()\}$. For simplicity (without loss of generality) and identifiability, we assume:
\begin{enumerate}[]
\item{\textbf{(A1)}}\label{A1} The density $h(\textbf{x})$ is of the continuous type and has support in $[0,1]^{p}$. Also, we assume $\exists \epsilon_{h}>0\,\,$ s.t. $\,h(\textbf{x})\geq \epsilon_{h}$ $\, \forall \textbf{x} \in [0,1]^{p}$.
\item{\textbf{(A2)}}\label{A2} For $k=1,...,p$, $\int_{0}^{1}f_{k}(x)dx_{k}=0$.
\item{\textbf{(A3)}}\label{A3} For $k=1,...,p$, $\mathop{\sup}\limits_{x\in [0,1]}|f_{k}(x)|\leq M_{k}< \infty$ and $\mathop{\inf}\limits_{x\in [0,1]}\left\{f_{k}(x)\right\}\geq m_{k}> -\infty$. This implies that for $k=1,...,p$, $f_{k}\in \mathbb{L}_{2}([0,1])$.
\item{\textbf{(A4)}}\label{A6} The density $h(\textbf{x})$ is uniformly bounded in $[0,1]^{p}$, that is, $\forall \textbf{x}\in [0,1]^{p}, \,  |h(\textbf{x})|\leq M$, $M<\infty$.
\end{enumerate}
\medskip

\medskip
Furthermore, since $\left\{ \phi^{per}_{j,k}(x), 0\leq k \leq 2^{j}-1 , j\geq 0\right\}$ spans $\mathbb{L}_{2}([0,1])$, each of the functions in (\ref{eq:3.1}) can be represented as:
\begin{equation}\label{eq:3.3}
f_{l}(x)=\sum_{j\geq 0}\sum_{k=0}^{2^{j}-1}c_{jk}^{(l)}\cdot\phi_{jk}^{per}(x),\quad l=1,...,p\,,
\end{equation}
where $c_{jk}^{(l)}$ denotes the $j,k-$th wavelet coefficient of the $l-$th function in the model. Similarly, for some fixed $J$, $f_{l,J}(x),\, l=1,...,p$ represents the orthogonal projection of $f_{l}(x)$ onto the multiresolution space $V_{J}$. Therefore, $f_{l,J}(x)$ can be expressed as:
\begin{equation}\label{eq:3.4}
f_{l,J}(x)=\sum_{k=0}^{2^{J}-1}c_{Jk}^{(l)}\cdot\phi_{Jk}^{per}(x),\quad l=1,...,p\,,
\end{equation}
where:
\begin{equation}\label{eq:3.5}
c_{Jk}^{(l)}=\langle f_{l}(x),\phi_{Jk}^{per}(x)\rangle=\int_{0}^{1}f_{l}(x)\phi_{Jk}^{per}(x)dx,\quad l=1,...,p\,.
\end{equation}
Based on the model (\ref{eq:3.1}) and (\ref{eq:3.4}), it is possible to approximate $f(\textbf{x})$ by an orthogonal projection $f_{J}(\textbf{x})$ onto the multiresolution space
spanned by the set of scaling functions $\left\{ \phi^{per}_{J,k}(x), 0\leq k \leq 2^{J}-1\right\}$, by approximating each of the functions $f_{l}()$ as described above. Therefore, $f_{J}(\textbf{x})$ can be expressed as:

\begin{equation}\label{eq:3.6}
f_{J}(\textbf{x})=\beta_{0}+\sum_{l=1}^{p}\sum_{k=0}^{2^{J}-1}c_{Jk}^{(l)}\phi_{Jk}^{per}(x)\,.
\end{equation}
Now, the goal is for a pre-specified multiresolution index $J$, to use the observed samples to estimate the unknown constant $\beta_{0}$ and the orthogonal projections of the functions $f_{l,J}(x),\, l=1,...,p$.

\subsubsection*{Remarks}
\begin{enumerate}[(i)]
\item \label{A7b} Also, from the above conditions, the variance of the response $y(\textbf{x})$ is bounded for every $\textbf{x}\in \mathbb{R}^{p}$.
\item The assumption that the support of the random vector $\textbf{X}$ is $[0,1]^{p}$ can be always satisfied by carrying out appropriate monotone increasing transformations of each dimensional component, even in the case when the support before transformation is unbounded. In practice, it would be sufficient to transform the empirical support to $[0,1]^{p}$.
\end{enumerate}

\section{A Least Squares approach for non-linear Additive model estimation using orthogonal wavelet basis}\label{LSApproach}

As it is shown in Chapter 22 of \cite{Gyorfi2002}, it is possible to study the problem of additive regression using least squares. The empirical $\mathbb{L}_{2}$ risk is minimized over a linear spaced spanned by a defined orthogonal basis with dimension depending on the sample size. In this setting, consider the unknown functions $\left\{f_{1},...,f_{p} \right\}$ to be approximated by their respective orthogonal projections onto the multiresolution space $V_{J}$ spanned by a given set of scaling functions $\left\{ \phi^{per}_{J,k}(x), k =0,...,2^{J}-1\right\}$. Consequently, the projection of the function $f_{A}(\textbf{x})=\sum_{j=1}^{p}f_{j}(x_{j})$ onto $V_{J}$ belongs to the linear space defined as:
\begin{equation} \label{eqProp8:1}
\mathcal{F}_{n}=\left\{f: [0,1]^{p} \rightarrow \mathbb{R}\, | \, f(\textbf{x})=\sum_{j=1}^{p}\sum_{k=0}^{2^{J(n)}-1}c_{J(n),k}^{(j)}\phi_{J(n),k}^{per}(x_{j})\,,\,\textbf{x}\in[0,1]^{p}\right\}\,,
\end{equation}
\medskip
where $x_{j}\,,j=1,...,p$ corresponds to the $j$-th component of the vector $\textbf{x}\in[0,1]^{p}$. Thus, this projection of $f_{A}(\textbf{x})$ onto $\mathcal{F}_{n}$ is defined by the set of coefficients:
\begin{center}
$ \left\{c_{J,k}^{(j)}\,,\,j=1,...,p\,;\,k=0,...,2^{J(n)}-1\right\} $\,.
\end{center}

\medskip

As it is shown in \cite{Daubechies1992}, by the properties of MRA, $\cup_{j\geq 0} V_{j}$ is dense in $\mathbb{L}_{2}([0,1])$, where $V_{j}$ is the space spanned by the orthonormal basis $\left\{ \phi^{per}_{j,k}(x), k =0,...,2^{j}-1\,;\,j\geq 0\right\}$. Therefore, for any lebesgue measure $\mu(\cdot)$ in $\mathbb{R}$ that is bounded away from zero and infinity in its support, we have that $\cup_{j\geq 0} V_{j}$ is dense in $\mathbb{L}_{2}\left(\mu([0,1])\right)$, thus the following result holds:

\subsubsection*{Proposition 1} \label{prop8}

For any $f\in \mathbb{L}_{2}([0,1])$,  $\epsilon>0$ and bounded lebesgue measure $\mu(\textbf{x})$ in $\mathbb{R}^{p}$,  $\exists\,\left\{c_{J,0}^{(1)*},...,c_{J,2^{J}-1}^{(1)*},...,c_{J,0}^{(p)*},...,c_{J,2^{J}-1}^{(p)*} \right\}$ for which $J=J^{*}(n_{0}(\epsilon))$, such that:

\begin{equation} \label{eqProp8:2}
\int_{[0,1]^{p}}\left|\sum_{j=1}^{p}\left(\sum_{k=0}^{2^{J}-1}c_{J,k}^{(j)}\phi_{J,k}^{per}(x_{j})-f_{j}(x_{j})\right) \right|^{2}\mu(d\textbf{x})\leq \epsilon\,.
\end{equation}
The proof of the above assertion follows from the application of the inequality $(\sum_{j=1}^{d}a_{j})^{2}\leq d\cdot\sum_{j=1}^{d}a_{j}^{2}$, together with the fact that $\bigcup_{j\geq 0} V_{j}$ is dense in $\mathbb{L}_{2}\left(\mu([0,1])\right)$. This enables to find a multiresolution index $J$ as a function of the sample size $n$ sufficiently large, such that it is possible to approximate each of the functions $f_{j}$ with a precision $\epsilon_{j}\leq\frac{\epsilon}{p\cdot||\mu||_{\infty}}\,,\,j=1,...,p$, for $||\mu||_{\infty}$ defined as the infinity norm of the lebesgue measure $\mu$.

\subsection{Least Squares problem formulation.}

Following (\ref{eq:3.1}), suppose a model of the form:
\begin{equation}\label{eq:LSa1}
y(\textbf{x})=f_{A}(\textbf{x})+\sigma\cdot \epsilon\,.
\end{equation}
Assume conditions stated in \ref{LSApproach} are satisfied. From (\ref{eqProp8:2}), for a sample $\left\{(\textbf{X}_{i},Y_{i}) \right\}_{i=1}^{n}$ it is possible to define a least squares estimator of $f(\textbf{x})$ over the space of functions defined by $\mathcal{F}_{n}$ in (\ref{eqProp8:1}), as follows:

\begin{eqnarray}
\nonumber
\hat{f}_{J(n)}&=&\mathop{\arg\inf}\limits_{f\in\mathcal{F}_{n}}\frac{1}{n}\sum_{i=1}^{n}\left|f(\textbf{X}_{i})-Y_{i} \right|^{2}\,, \\
&=&\mathop{\arg\min}\limits_{\left\{c_{J,k}^{(j)}\,,\,j=1,...,p\,;\,k=0,...,2^{J(n)}-1\right\}}\frac{1}{n}\sum_{i=1}^{n}\left|\sum_{j=1}^{p}\sum_{k=0}^{2^{J}-1}c_{J,k}^{(j)}\phi_{J,k}^{per}(X_{ij})-Y_{i} \right|^{2}\,. \label{eq:LS1}
\end{eqnarray}

Define:
\begin{equation}\label{eqLS:2}
\begin{aligned}[c]
    \textbf{c} = \begin{bmatrix}
           c_{J,0}^{(1)} \\
           \vdots \\
           c_{J,2^{J}-1}^{(1)} \\
           \vdots \\
           c_{J,0}^{(p)} \\
           \vdots \\
           c_{J,2^{J}-1}^{(p)}
         \end{bmatrix}_{p\cdot2^{J(n)}\times 1}\,,
\end{aligned}
\qquad
\begin{aligned}[c]
    \textbf{B}(\textbf{x}_{i}) = \begin{bmatrix}
           \phi_{J,0}^{per}(x_{i1}) \\
           \vdots \\
           \phi_{J,2^{J}-1}^{per}(x_{i1}) \\
           \vdots \\
           \phi_{J,0}^{per}(x_{ip}) \\
           \vdots \\
           \phi_{J,2^{J}-1}^{per}(x_{ip})
         \end{bmatrix}_{p\cdot2^{J(n)}\times 1}\,,
\end{aligned}
\qquad
\begin{aligned}[c]
    \textbf{B} = \begin{bmatrix}
           \textbf{B}(\textbf{x}_{1})^{T} \\
           \vdots \\
           \textbf{B}(\textbf{x}_{n})^{T}
         \end{bmatrix}_{n\times p\cdot2^{J(n)}}\,,
\end{aligned}
\qquad
\begin{aligned}[c]
    \textbf{Y} = \begin{bmatrix}
           Y_{1} \\
           \vdots \\
           Y_{n}
         \end{bmatrix}_{n\times 1}\,.
\end{aligned}
\end{equation}

Then, it is possible to represent (\ref{eq:LS1}) as:

\begin{eqnarray}
\hat{f}_{J(n)}&=&\mathop{\arg\min}\limits_{\textbf{c}\in\mathbb{R}^{p\cdot2^{J(n)}}}\frac{1}{n} \left\|\textbf{B}\cdot\textbf{c}-\textbf{Y} \right\|_{2}^{2}\,. \label{eqLS:3}
\end{eqnarray}
Assuming that $\textbf{X}_{1},...,\textbf{X}_{n}$ have continuous joint distribution and $p\cdot2^{J(n)}\leq n$, the matrix $\textbf{B}$ is non-singular (since the event in which $\textbf{X}_{1},...,\textbf{X}_{n}$ are all distinct happens with probability 1). Therefore, the problem defined by (\ref{eqLS:3}) has a unique solution given by:

\begin{equation}\label{eqLS:4}
\textbf{c}^{*}=\left(\textbf{B}^{T}\textbf{B}\right)^{-1}\textbf{B}^{T}\textbf{Y}\,.
\end{equation}
Therefore, for a new observation $\textbf{x}$, the estimator $\hat{f}_{J(n)}(\textbf{x})$ can be represented as:

\begin{equation}\label{eqLS:5}
\hat{f}_{J(n)}(\textbf{x}) = \textbf{B}(\textbf{x})^{T}\textbf{c}^{*}\,.
\end{equation}

\subsection{Strong consistency of the Linear Least Squares Estimator.}

In this section, we investigate the consistency property for the least squares estimator defined by equations (\ref{eqLS:4}) and (\ref{eqLS:5}). Throughout the analysis, we will use results and definitions contained in \ref{previousThms} of the appendix, which have been previously introduced in the statistical literature.

\medskip

\subsubsection{Theorem 1: Strong consistency of the Wavelet-based Least Squares Estimator}\label{Th1}

Suppose an orthonormal basis $\left\{ \phi^{per}_{j,k}(x), k =0,...,2^{j}-1,\,j\geq0\right\}$ which is dense in $\mathbb{L}_{2}(\nu([0,1]))$ for $\nu\in \Upsilon$, and let $\Upsilon$ be the set of bounded lebesgue measures in $[0,1]$. Suppose $\mu$ is a bounded lebesgue measure in $[0,1]^{p}$, and the following conditions are satisfied for the scaling function $\phi$:

\begin{enumerate}
  \item $\exists\,\Phi$, bounded and non-increasing function in $\mathbb{R}$ such that $\int\Phi(|u|)du<\infty$ and $|\phi(u)|\leq \Phi(|u|)$ almost everywhere (a.e.).
  \item In addition, $\int_{\mathbb{R}}|u|^{N+1}\Phi(|u|)du<\infty$ for some $N\geq0$.
  \item $\exists\,F$, integrable, such that $|K(x,y)|\leq F(x-y)$, $\forall x,y \in \mathbb{R}$, for $K(x,y)=\sum_{k}\phi(x-k)\phi(y-k) $.
  \item Suppose $\phi$ satisfies:
  \begin{enumerate}
    \item $\sum_{k}|\hat{\phi}(\xi+2k\pi)|^{2}=1$, a.e., where $\hat{\phi}$ denotes the Fourier transform of the scaling function $\phi$. \label{ass:w1}
    \item $\hat{\phi}(\xi)=\hat{\phi}(\frac{\xi}{2})m_{0}(\frac{\xi}{2})$, where $m_{0}(\xi)$ is a $2\pi$-periodic function and $m_{0}\,\in\,\mathbb{L}_{2}(0,2\pi)$.\label{ass:w2}
  \end{enumerate}
  \item $\int_{\mathbb{R}}x^{k}\psi(x)dx=0$, for $k=0,1,...,N$, $N\geq1$ where $\psi$ is the mother wavelet corresponding to $\phi$.
  \item The functions $\left\{f_{l}\right\}_{l=1}^{p}$, are such that $f_{l}\,\in\,L_{\infty}([0,1])$ and $f_{l}\,\in\,W_{\infty}^{m+1}([0,1])\,,\,m \geq N$, where $W_{\infty}^{m}([0,1])$ denotes the space of functions that are $m$-times weakly-differentiable and $f_{l}^{(k)}\,\in\,L_{\infty}([0,1])\,,\, k=1,...,m$.
  \item $\theta_{\phi}(x):=\sum_{k}|\phi(x-k)|$ such that $||\theta_{\phi}||_{\infty}<\infty$.
\end{enumerate}
According to corollary 8.2 \cite{Hardle1998}, if $f\in W_{\infty}^{N+1}([0,1])$ then $||K_{J}f-f||_{\infty}^{p}=\mathcal{O}\left(2^{-pJ(N+1)}\right)\,,\,p\geq1$. Furthermore, assume condition (\textbf{A3}) is satisfied. Define the set of functions:
\begin{equation} \mathcal{F}_{n}=\left\{f:[0,1]^{p}\rightarrow\mathbb{R}\,|\,f(\textbf{x})=\sum_{j=1}^{p}\sum_{k=0}^{2^{J}-1}c_{Jk}^{(j)}\phi_{Jk}^{per}(x_{j})\,;\,J=J(n)\right\}\,, \label{eqTh1:3} \end{equation} %
where $x_{j}\,,j=1,...,p$ corresponds to the $j$-th component of the vector $\textbf{x}\in[0,1]^{p}$. Also, let $\beta_{n}>0$ be a parameter depending on the sample and assume $\mathbb{E}\left[Y^{2}\right]<\infty$. Define $\hat{f}_{J(n)}$ as in (\ref{eqLS:3}) and let $f_{J(n)}=T_{\beta_{n}}\hat{f}_{J(n)}:=\hat{f}_{J(n)}\mathbbm{1}_{\left\{|\hat{f}_{J(n)}|\leq \beta_{n}\right\}}+\text{sign}(\hat{f}_{J(n)})\beta_{n}\mathbbm{1}_{\left\{|\hat{f}_{J(n)}|> \beta_{n}\right\}}$, $\mathcal{K}_{n}=2^{J(n)}$. Assume the following conditions hold:
\begin{enumerate}[(i)]
\item $\beta_{n}\rightarrow\infty$ as $n\rightarrow\infty$.
\item $\frac{\mathcal{K}_{n}\beta_{n}^{4}\log\left(\beta_{n}\right)}{n}\rightarrow0$ as $n\rightarrow\infty$.
\item For some $\delta>0$ as $n\rightarrow\infty$ $\frac{n^{1-\delta}}{\beta_{n}^{4}}\rightarrow\infty$.
\end{enumerate}
Then:
\begin{eqnarray}
\label{eqTh1:1}
\mathop{\lim}\limits_{n\rightarrow\infty} \int \left|f_{J(n)}(\textbf{x})-f_{A}(\textbf{x}) \right|^{2}\mu(d\textbf{x})=0\,\,\,\, \text{(a.s.)}\,, \\
\label{eqTh1:2}
\mathop{\lim}\limits_{n\rightarrow\infty} \mathbb{E}\left\{\int \left|f_{J(n)}(\textbf{x})-f_{A}(\textbf{x}) \right|^{2}\mu(d\textbf{x})\right\}=0\,.
\end{eqnarray}
The corresponding proof can be found in section \ref{proof:Theorem1} of the appendix.

\subsubsection*{Remarks}
\begin{enumerate}[(i)]
\item \label{A7a} Note that the scaling function $\phi(x)$ for the wavelet basis $\left\{ \phi^{per}_{j,k}(x), 0\leq k \leq 2^{j} , j\geq 0\right\}$ is absolutely integrable in $\mathbb{R}$. Therefore, $\int_{\mathbb{R}}|\phi(x)|dx=C_{\phi}<\infty$.
\end{enumerate}

\subsubsection*{Corollary 1} \label{corollary1}
Note that if $|Y|\leq B$, $B<\infty$ (known), to guarantee strong consistency of the least squares estimator it suffices to verify the following conditions are satisfied:
\begin{enumerate}[(a)]
  \item For some $\delta>0$, $n^{1-\delta}\rightarrow\infty$, as $n\rightarrow\infty$.
  \item $\frac{\mathcal{K}_{n}}{n}\rightarrow 0$, as $n\rightarrow\infty$.
\end{enumerate}
\subsubsection*{Remarks and comments} \label{remarks8}

\begin{enumerate}[(i)]
\item This theorem is similar to theorem 10.3 of \cite{Gyorfi2002}. In our case, we investigated the statistical properties possible to be obtained using a wavelet framework, in the set of functions $\mathcal{F}_{n}$ defined by (\ref{eqTh1:3}), and assuming conditions stated in \ref{Th1} for the scaling function $\phi$ hold, when the unknown regression function is additive and given by $m(\textbf{x})=\sum_{j=1}^{p}m_{j}(x_{j})$.
\item From this theorem it is possible to conclude that the estimator defined in (\ref{eqLS:3}) results from the application of the wavelet framework directly to the NESD generated by the observations $\textbf{X}_{1},...,\textbf{X}_{n}$. As was shown, this approach provides good statistical properties which suggests that it is possible to ignore the NESD condition without compromising the robustness and efficiency of the estimator.
\item As was presented, the strong consistency of (\ref{eqLS:3}) relies on parameters $\beta_{n}$ and $\mathcal{K}_{n}=2^{J(n)}$ that need to be selected. In the next section, optimal choices for both are proposed.
\end{enumerate}

\subsection{Convergence rate of the Wavelet-based Least Squares Estimator.}

As was seen in the previous section, theorem 1 shows that the least squares (LS) wavelet-based estimator is strongly consistent for all bounded lebesgue measures in $[0,1]^{p}$ when the set of assumptions for the unknown functions and wavelet basis are satisfied. In this section, we investigate the convergence rates that are possible to attain with this estimator. In particular, we are interested in studying the rate at which:
\begin{equation}
\nonumber
\mathbb{E}\left[\int_{[0,1]^{p}}\left|f_{J(n)}(\textbf{x})-f_{A}(\textbf{x})\right|^{2}\mu(d\textbf{x})\right]\mathop{\longrightarrow}\limits_{n\rightarrow\infty} 0\,,
\end{equation}
where $f_{J(n)}=T_{\beta_{n}}\hat{f}_{J(n)}$ for $\beta_{n}>0$ and $\hat{f}_{J(n)}$ defined as in (\ref{eqLS:3}).
\medskip
Similarly as in the previous section, to investigate the convergence properties of the LS estimator, we use theorem \ref{ThPollard2}, introduced by Pollard (1984), detailed in \ref{previousThms} of the appendix.

\medskip

\subsubsection{Lemma 1}\label{Lemma1}

Suppose an orthonormal basis $\left\{ \phi^{per}_{j,k}(x), k =0,...,2^{j}-1,\, j\geq0\right\}$ which is dense in $\mathbb{L}_{2}(\nu([0,1]))$ for $\nu\in \Upsilon$, where $\Upsilon$ represents the set of bounded lebesgue measures in $[0,1]$. Suppose $\mu$ is a bounded lebesgue measure in $[0,1]^{p}$ and conditions stated in Theorem 1 for the scaling function $\phi$, and assumptions (\textbf{A1})-(\textbf{A4}) defined in \ref{ARM} are satisfied. Define the set of functions $\mathcal{F}_{n}$ as in (\ref{eqTh1:3}). Also, let $\beta_{n}>0$ be a parameter depending on the sample and assume $\mathbb{E}\left[Y^{2}\right]<\infty$. Define $\hat{f}_{J(n)}$ as in (\ref{eqLS:3}) and let $f_{J(n)}=T_{\beta_{n}}\hat{f}_{J(n)}$, let $\mathcal{K}_{n}=p\,2^{J(n)}$. Furthermore, assume the following condition holds:
\begin{enumerate}[(i)]
  \item $\sum_{j=1}^{p}||f_{j}||_{\infty}<L$, for some $L<\beta_{n}$.
\end{enumerate}
Then:
\begin{equation}\label{eqLSlemma1:1}
\mathbb{E}\left[\frac{1}{n}\sum_{i=1}^{n}\left|f_{J(n)}(\textbf{x}_{i})-f_{A}(\textbf{x}_{i}) \right|^{2} \mid \textbf{X}_{1}^{n}\right]\leq \mathop{\min}\limits_{f\in\mathcal{F}_{n}}\left\{||f-f_{A} ||_{n}^{2} \right\}+\frac{\sigma^{2}}{n}\mathcal{K}_{n}\,,
\end{equation}
where $||f||_{n}^{2}=\frac{1}{n}\sum_{i=1}^{n}f(x_{i})^{2}$. The corresponding proof can be found in Appendix \ref{proof:Lemma1}.

\subsubsection{Lemma 2}\label{Lemma2}
Suppose an orthonormal basis $\left\{ \phi^{per}_{j,k}(x), k =0,...,2^{j}-1,\,j\geq0\right\}$ which is dense in $\mathbb{L}_{2}(\nu([0,1]))$ for $\nu\in \Upsilon$,  where $\Upsilon$  represents the set of bounded lebesgue measures in $[0,1]$. Suppose assumptions stated in theorem 1 for the scaling function $\phi$, and conditions (\textbf{A1})-(\textbf{A4}) defined in \ref{ARM} hold. Let the set of functions $\mathcal{F}_{n}$ to be defined as in (\ref{eqTh1:3}).

\medskip
Then it follows:
\begin{equation}\label{eqLSlemma2:1}
\mathop{\inf}\limits_{f\in\mathcal{F}_{n}}\int_{[0,1]^{p}}\left|f(\textbf{x})-f_{A}(\textbf{x}) \right|^{2}\mu(d\textbf{x}) \leq p^{2}\,C_{2}^{2}\,2^{-2(N+1)\,J(n)}\,,
\end{equation}
for a constant $C_{2}>0$, independent of $n,J$. The corresponding proof can be found in Appendix \ref{proof:Lemma2}.

\subsubsection{Theorem 2}\label{Theorem1}
Consider assumptions stated for Lemma 1 and conditions (i)-(iii) for Theorem 1 hold . Define $\hat{f}_{J(n)}$ as in (\ref{eqLS:3}) and let $f_{J(n)}=T_{\beta_{n}}\hat{f}_{J(n)}$, let $\mathcal{K}_{n}=2^{J(n)}$. Then:
\begin{equation}\label{eqLSTheorem2:1}
\mathbb{E}\left[\int_{[0,1]^{p}}\left|f_{J(n)}(\textbf{x})-f_{A}(\textbf{x})\right|^{2}\mu(d\textbf{x})\right] \leq \tilde{C}\max\left\{\beta_{n}^{2},\sigma^{2}\right\}\frac{p\,2^{J(n)}}{n}\left(\log(n)+1\right)
+ 8\,C_{2}^{2}\,p^{2}\,2^{-2(N+1)J(n)}\,,
\end{equation}
for proper constants $\tilde{C}>0$ and $C_{2}>0$ independent of $n,N,p$. The corresponding proof is based on the application of Lemma 1, Lemma 2 and Theorem P2 and can be found in Appendix \ref{proof:Theorem2}.

\subsection{Optimal choice of Estimator parameters $J(n)$ and $\beta_{n}$.}

In this section we propose choices for the parameters $J(n)$ and $\beta_{n}$ used in the estimator. First, we look at the selection of the truncating parameter $\beta_{n}$.

\subsubsection{Lemma 3}\label{Lemma3}

Suppose a model of the form (\ref{eq:LSa1}), with $0<\sigma<\infty$. Assume $\epsilon$ is a sub-gaussian random variable independent of $\textbf{x}$, such that $\mathbb{E}[\epsilon]=0$, $\mathbb{E}[\epsilon^{2}]=1$. Let $\left\{Y_{1},...,Y_{n} \right\}$ be the response observations in the sample $\left\{Y_{i},\textbf{X}_{i} \right\}_{i=1}^{n}$.

\medskip
Then, for $\beta_{n}=4\sigma\sqrt{\log(n)}$ it follows:

\begin{equation}\label{eqLSlemma3:1}
\mathbb{P}\left\{\max\left\{Y_{1},...,Y_{n} \right\}>\beta_{n} \right\}=\mathcal{O}\left(\frac{1}{n}\right)\,,
\end{equation}
which implies that $\mathop{\lim}\limits_{n\rightarrow\infty}\mathbb{P}\left\{\max\left\{Y_{1},...,Y_{n} \right\}>\beta_{n} \right\}\rightarrow0$ at a rate $\frac{1}{n}$. The corresponding proof can be found in Appendix \ref{proof:Lemma3}.

\paragraph*{Remarks}
\begin{enumerate}[(i)]
  \item In practice, the value of $\sigma$ is not known and it can be estimated by the sample variance $\hat{\sigma^{2}}$ of the response. Assuming independence between the random error $\epsilon$ and predictors $\textbf{X}$, this is a suitable choice. However, this in practice could lead to a larger than optimal truncating parameter, since $Var(f(\textbf{x}))\geq \sigma^{2}$.
  \item Another possibility for choosing $\sigma$ could be the one proposed by Donoho and Johnstone (1994), which is given by $\hat{\sigma}=\frac{\text{median}\left(\left\{\left|\hat{d}_{J-1,k} \right|\,:\,k=0,...,2^{J}-1 \right\}\right)}{0.6745}$, where $\hat{d}_{J-1,k}$ are the discrete wavelet coefficients resulting from the DWT of the observed response $\textbf{y}$.
\end{enumerate}

\subsubsection{Lemma 4}\label{Lemma4}

Define $\hat{f}_{J(n)}$ as in (\ref{eqLS:3}) and let $f_{J(n)}=T_{\beta_{n}}\hat{f}_{J(n)}$. Suppose assumptions for Theorem 2 hold. Then, for $\beta_{n}=4\sigma\sqrt{\log(n)}$ ($n\geq2$), setting the multiresolution level $J(n)$ as:

\begin{equation}\label{eqLSlemma4:1}
J^{*}(n)=\mathcal{K}_{1}+\frac{1}{2\,N+3}\log_{2}\left(\frac{n}{\log(n)\left(\log(n)+1\right)} \right)\,,
\end{equation}
minimizes the $\mathbb{L}_{2}$-risk upper bound given by (\ref{eqLSTheorem2:1}) and guarantees the strong consistency of the estimator $\hat{f}_{J(n)}$, where $\mathcal{K}_{1}=\frac{1}{2\,N+3}\log_{2}\left(\frac{(N+1)\,C_{2}^{2}\,p}{\tilde{C}\,\sigma^{2}}\right)$.

\medskip

The proof of this Lemma consists in the minimization of the upper bound (\ref{eqLSTheorem2:1}) with respect $\mathcal{\tilde{K}}_{n}=2^{J(n)}$. Note that the minimun exists and is unique due to the convexity of the objective function defined by (\ref{eqLSTheorem2:1}). Similarly, it is possible to guarantee conditions (i)-(iii) of Theorem 1 are satisfied since:
\begin{equation}
\nonumber
\mathop{\lim}\limits_{n\rightarrow\infty}\left(\frac{\log(n)^{\gamma+t}}{n^{\gamma}} \right)=0\,,
\end{equation}
$\forall\, \gamma \geq1\,,t>0$ (integers) which can be proved by applying L'Hopital's rule.

\subsubsection{Theorem 3}\label{Theorem3}

Suppose assumptions and results for Theorems 1, 2 and Lemmas 3 and 4 hold. Then, the estimator defined by in (\ref{eqLS:3}), and $f_{J(n)}=T_{\beta_{n}}\hat{f}_{J(n)}$ attains the following convergence rate for the $\mathbb{L}_{2}$-risk:
\begin{equation}\label{eqLSTheorem3:1}
\mathbb{E}\left[\int_{[0,1]^{p}}\left|f_{J(n)}(\textbf{x})-f_{A}(\textbf{x})\right|^{2}\mu(d\textbf{x})\right] \leq \tilde{K}\,\left(\frac{\beta_{n}^{2}\,\log(n)}{n} \right)^{\frac{2\gamma}{2\gamma+1}}\,,
\end{equation}
where $\gamma=N+1$, $\tilde{K}=\left(2\,\gamma\,\tilde{C}\,p \right)^{\frac{2\gamma}{\gamma+1}}\left(8\,C_{2}^{2}\,p^{2}\right)^{\frac{1}{2\gamma+1}}$.
\medskip

From (\ref{eqLSTheorem3:1}), it is possible to distinguish 2 cases:

\begin{enumerate}[(i)]
  \item From Corollary 1,  if $|Y|\leq B$, $B<\infty$ (known) it follows:
\begin{equation}\label{eqLSTheorem3:2}
\mathbb{E}\left[\left\|f_{J(n)}-f_{A}\right\|^{2}\right]=\mathcal{O}\left(\frac{\log(n)}{n} \right)^{\frac{2\gamma}{2\gamma+1}}\,.
\end{equation}
  \item If the upper bound of $Y$ is not known, choosing $\beta_{n}$ as in Lemma 3, the convergence rate takes the form of:
\begin{equation}\label{eqLSTheorem3:3}
\mathbb{E}\left[\left\|f_{J(n)}-f_{A}\right\|^{2}\right]=\mathcal{O}\left(\frac{\log(n)^{2}}{n} \right)^{\frac{2\gamma}{2\gamma+1}}\,.
\end{equation}
\end{enumerate}
The proof of the above assertions follows from Lemmas 3 and 4 applied to Theorem 2.

\paragraph*{Remarks}
\begin{enumerate}[(i)]
  \item Note that results (i) and (ii) show that the LS estimator defined by $\hat{f}_{J(n)}$ as in (\ref{eqLS:3}) does not suffer from the curse of dimensionality. Moreover, its convergence rate is optimal up to a logarithmic factor. This implies that is possible to apply the wavelet framework directly over non-equally spaced designs without compromising desirable statistical properties such as strong consistency and optimal $\mathbb{L}_{2}$ convergence rates.
\end{enumerate}

\newpage

\subsection{Simulation Study}\label{Simulation_study}

In the last section, we introduced a wavelet based least squares estimator for the additive regression model and proved its statistical properties. In this section, we investigate the performance of $\hat{f}_{n}(\textbf{x})$ with respect to the AMSE (Average Mean Squared Error) of estimation, via a simulation study. For this objective, we choose a set of exemplary baseline functions that combine different smoothness and spectral properties and are aimed to challenge the estimation process.
\medskip

To simplify the implementation, we select specific functions that are supported in the [0,1] and also satisfy assumptions (\textbf{A1})-(\textbf{A4}). These functions are defined as follows:

\begin{eqnarray}
\nonumber
f_{1}(x)  =  \frac{1}{\sqrt{2}\sin\left(2\pi\,x \right)} & & f_{2}(x) =  1-4\,|x-\frac{1}{2}| \\
\nonumber
f_{3}(x) = -\cos\left(4\pi\,x +1\right) & &  f_{4}(x) =  8\,\left(x-\frac{1}{2} \right)^{2}-\frac{2}{3} \\
\nonumber
f_{5}(x) =  \frac{1}{\sqrt{2}}\cos\left(2\pi\,x \right) & & f_{6}(x) = \frac{1}{\sqrt{2}}\cos\left(4\pi\,x \right)  \\
\nonumber
f_{7}(x) =  -0.5275 + 4\,e^{-500(x-0.23)^{2}} + 2\,e^{-2000(x-0.33)^{2}} & & + 4\,e^{-8000(x-0.47)^{2}} + 3\,e^{-16000(x-0.69)^{2}} + e^{-32000(x-0.83)^{2}} \\
\nonumber
f_{8}(x) = 0.2\,\cos\left(4\pi\,x +1\right) + 0.1\,\cos\left(24\pi\,x +1\right) & & f_{9}(x) = -0.1744 + 2\,x^{3}\,\mathbf{1}_{(0.5<x\leq0.8)} + 2\,\left(x-1 \right)^{3}\mathbf{1}_{(0.8<x\leq1)}
\end{eqnarray}
\begin{figure}[!htb]
\begin{center}
 \includegraphics[width=0.7\textwidth]{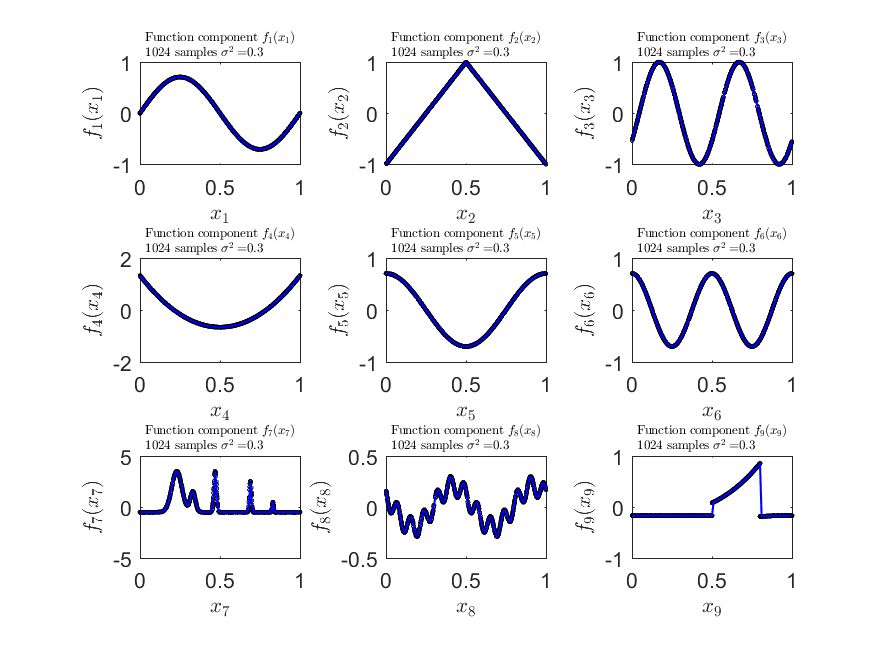}
\caption{Graphic representation of the testing functions for the Additive Model.}
\label{fig:Functions}
\end{center}
\end{figure}
In this simulation study, we investigate the performance of the estimator for different sample sizes, noise variances $\sigma^2$, wavelet filters and distribution of the predictors $\textbf{X}$. To quantify the estimator performance, we use the following global error measure:

\begin{equation}\label{eq:Sim1}
\hat{RMSE}=\sqrt(\frac{1}{B}\sum_{b=1}^{B}\frac{1}{n}\sum_{i=1}^{n}\left(f(\textbf{x}_{i})-\hat{f}_{n,b}(\textbf{x}_{i}) \right)^{2})\,,
\end{equation}
where $B$ is the number of replications of the experiment and $n$ is the number of samples. For all experiments we choose $B=200$.
\medskip

While implementing the simulations, we considered the following settings in a \ttmat{matlab}-based script:

\begin{enumerate}[(i)]
\item We generated independent random numbers $\left\{\textbf{X}_{i}\right\}_{i=1}^{N}$ from the $\left\{\mathcal{U}[0,1]\right\}^{9}$ and $\left\{Beta(\frac{3}{2},\frac{3}{2})\right\}^{9}$ joint distributions (satisfying assumptions (\textbf{A1})-(\textbf{A4})), and constructed the model defined in (\ref{eq:LSa1}).
\item For the noise variance, we used $\sigma^{2}=0.75$ and $\sigma^{2}=0.25$, which produced different signal-to-noise ratios (SNR) used to assess the estimator robustness against noisy observations.
\item For the computation of the least squares estimator, we chose the scaling functions generated by the wavelet filters \verb"Coiflets" and \verb"Daubechies" with 24 and 4 coefficients respectively.
\item Both of the chosen wavelet filters satisfy conditions 1-6 listed in theorem 1. For Coiflets, the wavelet is near symmetric with compact support and has $N/3$ vanishing moments ($N$ is the number of filter taps); in the case of Daubechies, the wavelet does not have the near-symmetry property but it has compact support and $N$ vanishing moments.
\item For the evaluation of the scaling functions $\phi_{Jk}^{per}$ (and construction of matrix $B$) we used Daubechies-Lagarias's algorithm.
\item The multiresolution level $J$ was chosen to be $J(n)=1+\lfloor\log_{2}(n)-\log_{2}\left(\log(n)\,(\log(n)+1)\right)\rfloor$.
\item The truncating parameter $\beta_{n}$ was selected using the proposition detailed in remark (ii) of Lemma 3.
\end{enumerate}

\subsubsection{Simulation Results.}\label{SimResults}

\begin{figure}[!htb]
   \centering
   \begin{subfigure}[b]{0.5\textwidth}
       \includegraphics[width=\textwidth,height=70mm]{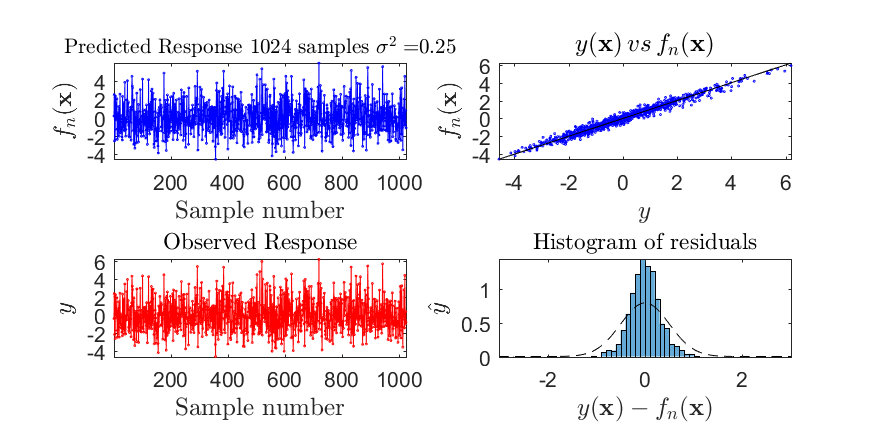}
       \caption{}
       \label{fig:Yhat_1024UnifDesign_Sigma2_025}
   \end{subfigure}\hfill
   \begin{subfigure}[b]{0.5\textwidth}
       \includegraphics[width=\textwidth,height=70mm]{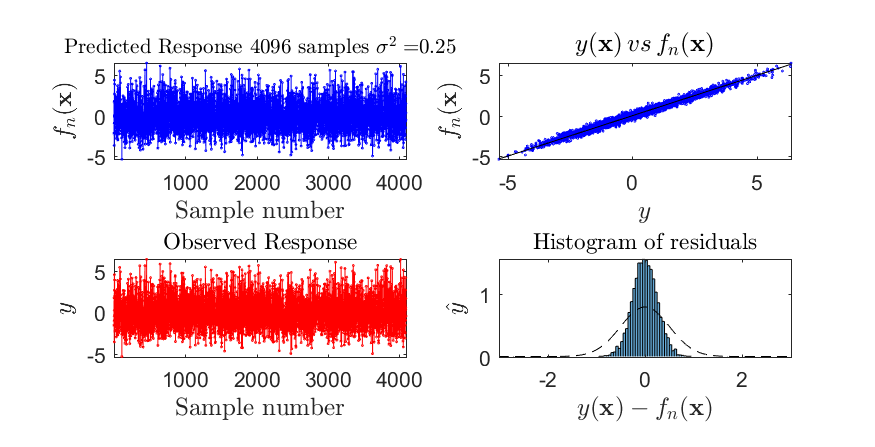}
       \caption{}
       \label{fig:Yhat_4096UnifDesign_Sigma2_025}
    \end{subfigure}\hfill
\caption{Estimation summary plots using Uniform Design and Coiflets filter.}
\label{fig:EstPlotsUnif}
\end{figure}

\begin{figure}[!htb]
   \centering
   \begin{subfigure}[b]{0.25\textwidth}
       \includegraphics[width=\textwidth]{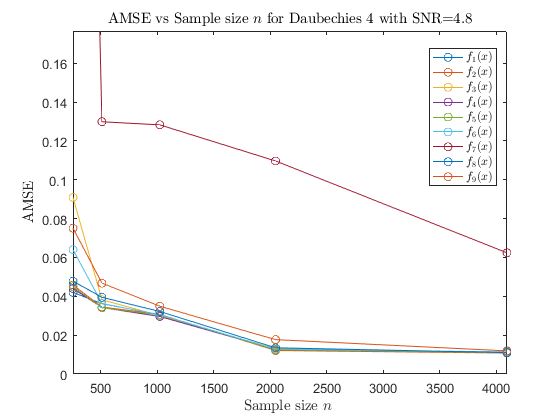}
       \caption{Daubechies filter, $\sigma^{2}=0.25$}
       \label{fig:AMSE_Daubechies_Unif_sigma1v2}
    \end{subfigure}\hfill
    \begin{subfigure}[b]{0.25\textwidth}
       \includegraphics[width=\textwidth]{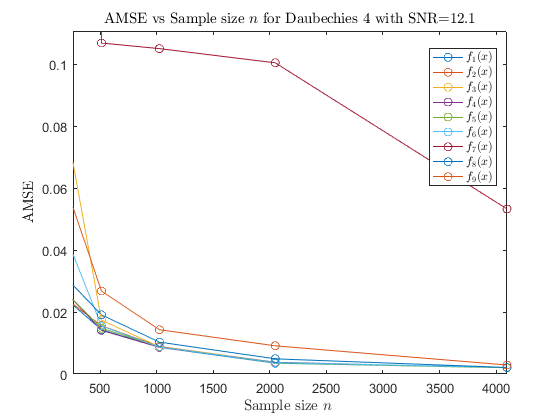}
       \caption{Daubechies filter, $\sigma^{2}=0.75$}
       \label{fig:AMSE_Daubechies_Unif_sigma2v2}
    \end{subfigure}\hfill
    \begin{subfigure}[b]{0.25\textwidth}
       \includegraphics[width=\textwidth]{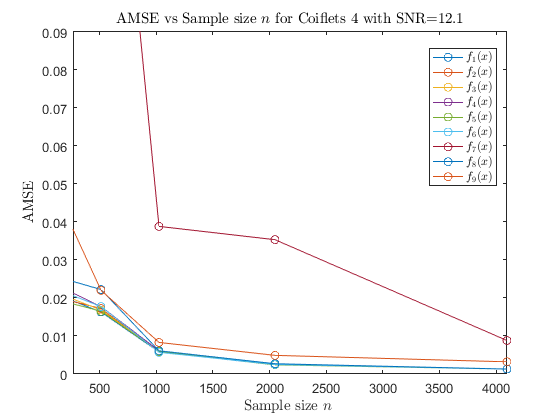}
       \caption{Coiflets filter, $\sigma^{2}=0.25$}
       \label{fig:AMSE_Coif_Unif_sigma1v2}
    \end{subfigure}\hfill
    \begin{subfigure}[b]{0.25\textwidth}
       \includegraphics[width=\textwidth]{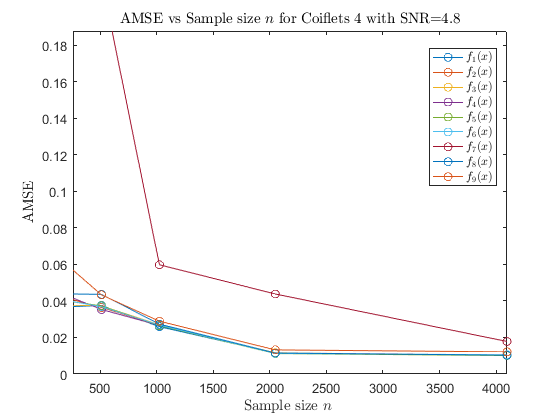}
       \caption{Daubechies filter, $\sigma^{2}=0.75$}
       \label{fig:AMSE_Coif_Unif_sigma2v2}
    \end{subfigure}
\caption{RMSE results for Uniform Design using Daubechies and Coiflets filter, for values of $\sigma^{2}=0.25\,,\,0.75$.}
\label{fig:UnifDaub}
\end{figure}

\begin{figure}[!htb]
   \centering
      \begin{subfigure}[b]{0.25\textwidth}
       \includegraphics[width=\textwidth]{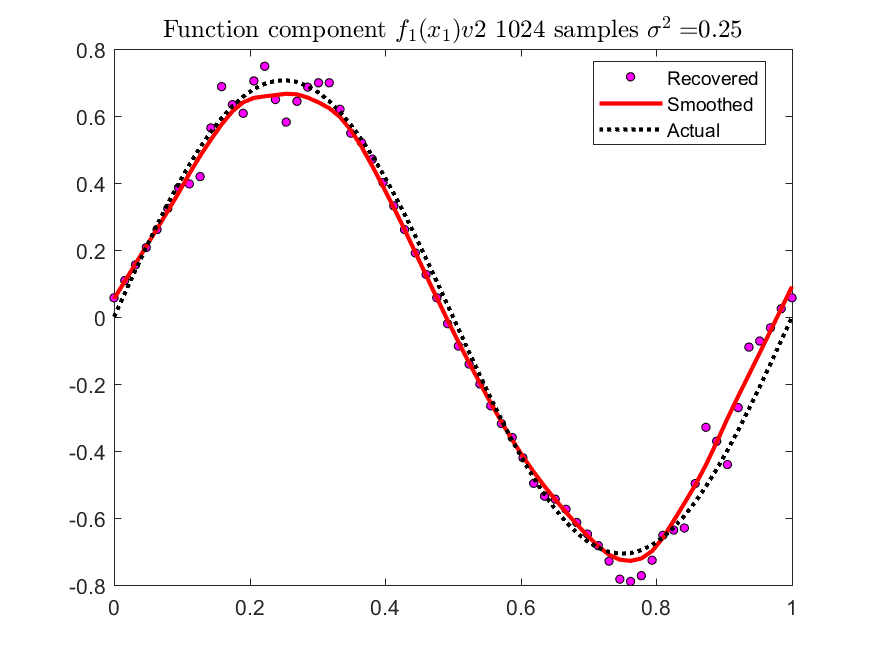}
       \caption{}
       \label{fig:f1V2_1024_UnifDesign_Sigma_025}
    \end{subfigure}\hfill
    \begin{subfigure}[b]{0.25\textwidth}
       \includegraphics[width=\textwidth]{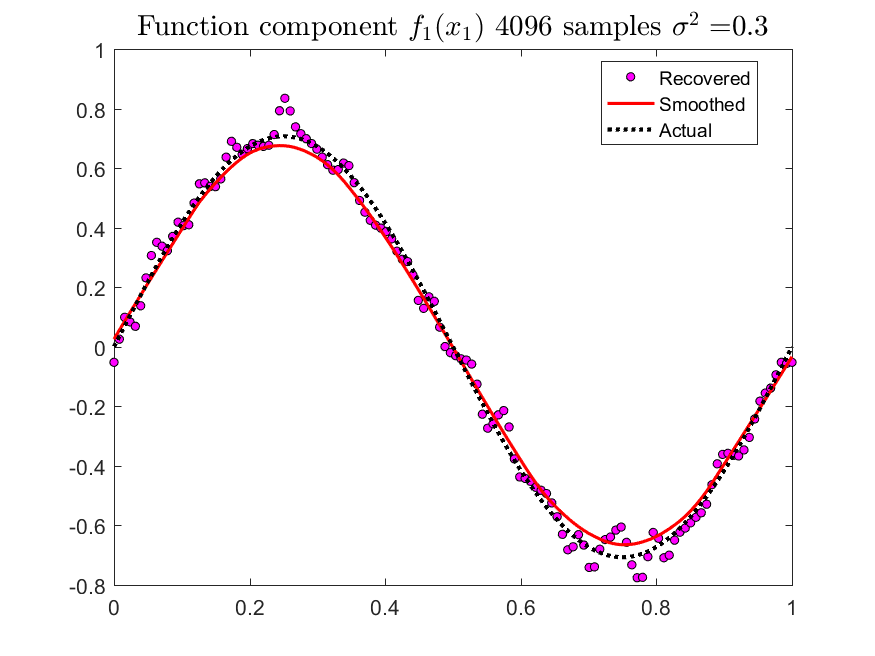}
       \caption{}
       \label{fig:f1V2_4096_UnifDesign_Sigma_03}
    \end{subfigure}\hfill
    \begin{subfigure}[b]{0.25\textwidth}
       \includegraphics[width=\textwidth]{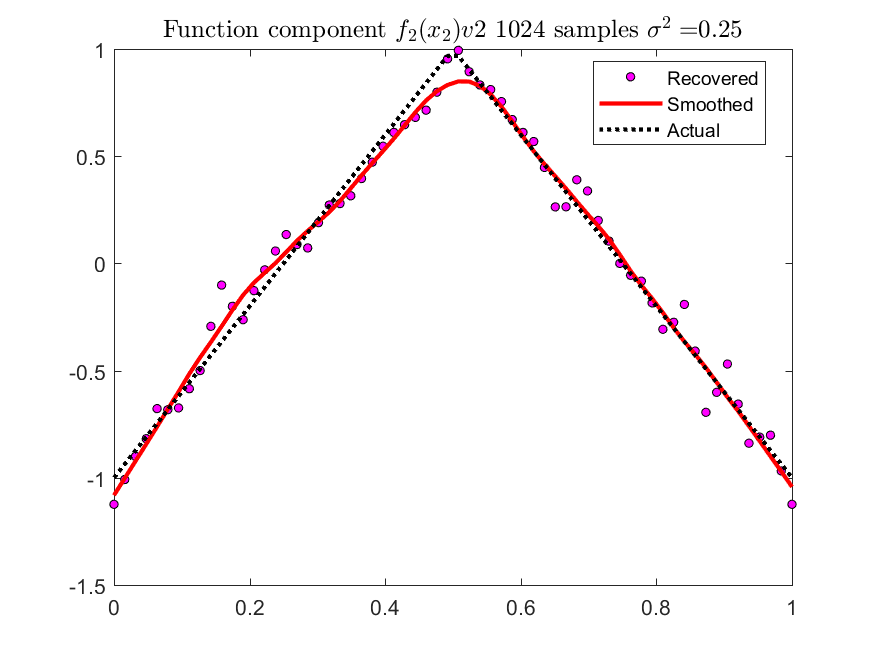}
       \caption{}
       \label{fig:f2V2_1024_UnifDesign_Sigma_025}
    \end{subfigure}\hfill
    \begin{subfigure}[b]{0.25\textwidth}
       \includegraphics[width=\textwidth]{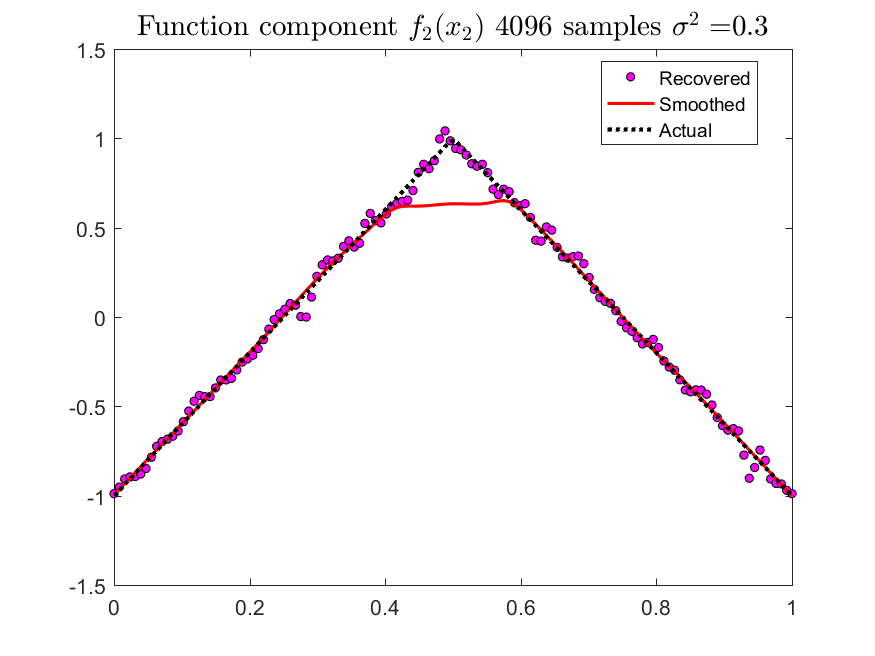}
       \caption{}
       \label{f2V2_4096_UnifDesign_Sigma_03}
    \end{subfigure}
\caption{Estimated $f_{1}(x)$ and $f_{2}(x)$ using Uniform Design and Coiflets filter.}
\label{fig:F1UnifCoif}
\end{figure}

\begin{figure}[!htb]
   \centering
   \begin{subfigure}[b]{0.25\textwidth}
       \includegraphics[width=\textwidth]{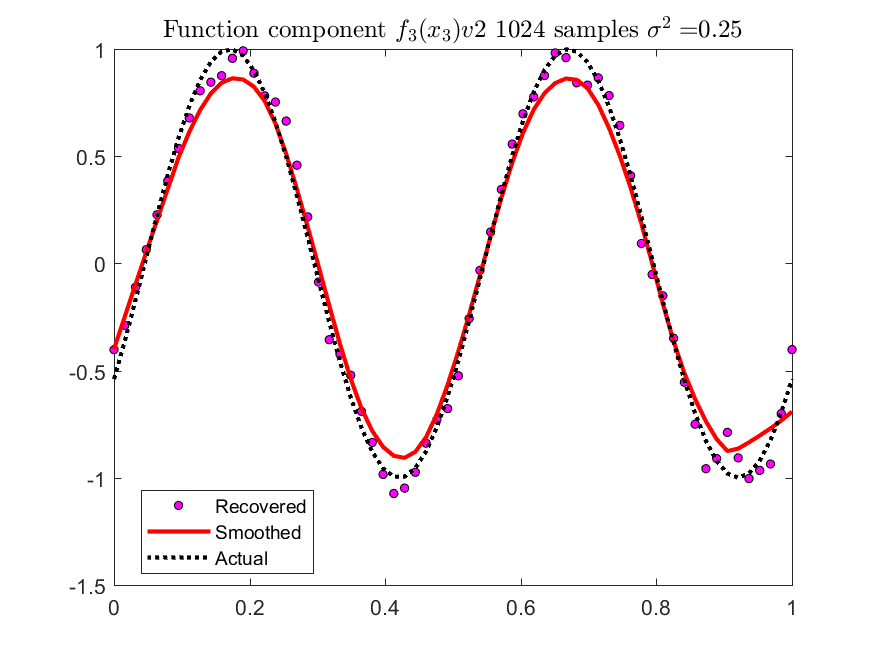}
       \caption{}
       \label{fig:f3V2_1024_UnifDesign_Sigma_025}
    \end{subfigure}\hfill
    \begin{subfigure}[b]{0.25\textwidth}
       \includegraphics[width=\textwidth]{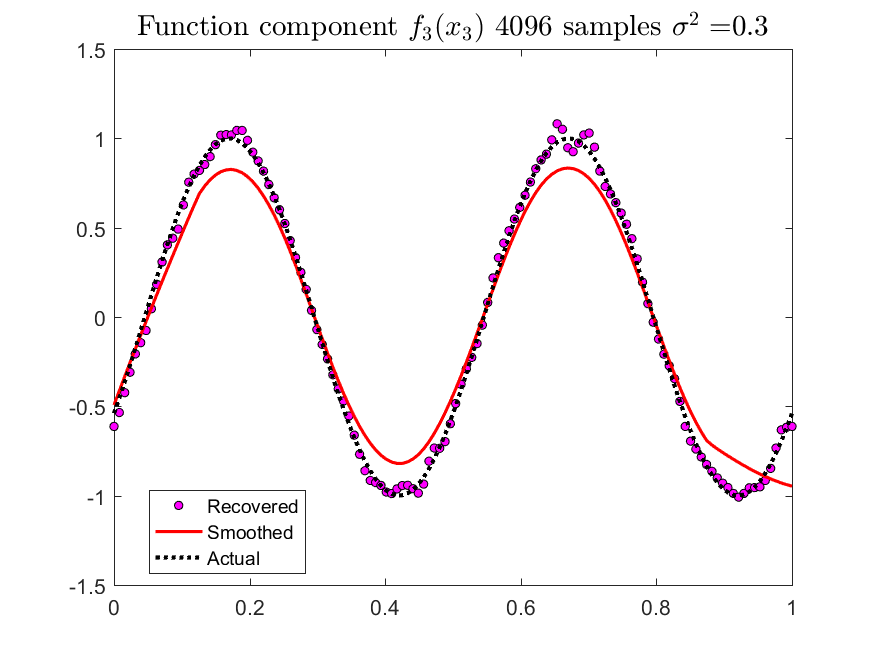}
       \caption{}
       \label{f3V2_4096_UnifDesign_Sigma_03}
    \end{subfigure}\hfill
    \begin{subfigure}[b]{0.25\textwidth}
       \includegraphics[width=\textwidth]{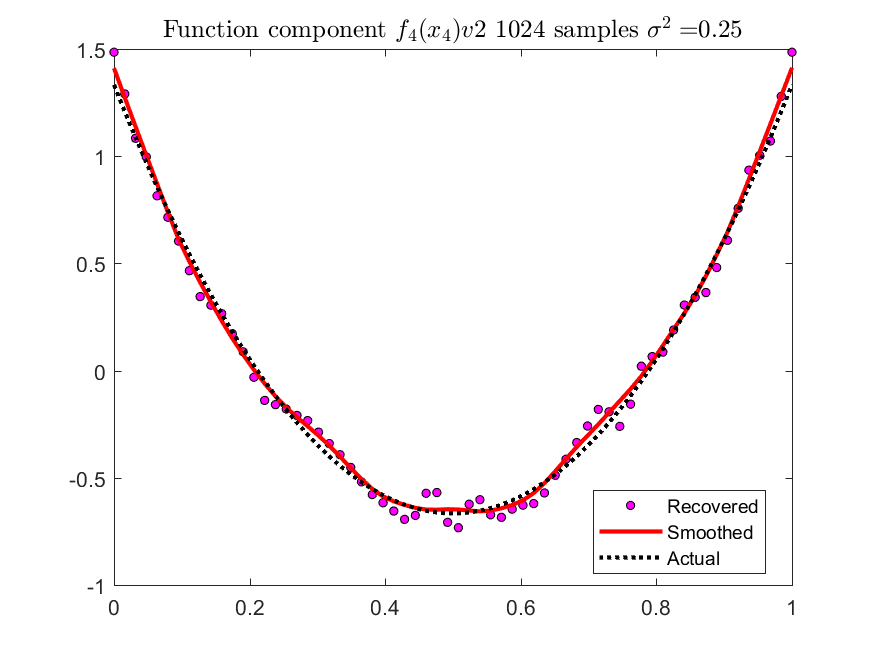}
       \caption{}
       \label{fig:f4V2_1024_UnifDesign_Sigma_025}
    \end{subfigure}\hfill
    \begin{subfigure}[b]{0.25\textwidth}
       \includegraphics[width=\textwidth]{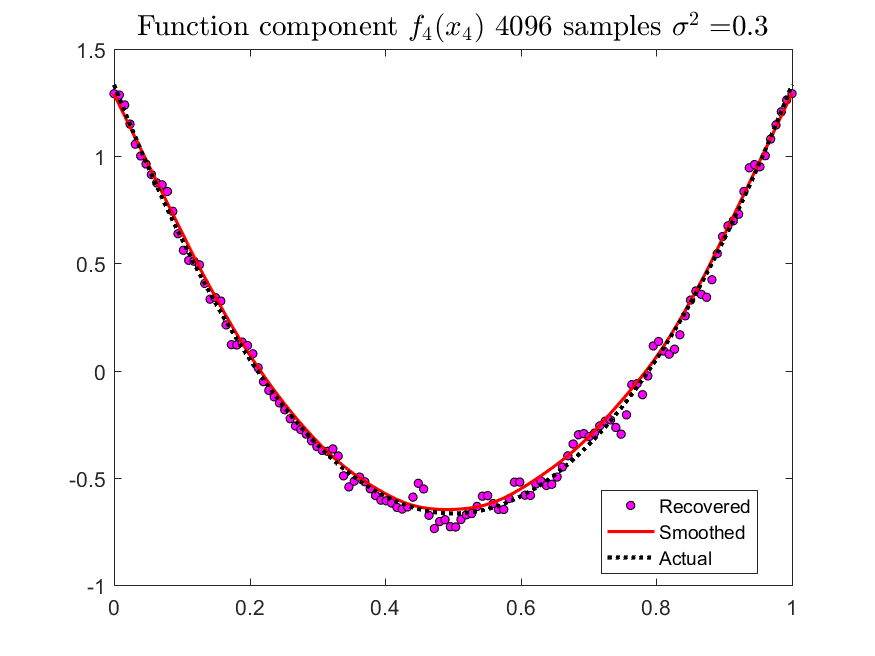}
       \caption{}
       \label{f4V2_4096_UnifDesign_Sigma_03}
    \end{subfigure}
\caption{Estimated $f_{3}(x)$ and $f_{4}(x)$ using Uniform Design and Coiflets filter.}
\label{fig:F3UnifCoif}
\end{figure}

\begin{figure}[!htb]
   \centering
   \begin{subfigure}[b]{0.25\textwidth}
       \includegraphics[width=\textwidth]{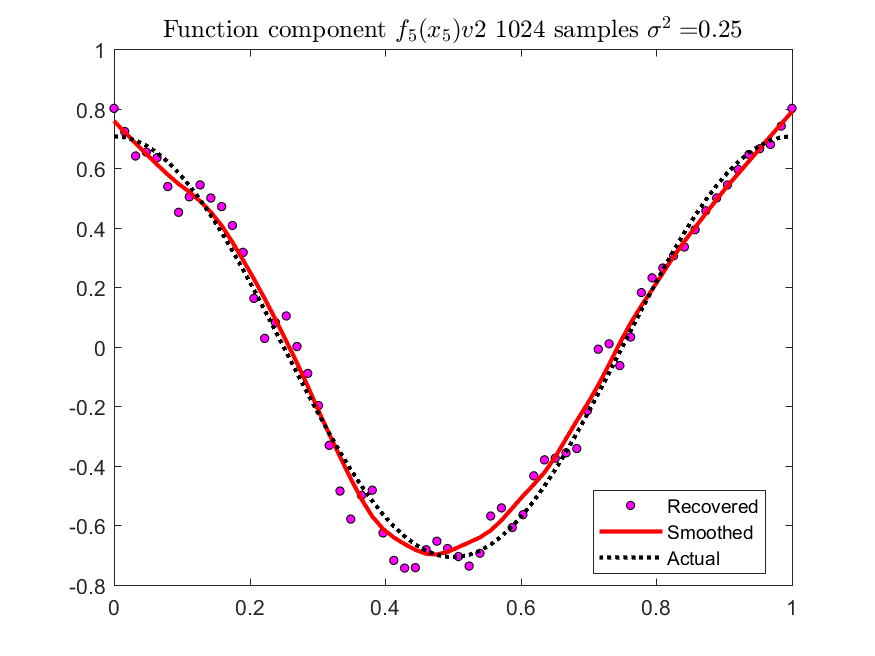}
       \caption{}
       \label{fig:f5V2_1024_UnifDesign_Sigma_025}
    \end{subfigure}\hfill
    \begin{subfigure}[b]{0.25\textwidth}
       \includegraphics[width=\textwidth]{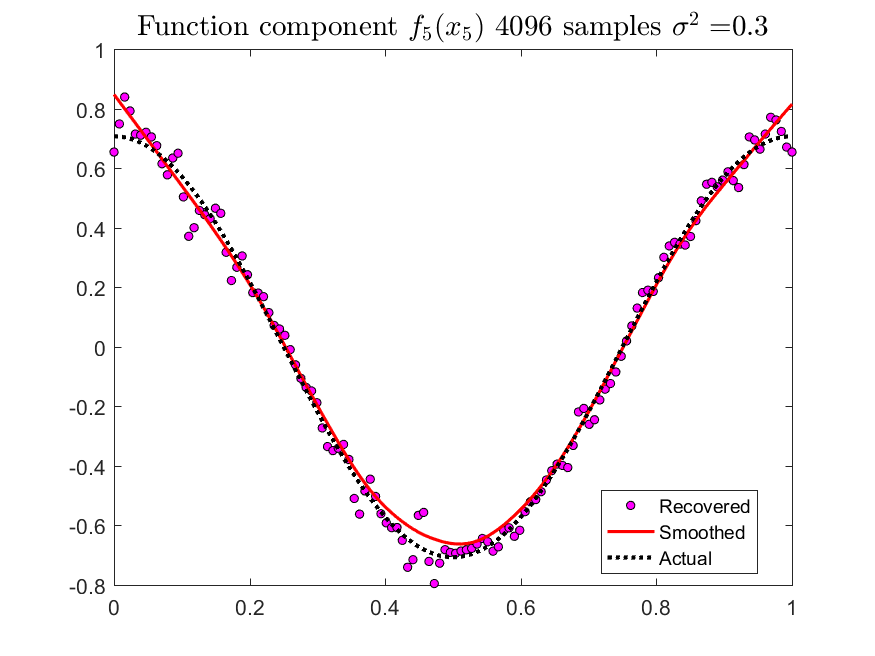}
       \caption{}
       \label{f5V2_4096_UnifDesign_Sigma_03}
    \end{subfigure}\hfill
    \begin{subfigure}[b]{0.25\textwidth}
       \includegraphics[width=\textwidth]{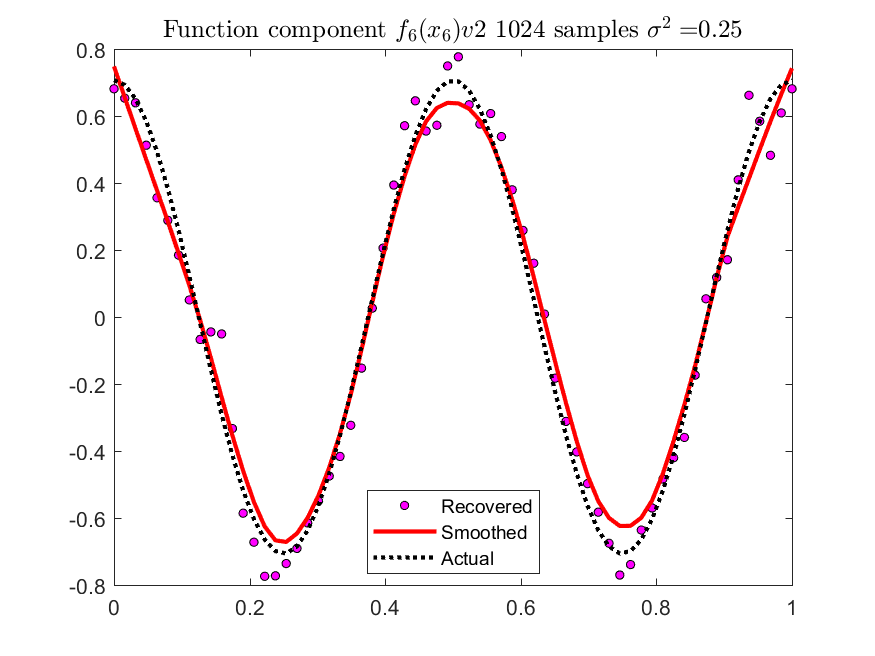}
       \caption{}
       \label{fig:f6V2_1024_UnifDesign_Sigma_025}
    \end{subfigure}\hfill
    \begin{subfigure}[b]{0.25\textwidth}
       \includegraphics[width=\textwidth]{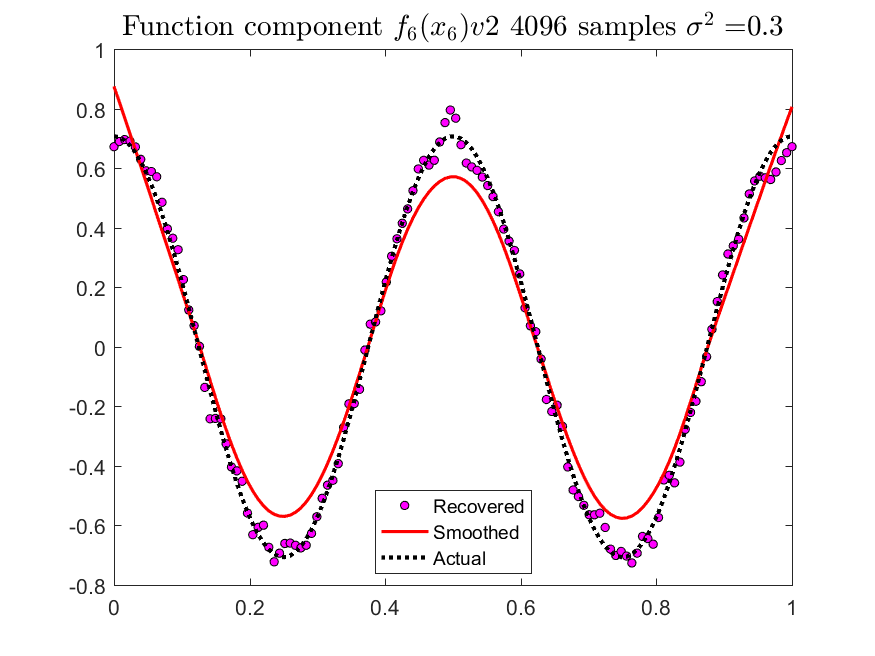}
       \caption{}
       \label{f6V2_4096_UnifDesign_Sigma_03}
    \end{subfigure}
\caption{Estimated $f_{5}(x)$ and $f_{6}(x)$ using Uniform Design and Coiflets filter.}
\label{fig:F5UnifCoif}
\end{figure}

\begin{figure}[!htb]
   \centering
   \begin{subfigure}[b]{0.25\textwidth}
       \includegraphics[width=\textwidth]{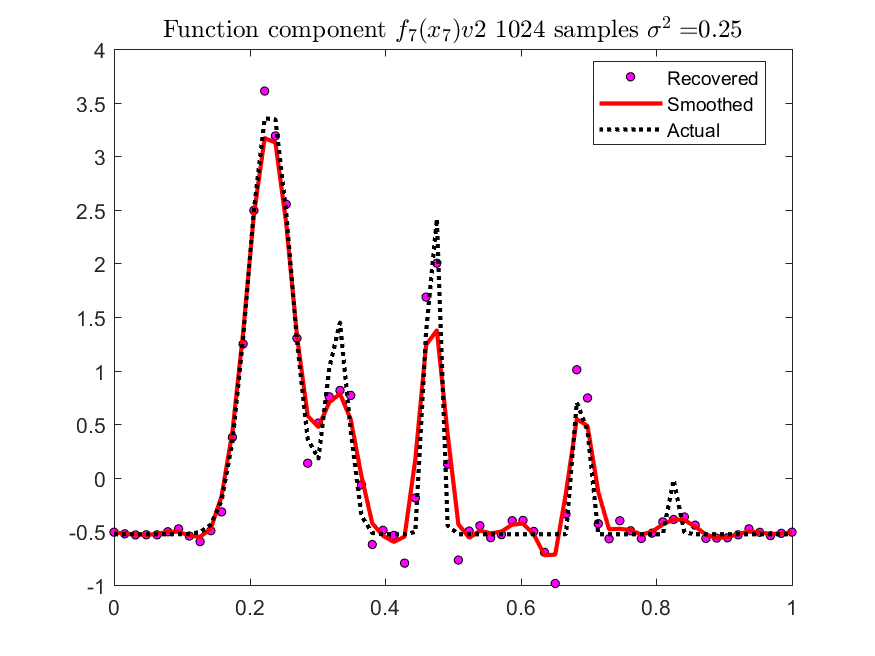}
       \caption{}
       \label{fig:f7V2_1024_UnifDesign_Sigma_025}
    \end{subfigure}\hfill
    \begin{subfigure}[b]{0.25\textwidth}
       \includegraphics[width=\textwidth]{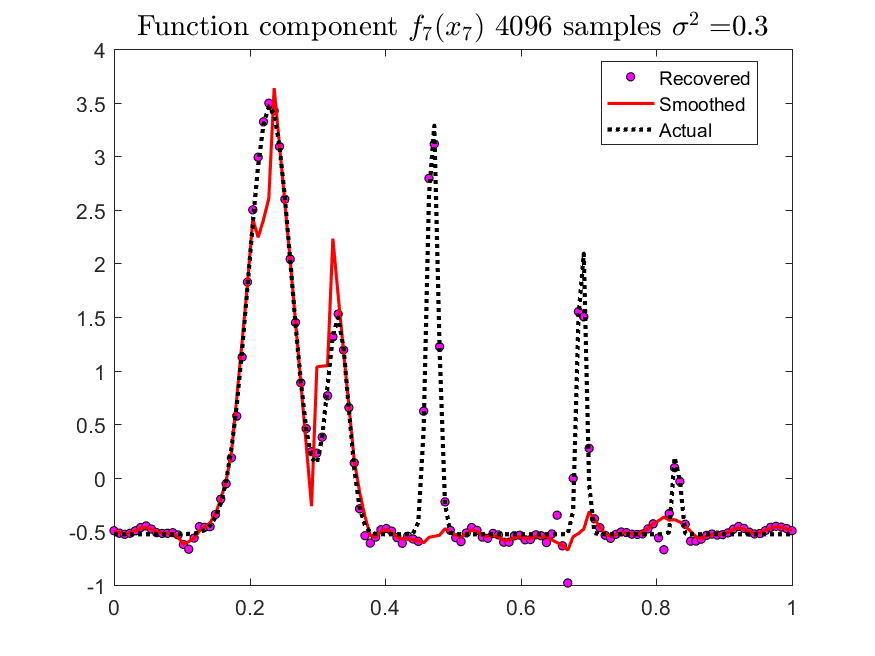}
       \caption{}
       \label{f7V2_4096_UnifDesign_Sigma_03}
    \end{subfigure}\hfill
    \begin{subfigure}[b]{0.25\textwidth}
       \includegraphics[width=\textwidth]{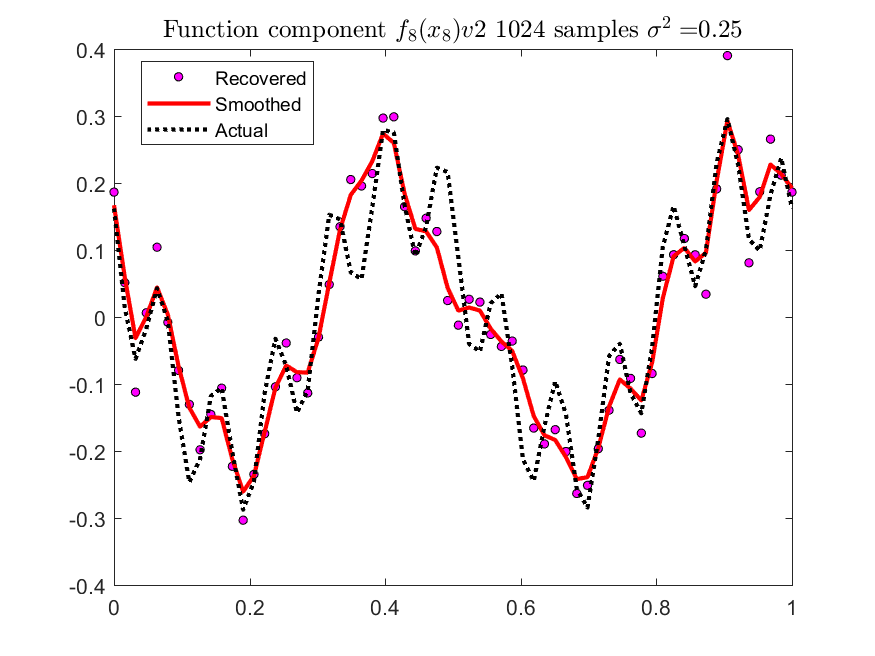}
       \caption{}
       \label{fig:f8V2_1024_UnifDesign_Sigma_025}
    \end{subfigure}\hfill
    \begin{subfigure}[b]{0.25\textwidth}
       \includegraphics[width=\textwidth]{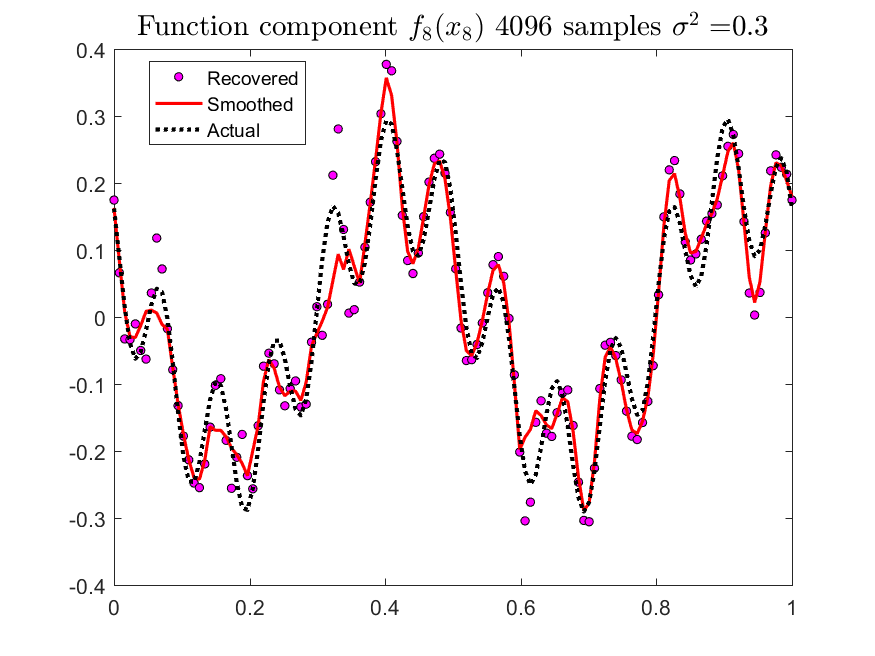}
       \caption{}
       \label{f8V2_4096_UnifDesign_Sigma_03}
    \end{subfigure}
\caption{Estimated $f_{7}(x)$ and $f_{8}(x)$ using Uniform Design and Coiflets filter.}
\label{fig:F7UnifCoif}
\end{figure}

\begin{figure}[!htb]
   \centering
   \begin{subfigure}[b]{0.3\textwidth}
   \centering
       \includegraphics[width=\textwidth]{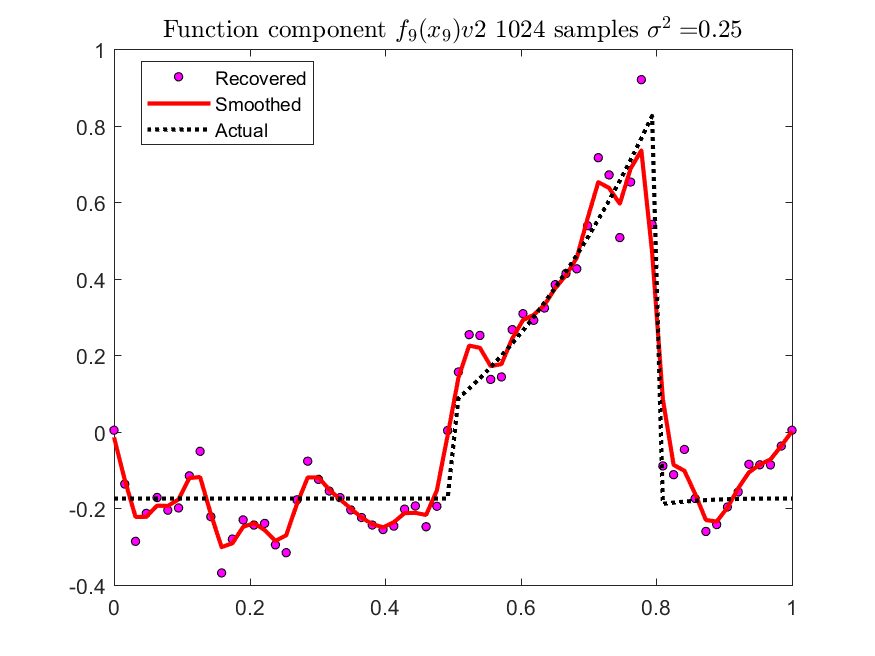}
       \caption{}
       \label{fig:f9V2_1024_UnifDesign_Sigma_025}
    \end{subfigure}
    \begin{subfigure}[b]{0.3\textwidth}
    \centering
       \includegraphics[width=\textwidth]{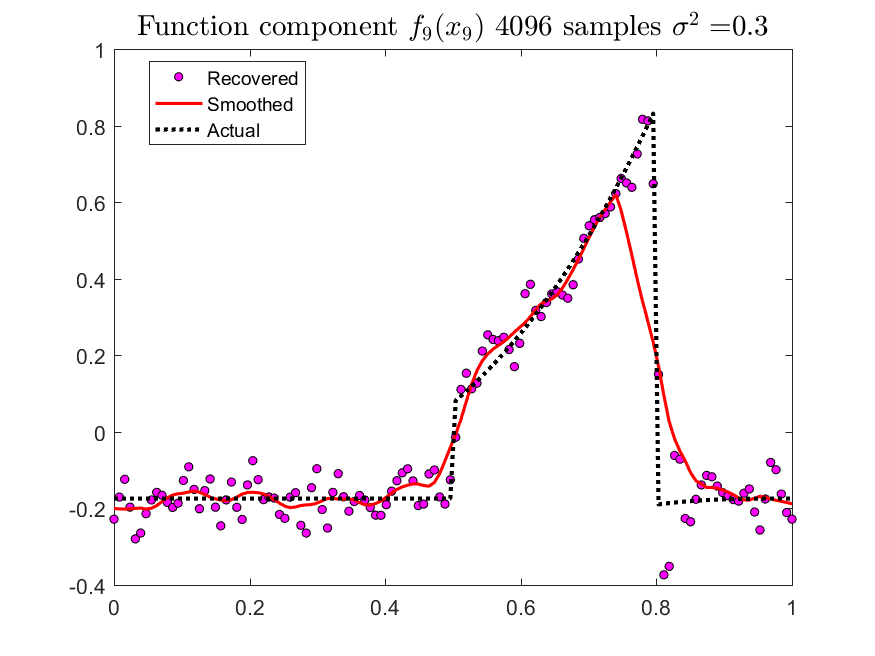}
       \caption{}
       \label{fig:f9V2_4096_UnifDesign_Sigma_03}
    \end{subfigure}
\caption{Estimated $f_{9}(x)$ using Uniform Design and Coiflets filter.}
\label{fig:F9UnifCoif}
\end{figure}

\begin{figure}[!htb]
   \centering
   \begin{subfigure}[b]{0.5\textwidth}
       \includegraphics[width=\textwidth]{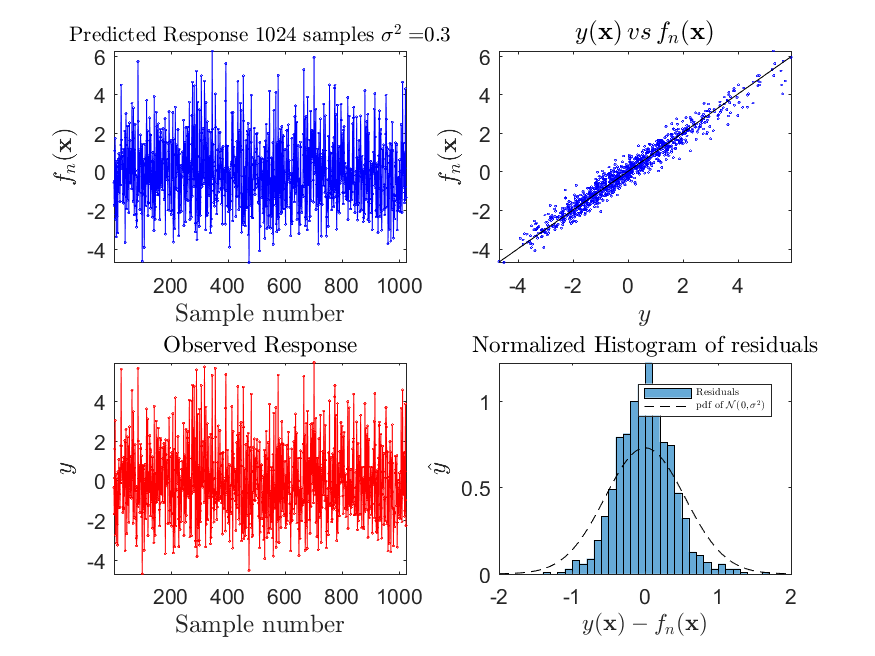}
       \caption{}
       \label{fig:Yhat_1024BetaDesign_Sigma2_03}
   \end{subfigure}\hfill
   \begin{subfigure}[b]{0.5\textwidth}
       \includegraphics[width=\textwidth]{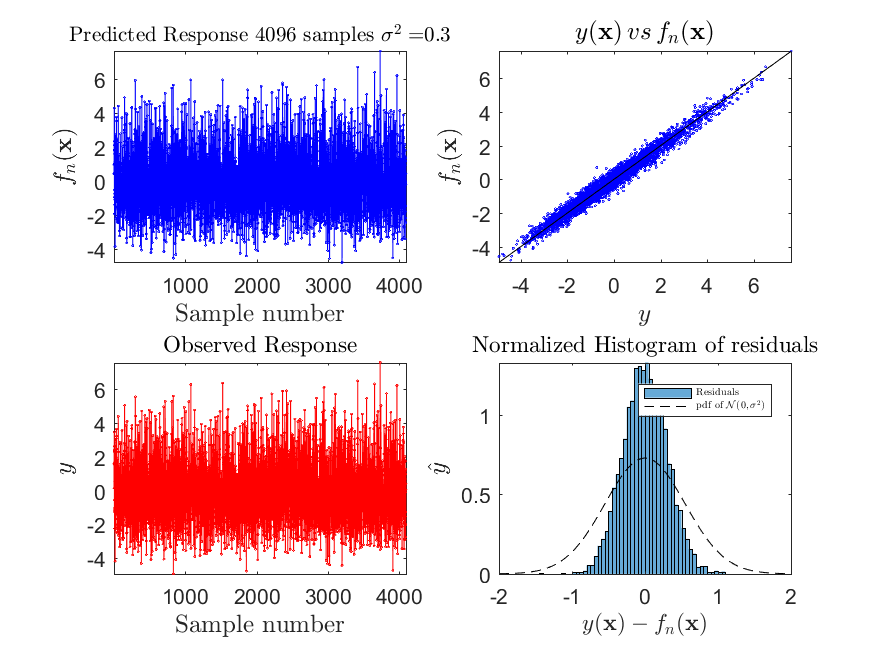}
       \caption{}
       \label{fig:Yhat_4096BetaDesign_Sigma2_03}
    \end{subfigure}\hfill
\caption{Estimation summary plots using Beta Design and Coiflets filter.}
\label{fig:EstPlotsBeta}
\end{figure}

\begin{figure}[!htb]
   \centering
   \begin{subfigure}[b]{0.25\textwidth}
       \includegraphics[width=\textwidth]{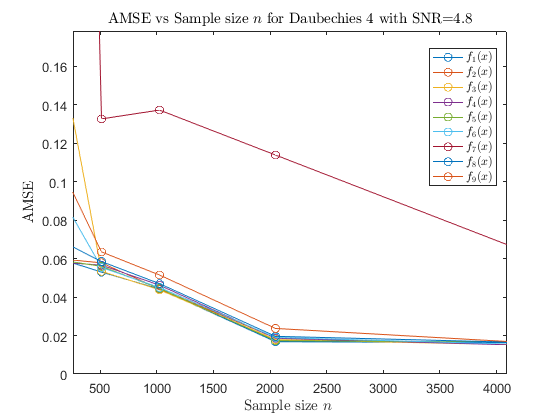}
       \caption{Daubechies filter, $\sigma^{2}=0.25$}
       \label{fig:AMSE_Daub_Beta_sigma1_v2}
    \end{subfigure}\hfill
    \begin{subfigure}[b]{0.25\textwidth}
       \includegraphics[width=\textwidth]{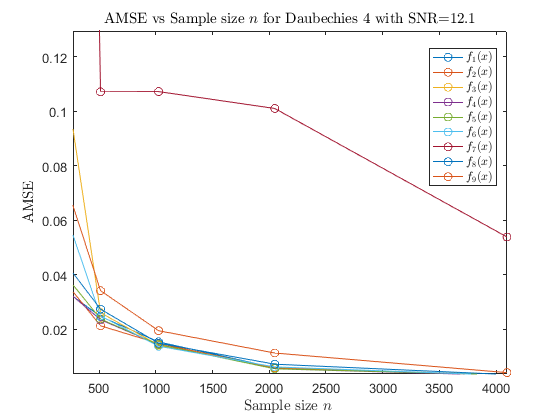}
       \caption{Daubechies filter, $\sigma^{2}=0.75$}
       \label{fig:AMSE_Daub_Beta_sigma2_v2}
    \end{subfigure}\hfill
    \begin{subfigure}[b]{0.25\textwidth}
       \includegraphics[width=\textwidth]{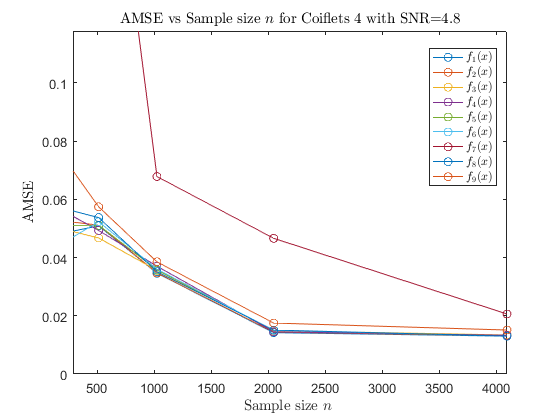}
       \caption{Coiflets filter, $\sigma^{2}=0.25$}
       \label{fig:AMSE_Coiflets_Beta_sigma1_v2}
    \end{subfigure}\hfill
    \begin{subfigure}[b]{0.25\textwidth}
       \includegraphics[width=\textwidth]{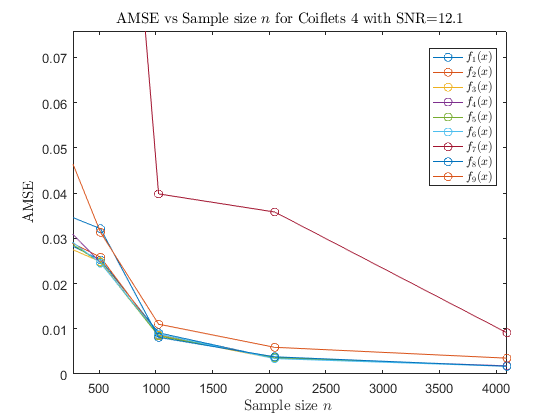}
       \caption{Coiflets filter, $\sigma^{2}=0.75$}
       \label{fig:AMSE_Coiflets_Beta_sigma2_v2}
    \end{subfigure}
\caption{RMSE results for Beta Design using Daubechies and Coiflets filters, for values of $\sigma^{2}=0.25\,,\,0.75$..}
\label{fig:BetaDaub}
\end{figure}

\begin{figure}[!htb]
   \centering
   \begin{subfigure}[b]{0.25\textwidth}
       \includegraphics[width=\textwidth]{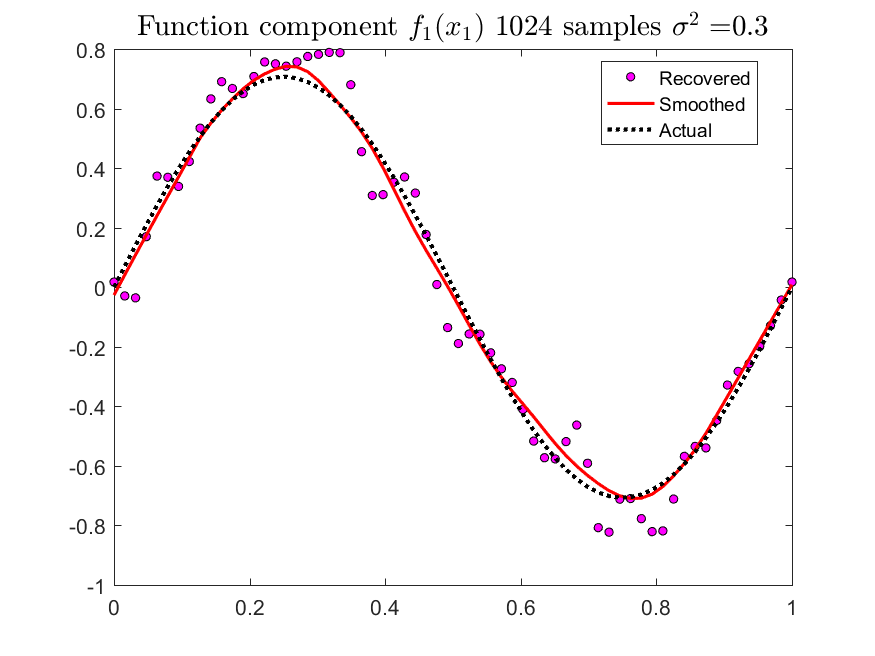}
       \caption{}
       \label{fig:f1V2_1024_BetaDesign_Sigma_03}
   \end{subfigure}\hfill
   \begin{subfigure}[b]{0.25\textwidth}
       \includegraphics[width=\textwidth]{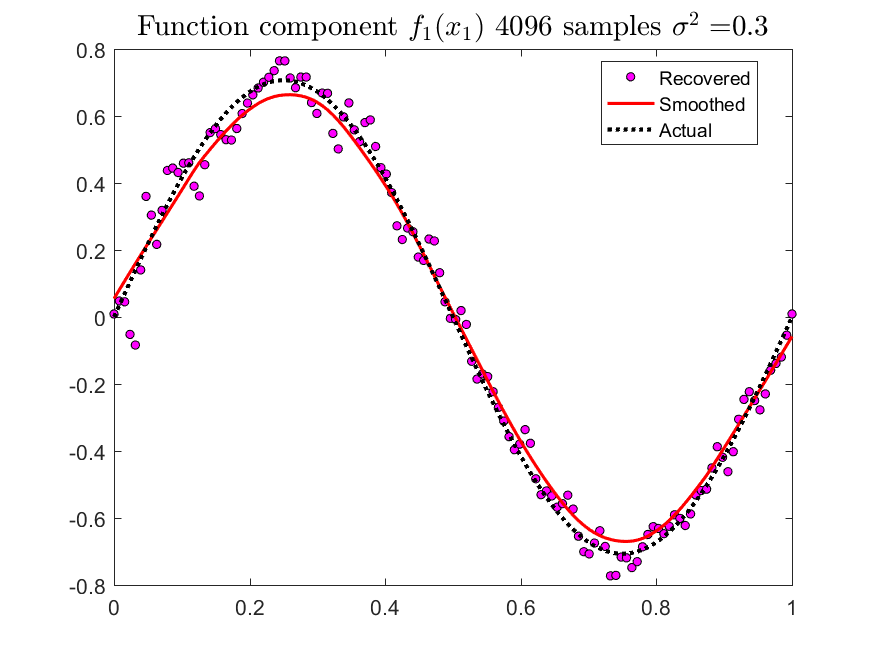}
       \caption{}
       \label{fig:f1V2_4096_BetaDesign_Sigma_03}
    \end{subfigure}\hfill
    \begin{subfigure}[b]{0.25\textwidth}
       \includegraphics[width=\textwidth]{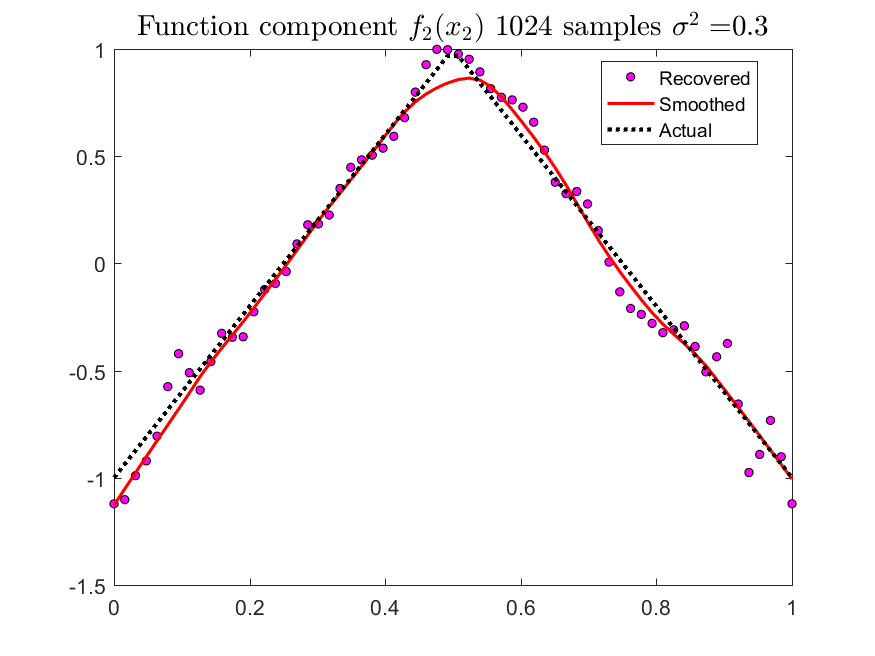}
       \caption{}
       \label{fig:f2V2_1024_BetaDesign_Sigma_03}
   \end{subfigure}\hfill
   \begin{subfigure}[b]{0.25\textwidth}
       \includegraphics[width=\textwidth]{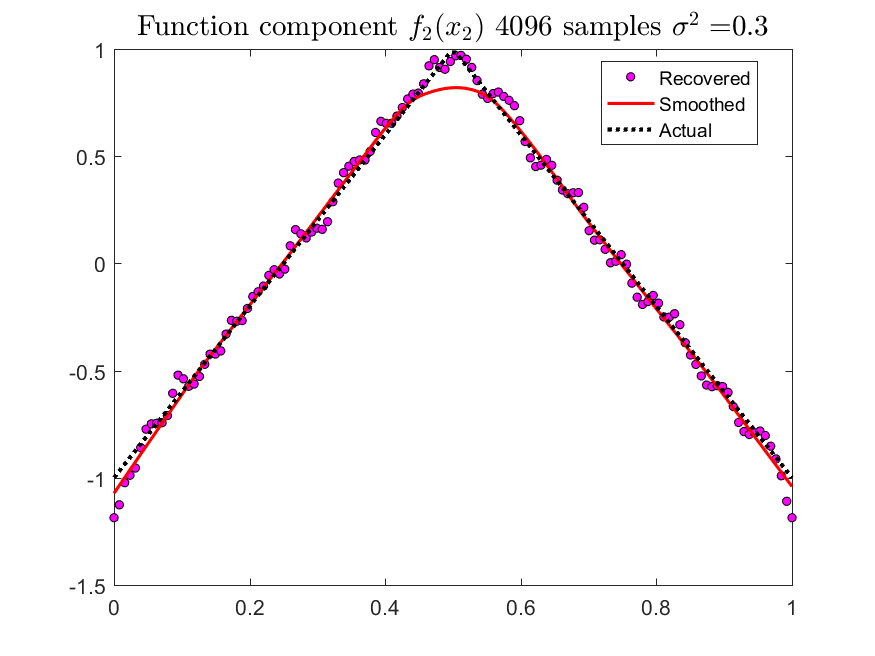}
       \caption{}
       \label{fig:f2V2_4096_BetaDesign_Sigma_03}
    \end{subfigure}
\caption{Estimated $f_{1}(x)$ and $f_{2}(x)$ using Beta Design and Coiflets filter.}
\label{fig:F1BetaCoif}
\end{figure}

\begin{figure}[!htb]
   \centering
   \begin{subfigure}[b]{0.25\textwidth}
       \includegraphics[width=\textwidth]{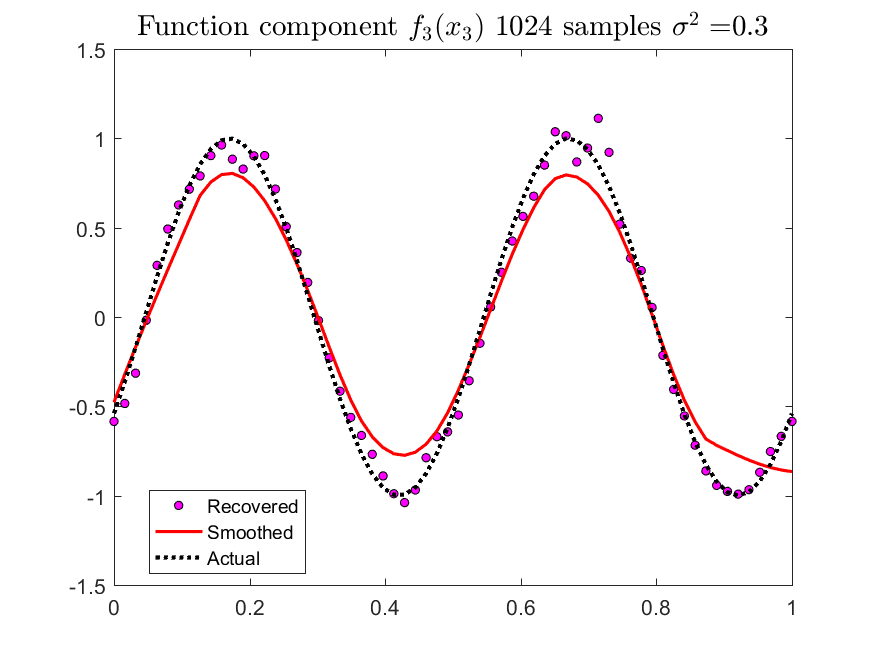}
       \caption{}
       \label{fig:f3V2_1024_BetaDesign_Sigma_03}
   \end{subfigure}\hfill
   \begin{subfigure}[b]{0.25\textwidth}
       \includegraphics[width=\textwidth]{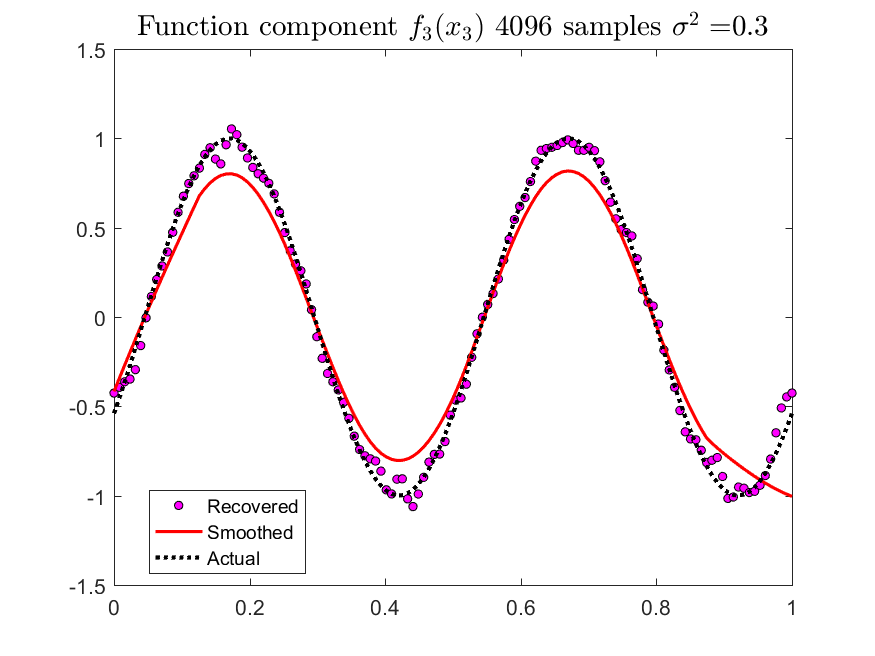}
       \caption{}
       \label{fig:f3V2_4096_BetaDesign_Sigma_03}
    \end{subfigure}\hfill
    \begin{subfigure}[b]{0.25\textwidth}
       \includegraphics[width=\textwidth]{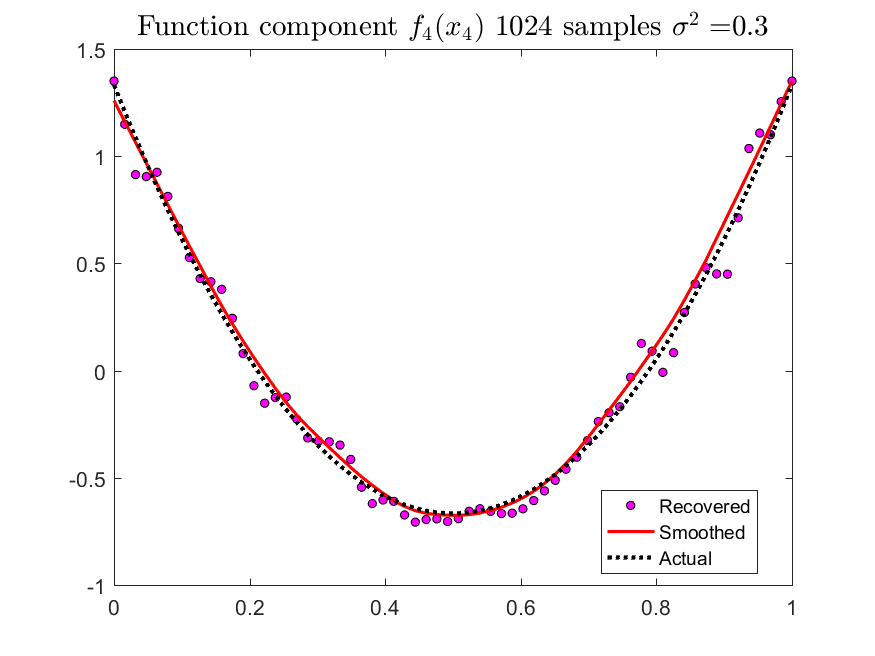}
       \caption{}
       \label{fig:f4V2_1024_BetaDesign_Sigma_03}
   \end{subfigure}\hfill
   \begin{subfigure}[b]{0.25\textwidth}
       \includegraphics[width=\textwidth]{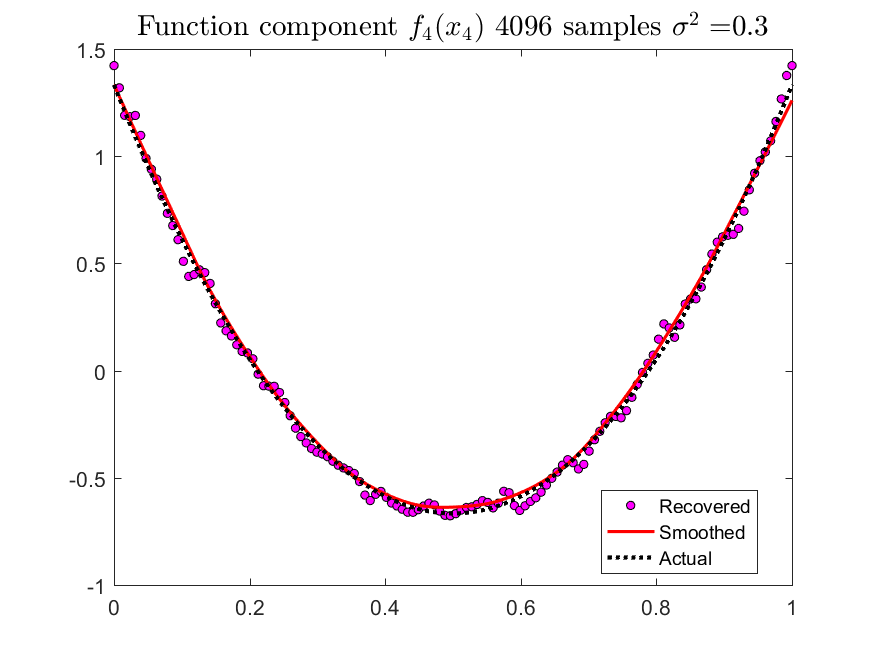}
       \caption{}
       \label{fig:f4V2_4096_BetaDesign_Sigma_03}
    \end{subfigure}
\caption{Estimated $f_{3}(x)$ and $f_{4}(x)$ using Beta Design and Coiflets filter.}
\label{fig:F3BetaCoif}
\end{figure}

\begin{figure}[!htb]
   \centering
   \begin{subfigure}[b]{0.25\textwidth}
       \includegraphics[width=\textwidth]{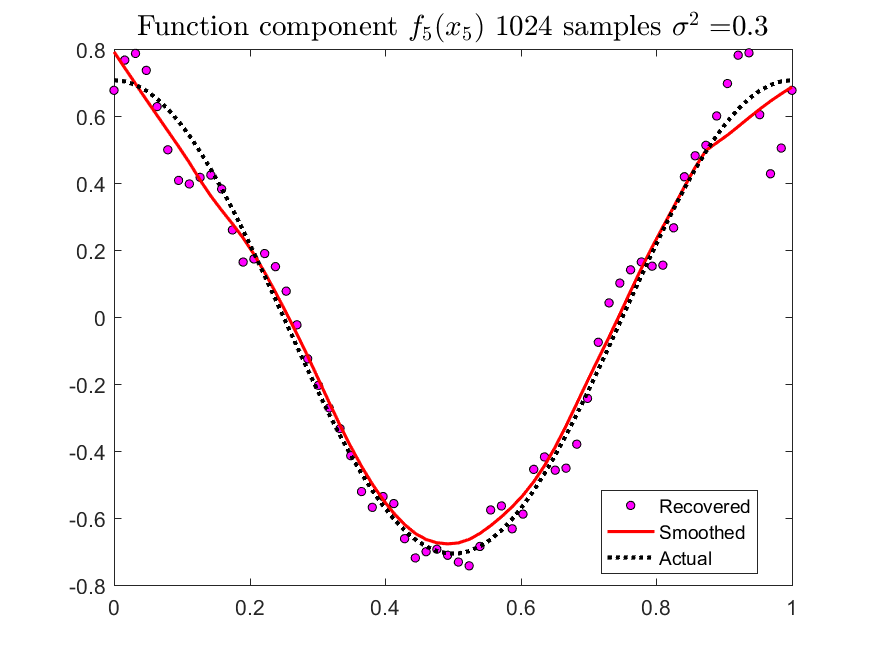}
       \caption{}
       \label{fig:f5V2_1024_BetaDesign_Sigma_03}
   \end{subfigure}\hfill
   \begin{subfigure}[b]{0.25\textwidth}
       \includegraphics[width=\textwidth]{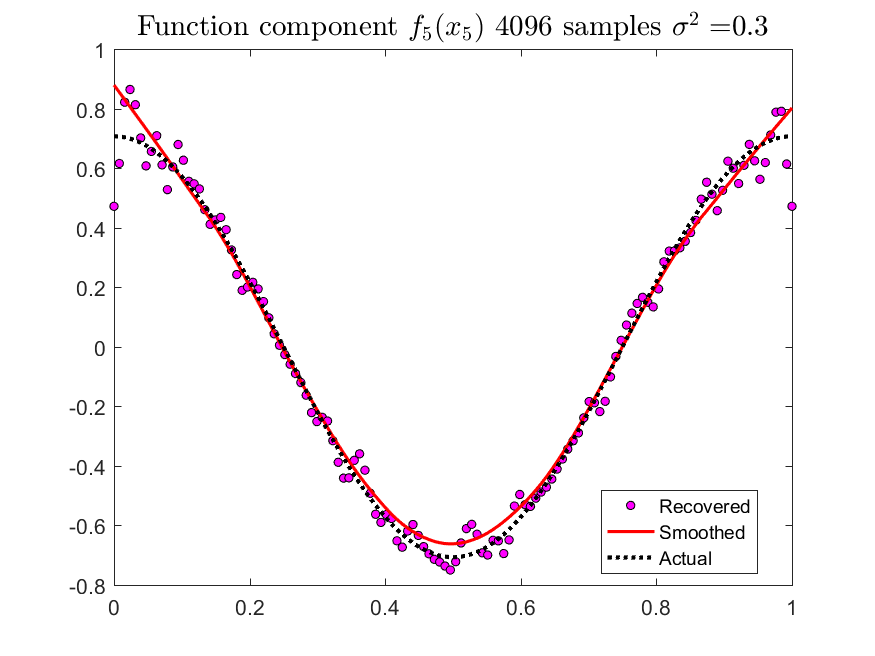}
       \caption{}
       \label{fig:f5V2_4096_BetaDesign_Sigma_03}
    \end{subfigure}\hfill
    \begin{subfigure}[b]{0.25\textwidth}
       \includegraphics[width=\textwidth]{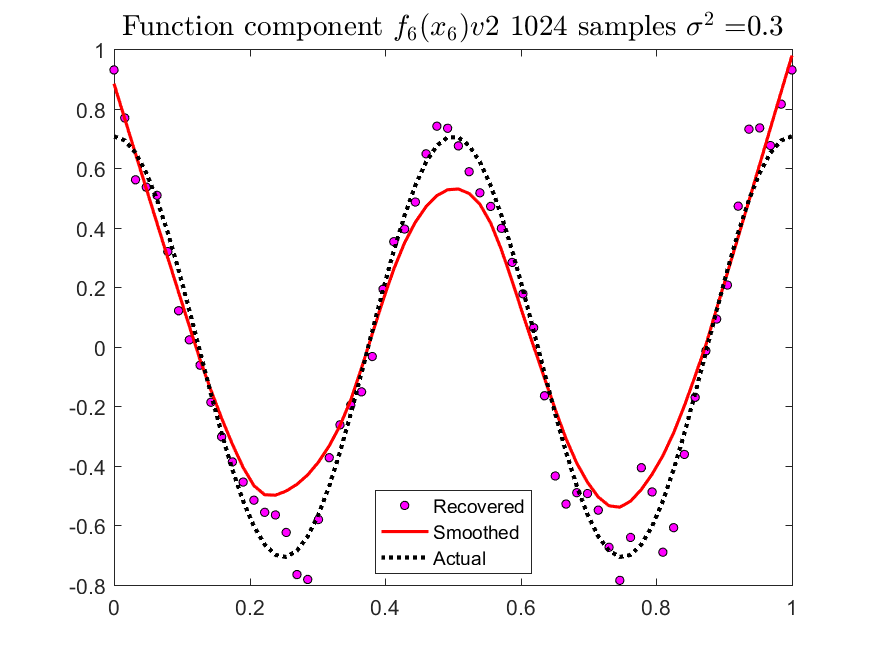}
       \caption{}
       \label{fig:f6V2_1024_BetaDesign_Sigma_03}
   \end{subfigure}\hfill
   \begin{subfigure}[b]{0.25\textwidth}
       \includegraphics[width=\textwidth]{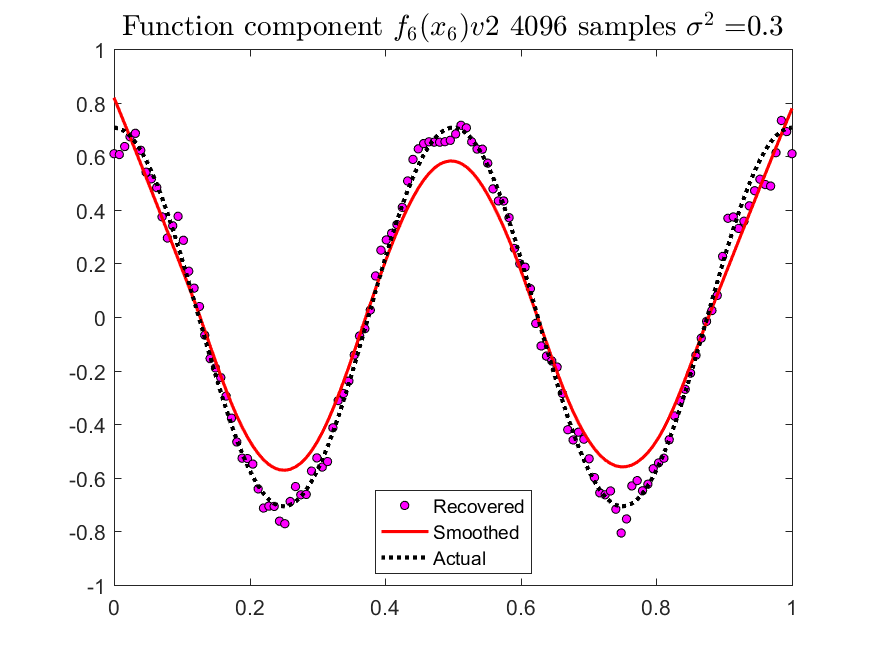}
       \caption{}
       \label{fig:f6V2_4096_BetaDesign_Sigma_03}
    \end{subfigure}
\caption{Estimated $f_{5}(x)$ and $f_{6}(x)$ using Beta Design and Coiflets filter.}
\label{fig:F5BetaCoif}
\end{figure}

\begin{figure}[!htb]
   \centering
   \begin{subfigure}[b]{0.25\textwidth}
       \includegraphics[width=\textwidth]{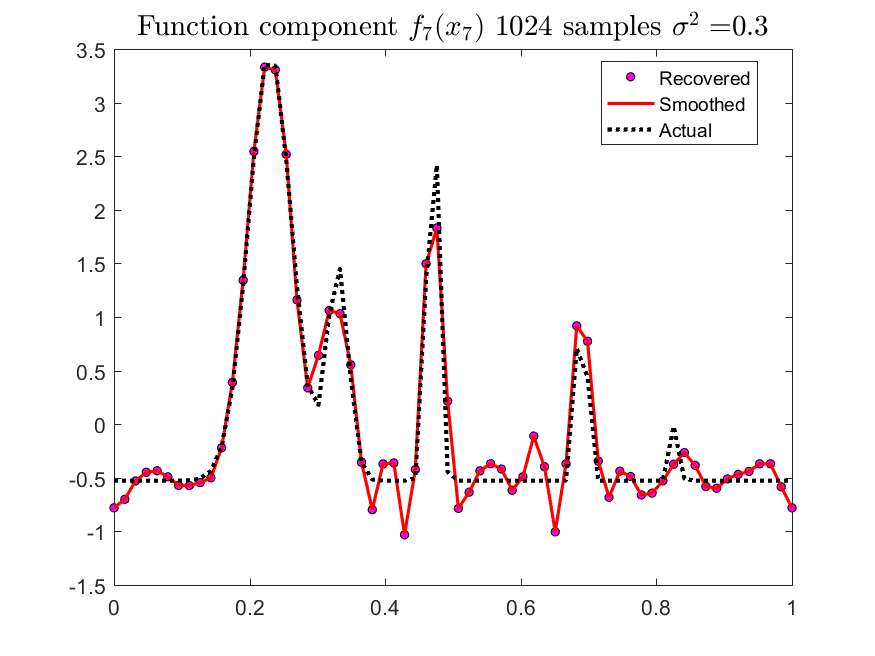}
       \caption{}
       \label{fig:f7V2_1024_BetaDesign_Sigma_03}
   \end{subfigure}\hfill
   \begin{subfigure}[b]{0.25\textwidth}
       \includegraphics[width=\textwidth]{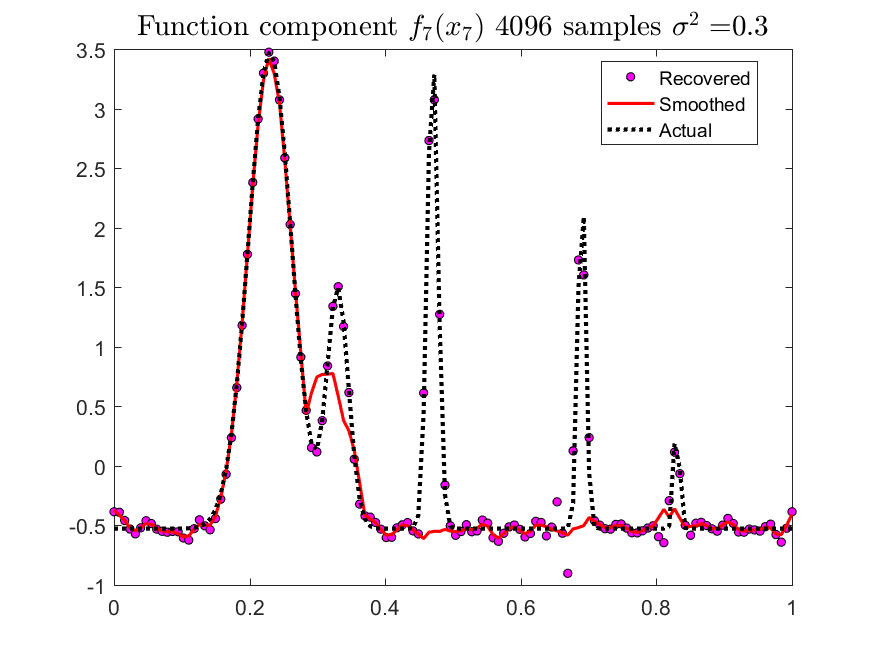}
       \caption{}
       \label{fig:f7V2_4096_BetaDesign_Sigma_03}
    \end{subfigure}\hfill
    \begin{subfigure}[b]{0.25\textwidth}
       \includegraphics[width=\textwidth]{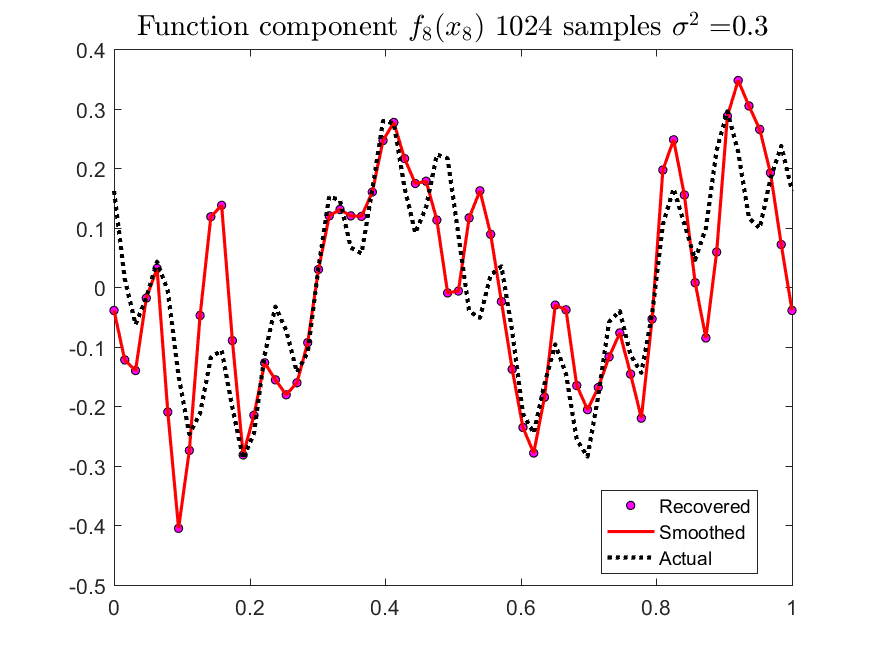}
       \caption{}
       \label{fig:f8V2_1024_BetaDesign_Sigma_03}
   \end{subfigure}\hfill
   \begin{subfigure}[b]{0.25\textwidth}
       \includegraphics[width=\textwidth]{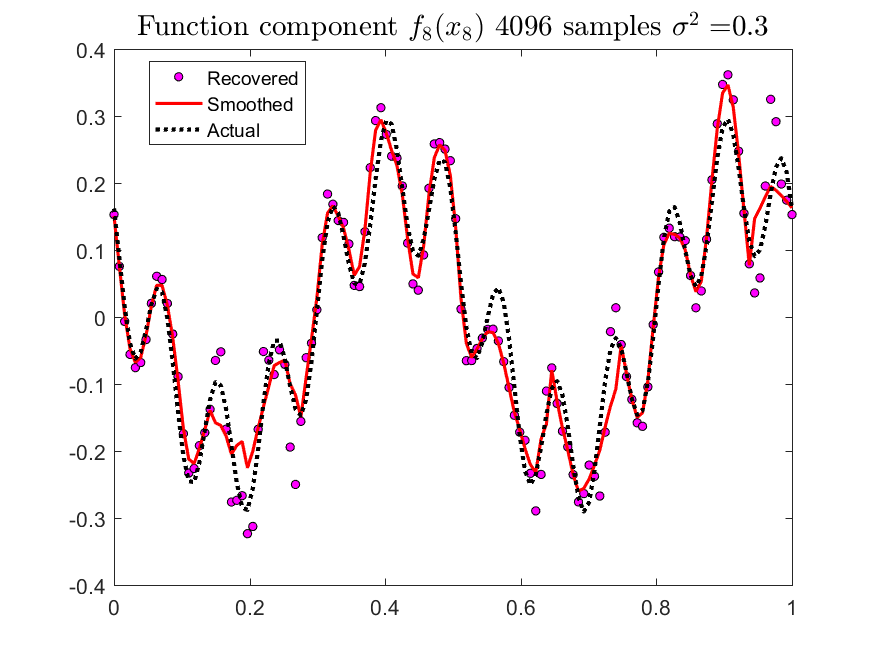}
       \caption{}
       \label{fig:f8V2_4096_BetaDesign_Sigma_03}
    \end{subfigure}
\caption{Estimated $f_{7}(x)$ and $f_{8}(x)$ using Beta Design and Coiflets filter.}
\label{fig:F7BetaCoif}
\end{figure}

\begin{figure}[!htb]
   \centering
   \begin{subfigure}[b]{0.25\textwidth}
       \includegraphics[width=\textwidth]{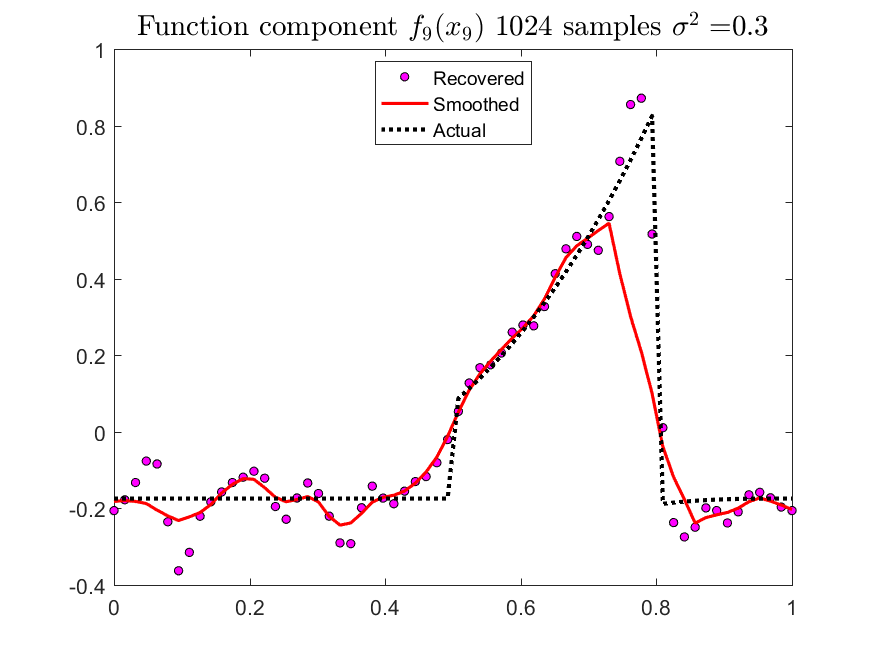}
       \caption{}
       \label{fig:f9V2_1024_BetaDesign_Sigma_03}
   \end{subfigure}
   \begin{subfigure}[b]{0.25\textwidth}
       \includegraphics[width=\textwidth]{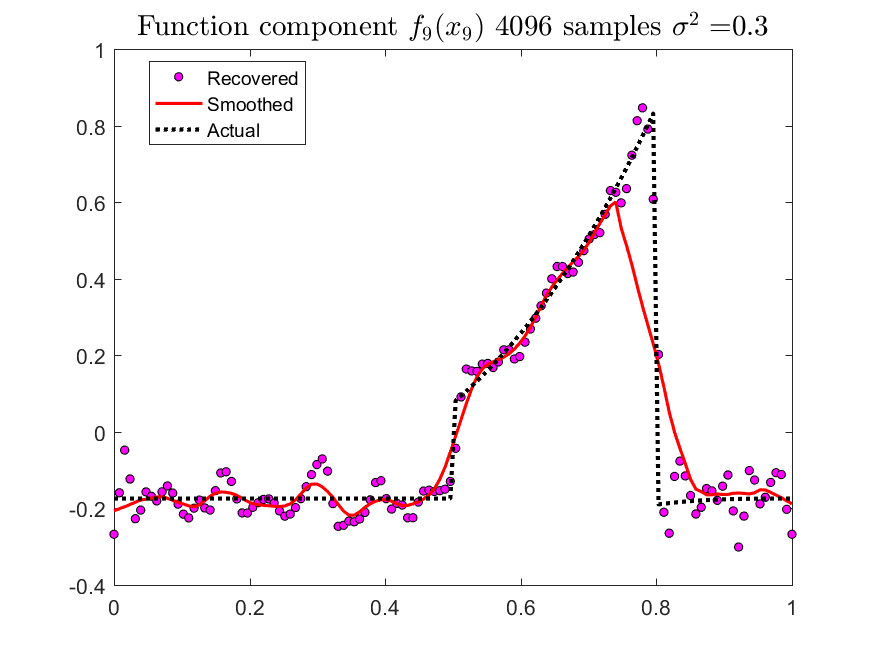}
       \caption{}
       \label{fig:f9V2_4096_BetaDesign_Sigma_03}
    \end{subfigure}
\caption{Estimated $f_{9}(x)$ using Beta Design and Coiflets filter.}
\label{fig:F9BetaCoif}
\end{figure}

In this section, we summarize the simulation results obtained for the baseline distributions previously defined. In particular, we present the following:

\begin{enumerate}[(i)]
  \item Tables \ref{tab:RMSEDaubUnifSigma1} to \ref{tab:RMSECoifUnifSigma2} present details for RMSE results obtained for each of the baseline distributions using a Uniform design $\left\{\mathcal{U}[0,1]\right\}^{9}$ for predictors. Similarly, in Tables \ref{tab:RMSEDaubBetaSigma1} to \ref{tab:RMSECoifBetaSigma2} present details for RMSE results obtained for each of the baseline functions using a $\left\{Beta(\frac{3}{2},\frac{3}{2})\right\}^{9}$ design.
  \item Figures \ref{fig:AMSE_Daubechies_Unif_sigma1v2} - \ref{fig:AMSE_Daubechies_Unif_sigma2v2} show the behavior of the RMSE for each of the functions $f_{1},...,f_{9}$ with respect to sample size and noise variance values $\sigma^{2}=0.75\,,\,0.25$, for the Uniform design $\left\{\mathcal{U}[0,1]\right\}^{9}$ using Daubechies filter.
  \item Figures \ref{fig:AMSE_Coif_Unif_sigma1v2} - \ref{fig:AMSE_Coif_Unif_sigma2v2} show the behavior of the RMSE for each of the functions $f_{1},...,f_{9}$ with respect to the sample size and the noise variance values $\sigma^{2}=0.75\,,\,0.25$ for the Uniform design $\left\{\mathcal{U}[0,1]\right\}^{9}$ using Coiflets 24 filter.
\item Figures \ref{fig:Yhat_1024UnifDesign_Sigma2_025} - \ref{fig:Yhat_4096UnifDesign_Sigma2_025} show the estimation summary plots (observed responses $Y_{1},...,Y_{n}$, estimated response values $\hat{Y}_{1},...,\hat{Y}_{n}$, histogram of residuals $e_{i}=Y_{i}-\hat{Y}_{i}$ and plot $Y_{i}\,\text{vs.}\,\hat{Y}_{i}$) for the Uniform design $\left\{\mathcal{U}[0,1]\right\}^{9}$ using Coiflets 24 filter, $\sigma^{2}=0.25$ and sample sizes $n=1024,4096$.
  \item Figures \ref{fig:f1V2_1024_UnifDesign_Sigma_025} - \ref{fig:f9V2_4096_UnifDesign_Sigma_03} show the recovered functions $f_{1},...,f_{9}$ for different sample sizes $n=512,1024,4096$ and values of the noise variance $\sigma^{2}=0.25,0.3$ for the Uniform design $\left\{\mathcal{U}[0,1]\right\}^{9}$ using a Coiflets 24 filter. The dashed lines (black) correspond to the actual function, computed at each data point $x$, whereas the magenta points show the estimated values of the function at each sample x. The red lines corresponds to a smoothed version of the estimated function values, computed using locally weighted scatterplot smoothing (lowess) with parameter 0.25 (this was done just for visualization purposes).
  \item Figures \ref{fig:AMSE_Daub_Beta_sigma1_v2} - \ref{fig:AMSE_Daub_Beta_sigma2_v2} show the behavior of the RMSE for each of the functions $f_{1},...,f_{9}$ with respect to the sample size and the noise variance values $\sigma^{2}=0.75\,,\,0.25$ for the Beta design $\left\{Beta(\frac{3}{2},\frac{3}{2})\right\}^{9}$ using Daubechies filter.
  \item Figures \ref{fig:AMSE_Coiflets_Beta_sigma1_v2} - \ref{fig:AMSE_Coiflets_Beta_sigma2_v2} show the behavior of the RMSE for each of the functions $f_{1},...,f_{9}$ with respect to the sample size and the noise variance values $\sigma^{2}=0.75\,,\,0.25$ for the Beta design $\left\{Beta(\frac{3}{2},\frac{3}{2})\right\}^{9}$ using Coiflets 24 filter. In each figure, plots (b) and (d) correspond to zoomed in versions of plots (a) and (c) respectively.
\item Figures \ref{fig:f1V2_1024_BetaDesign_Sigma_03} - \ref{fig:f9V2_4096_BetaDesign_Sigma_03} show the recovered functions $f_{1},...,f_{9}$ for different sample sizes $n=1024,4096$ and values of the noise variance $\sigma^{2}=0.3$ for the Beta design $\left\{Beta(\frac{3}{2},\frac{3}{2})\right\}^{9}$ using Coiflets 24 filter. The dashed lines (black) correspond to the actual function, computed at each data point $x$, whereas the magenta points show the estimated values of the function at each sample x. The red lines corresponds to a smoothed version of the estimated function values, computed using lowess smoothing with parameter 0.25 (this was done just for visualization purposes).
\item Figures \ref{fig:Yhat_1024BetaDesign_Sigma2_03} - \ref{fig:Yhat_4096BetaDesign_Sigma2_03} show the estimation summary plots (observed responses $Y_{1},...,Y_{n}$, estimated response values $\hat{Y}_{1},...,\hat{Y}_{n}$, histogram of residuals $e_{i}=Y_{i}-\hat{Y}_{i}$ and plot $Y_{i}\,\text{vs.}\,\hat{Y}_{i}$) for the Beta design $\left\{Beta(\frac{3}{2},\frac{3}{2})\right\}^{9}$ using Coiflets 24 filter, $\sigma^{2}=0.3$ and sample sizes $n=1024,4096$.
\end{enumerate}


\begin{table}[!htb]
\parbox{.50\linewidth}{
\centering
\scalebox{0.65}{
\begin{tabular}[h]{|c|c|c|c|c|c|}
\hline
 & Uniform Design & & $\sigma^{2}=0.25$ & & Daubechies 4 \\
\hline
 & $n=256$ & $n=512$ & $n=1024$ & $n=2048$ & $n=4096$ \\
\hline
\hline
$f_{1}(x)$ & 0.0224 &	0.0143 &	0.0086 &	0.0035 &	0.002   \\
\hline
$f_{2}(x)$  & 0.0227 &	0.0156 &	0.0089 &	0.0038 &	0.002   \\
\hline
$f_{3}(x)$  & 0.0692 &	0.0174 &	0.0088 &	0.0038 &	0.002   \\
\hline
$f_{4}(x)$  & 0.0241 &	0.0141 &	0.0086 &	0.0038 &	0.002   \\
\hline
$f_{5}(x)$  & 0.0242 &	0.0148 &	0.0088 &	0.0036 &	0.002   \\
\hline
$f_{6}(x)$  & 0.0391 &	0.0155 &	0.0087 &	0.0037 &	0.0021  \\
\hline
$f_{7}(x)$  & 0.7327 &	0.1069 &	0.1051 &	0.1005 &	0.0533   \\
\hline
$f_{8}(x)$  & 0.0289 &	0.0191 &	0.0103 &	0.0049 &	0.0021   \\
\hline
$f_{9}(x)$  & 0.0543 &	0.0268 &	0.0143 &	0.0091 &	0.0029   \\
\hline
\end{tabular}}
\caption{RMSE results for Uniform distribution with $\sigma^{2}=0.25$ using Daubechies 4 wavelet filter.} \label{tab:RMSEDaubUnifSigma1}
}
\hfill
\parbox{.50\linewidth}{
\centering
\scalebox{0.65}{
\begin{tabular}[h]{|c|c|c|c|c|c|}
\hline
 & Uniform Design & & $\sigma^{2}=0.75$ & & Daubechies 4 \\
\hline
 & $n=256$ & $n=512$ & $n=1024$ & $n=2048$ & $n=4096$ \\
\hline
\hline
$f_{1}(x)$ & 0.042 &	0.0362 &	0.0306 &	0.0126 &	0.0114  \\
\hline
$f_{2}(x)$  & 0.0458 &	0.0345 &	0.0307 &	0.0121 &	0.0108  \\
\hline
$f_{3}(x)$  & 0.0909 &	0.0382 &	0.0301 &	0.013  &	0.0109  \\
\hline
$f_{4}(x)$  & 0.044 & 	0.0342 &	0.0296 &	0.0127 &	0.0113  \\
\hline
$f_{5}(x)$  & 0.0449 &	0.0341 &	0.0304 &	0.0125 &	0.0111  \\
\hline
$f_{6}(x)$  & 0.064 &	0.0363 &	0.0305 &	0.0128 &	0.0113  \\
\hline
$f_{7}(x)$  & 0.7577 &	0.1299 &	0.1283 &	0.1097 &	0.0624  \\
\hline
$f_{8}(x)$  & 0.0478 &	0.0395 &	0.0322 &	0.0135 &	0.011  \\
\hline
$f_{9}(x)$  & 0.0751 &	0.0468 &	0.0349 &	0.0177 &	0.0119  \\
\hline
\end{tabular}}
\caption{RMSE results for Uniform distribution with $\sigma^{2}=0.75$ using Daubechies 4 wavelet filter.} \label{tab:RMSEDaubUnifSigma2}
}\label{tab:UnifDaubechies}
\end{table}

\begin{table}[!htb]
\parbox{.50\linewidth}{
\centering
\scalebox{0.65}{
\begin{tabular}[h]{|c|c|c|c|c|c|}
\hline
 & Uniform Design & & $\sigma^{2}=0.25$ & & Coiflets 24 \\
\hline
 & $n=256$ & $n=512$ & $n=1024$ & $n=2048$ & $n=4096$ \\
\hline
\hline
$f_{1}(x)$ & 0.0193 &	0.0163 &	0.0058 &	0.0024 &	0.0013  \\
\hline
$f_{2}(x)$  & 0.0191 &	0.0172 &	0.0057 &	0.0025  &	0.0013  \\
\hline
$f_{3}(x)$  & 0.0198 &	0.0168 &	0.006 &	0.0025   &	0.0013  \\
\hline
$f_{4}(x)$  & 0.0214 &	0.0177 &	0.0061 &	0.0025  &	0.0013  \\
\hline
$f_{5}(x)$  & 0.0185 &	0.0165 &	0.0059 &	0.0024  &	0.0013   \\
\hline
$f_{6}(x)$  & 0.0207 &	0.0177 &	0.0057 &	0.0025  &	0.0013  \\
\hline
$f_{7}(x)$  & 0.7776 &	0.1946 &	0.0388 &	0.0353  &	0.0088  \\
\hline
$f_{8}(x)$  & 0.0244 &	0.0222 &	0.0061 &	0.0027  &	0.0013  \\
\hline
$f_{9}(x)$  & 0.0386 &	0.022 &	0.0083 &	0.0049  &	0.0032 \\
\hline
\end{tabular}}
\caption{RMSE results for Uniform distribution with $\sigma^{2}=0.25$ using Coiflets 24 wavelet filter.} \label{tab:RMSECoifUnifSigma1}
}
\hfill
\parbox{.50\linewidth}{
\centering
\scalebox{0.65}{
\begin{tabular}[h]{|c|c|c|c|c|c|}
\hline
 & Uniform Design & & $\sigma^{2}=0.75$ & & Coiflets 24 \\
\hline
 & $n=256$ & $n=512$ & $n=1024$ & $n=2048$ & $n=4096$ \\
\hline
\hline
$f_{1}(x)$ & 0.0369 &	0.0375 &	0.0259 &	0.0115 &	0.0102  \\
\hline
$f_{2}(x)$  & 0.0407 &	0.0364 &	0.0268 &	0.0112 &	0.0103  \\
\hline
$f_{3}(x)$  & 0.0377 &	0.0373 &	0.0269 &	0.0116 &	0.0102  \\
\hline
$f_{4}(x)$  & 0.0417 &	0.0353 &	0.0266 &	0.0115 &	0.0104  \\
\hline
$f_{5}(x)$  & 0.0395 &	0.0373 &	0.0265 &	0.0112 &	0.0101  \\
\hline
$f_{6}(x)$  & 0.0397 &	0.0368 &	0.0268 &	0.0113 &	0.0105  \\
\hline
$f_{7}(x)$  & 0.7796 &	0.2165 &	0.0598 &	0.0438 &	0.0178  \\
\hline
$f_{8}(x)$  & 0.0438 &	0.0436 &	0.0273 &	0.0115 &	0.0103  \\
\hline
$f_{9}(x)$  & 0.0571 &	0.0433 &	0.0289 &	0.0132 &	0.0121  \\
\hline
\end{tabular}}
\caption{RMSE results for Uniform distribution with $\sigma^{2}=0.75$ using Coiflets 24 wavelet filter.} \label{tab:RMSECoifUnifSigma2}
}
\end{table}

\begin{table}[!htb]
\parbox{.50\linewidth}{
\centering
\scalebox{0.65}{
\begin{tabular}[h]{|c|c|c|c|c|c|}
\hline
 & $Beta(\frac{3}{2},\frac{3}{2})$ Design & & $\sigma^{2}=0.25$ & & Daubechies 4 \\
\hline
 & $n=256$ & $n=512$ & $n=1024$ & $n=2048$ & $n=4096$ \\
\hline
\hline
$f_{1}(x)$ & 0.0324 &	0.0246 &	0.0153 &	0.0058 &	0.0031     \\
\hline
$f_{2}(x)$  & 0.0344 &	0.0212 &	0.0147 &	0.0057 &	0.003     \\
\hline
$f_{3}(x)$  & 0.0971 &	0.026 &	0.0141 &	0.006 &	0.0031     \\
\hline
$f_{4}(x)$  & 0.0325 &	0.0234 &	0.0143 &	0.0054 &	0.003     \\
\hline
$f_{5}(x)$  & 0.0369 &	0.0237 &	0.0143 &	0.0054 &	0.0032     \\
\hline
$f_{6}(x)$  & 0.0561 &	0.0248 &	0.0137 &	0.0061 &	0.003    \\
\hline
$f_{7}(x)$  & 0.7254 &	0.1071 &	0.1072 &	0.101 &	0.0538     \\
\hline
$f_{8}(x)$  & 0.0413 &	0.0273 &	0.0148 &	0.0071 &	0.0033     \\
\hline
$f_{9}(x)$  & 0.067 &	0.0341 &	0.0194 &	0.0112 &	0.004     \\
\hline
\end{tabular}}
\caption{RMSE results for $Beta(\frac{3}{2},\frac{3}{2})$ distribution with $\sigma^{2}=0.25$ using Daubechies 4 wavelet filter.} \label{tab:RMSEDaubBetaSigma1}
}
\hfill
\parbox{.50\linewidth}{
\centering
\scalebox{0.65}{
\begin{tabular}[h]{|c|c|c|c|c|c|}
\hline
 & $Beta(\frac{3}{2},\frac{3}{2})$ Design & & $\sigma^{2}=0.75$ & & Daubechies 4 \\
\hline
 & $n=256$ & $n=512$ & $n=1024$ & $n=2048$ & $n=4096$ \\
\hline
\hline
$f_{1}(x)$ & 0.0578 &	0.053 &	0.0442 &	0.0168 &	0.0163   \\
\hline
$f_{2}(x)$  & 0.0593 &	0.0578 &	0.0443 &	0.0187 &	0.0156    \\
\hline
$f_{3}(x)$  & 0.1342 &	0.0534 &	0.0438 &	0.0179 &	0.0153    \\
\hline
$f_{4}(x)$  & 0.0577 &	0.0566 &	0.0462 &	0.0186 &	0.0152    \\
\hline
$f_{5}(x)$  & 0.0583 &	0.056 &	0.0445 &	0.0173 &	0.0167   \\
\hline
$f_{6}(x)$  & 0.0819 &	0.0554 &	0.045 &	0.019 &	0.0156    \\
\hline
$f_{7}(x)$  & 0.7534 &	0.1327 &	0.1373 &	0.1139 &	0.0672   \\
\hline
$f_{8}(x)$  & 0.0662 &	0.0585 &	0.0470 &	0.0196 &	0.0166   \\
\hline
$f_{9}(x)$  &  0.0949 &	0.0635 &	0.0515 &	0.0237 &	0.0169   \\
\hline
\end{tabular}}
\caption{RMSE results for $Beta(\frac{3}{2},\frac{3}{2})$ distribution with $\sigma^{2}=0.75$ using Daubechies 4 wavelet filter.} \label{tab:RMSEDaubBetaSigma2}
}
\end{table}

\begin{table}[!htb]
\parbox{.50\linewidth}{
\centering
\scalebox{0.65}{
\begin{tabular}[h]{|c|c|c|c|c|c|}
\hline
 & $Beta(\frac{3}{2},\frac{3}{2})$ Design & & $\sigma^{2}=0.25$ & & Coiflets 24 \\
\hline
 & $n=256$ & $n=512$ & $n=1024$ & $n=2048$ & $n=4096$ \\
\hline
\hline
$f_{1}(x)$ & 0.0284 &	0.0252 &	0.0091 &	0.0035 &	0.0017    \\
\hline
$f_{2}(x)$  & 0.029 &	0.0258 &	0.0086 &	0.0036 &	0.0017    \\
\hline
$f_{3}(x)$  & 0.0276 &	0.0248 &	0.0085 &	0.0034 &	0.0018    \\
\hline
$f_{4}(x)$  & 0.0312 &	0.0246 &	0.0084 &	0.0036 &	0.0018    \\
\hline
$f_{5}(x)$  & 0.0288 &	0.0246 &	0.0084 &	0.0036 &	0.0017    \\
\hline
$f_{6}(x)$  & 0.0293 &	0.0245 &	0.0088 &	0.0034 &	0.0017   \\
\hline
$f_{7}(x)$  & 0.757 &	0.1977 &	0.0398 &	0.0358 &	0.0091    \\
\hline
$f_{8}(x)$  & 0.0347 &	0.0321 &	0.0081 &	0.0038 &	0.0017    \\
\hline
$f_{9}(x)$  & 0.047 &	0.0313 &	0.011 &	0.0059 &	0.0035    \\
\hline
\end{tabular}}
\caption{RMSE results for $Beta(\frac{3}{2},\frac{3}{2})$ distribution with $\sigma^{2}=0.25$ using Coiflets 24 wavelet filter.} \label{tab:RMSECoifBetaSigma1}
}
\hfill
\parbox{.50\linewidth}{
\centering
\scalebox{0.65}{
\begin{tabular}[h]{|c|c|c|c|c|c|}
\hline
 & $Beta(\frac{3}{2},\frac{3}{2})$ Design & & $\sigma^{2}=0.75$ & & Coiflets 24 \\
\hline
 & $n=256$ & $n=512$ & $n=1024$ & $n=2048$ & $n=4096$ \\
\hline
\hline
$f_{1}(x)$ & 0.0488 &	0.0509 &	0.0346 &	0.0142 &	0.013   \\
\hline
$f_{2}(x)$  & 0.0523 &	0.0511 &	0.0347 &	0.0144 &	0.0131    \\
\hline
$f_{3}(x)$  & 0.0492 &	0.0467 &	0.0356 &	0.0149 &	0.0134  \\
\hline
$f_{4}(x)$  & 0.0548 &	0.0493 &	0.037 &	0.0145 &	0.0133 \\
\hline
$f_{5}(x)$  & 0.051 &	0.0511 &	0.0357 &	0.015 &	0.013 \\
\hline
$f_{6}(x)$  & 0.0463 &	0.0523 &	0.036 &	0.015 &	0.013 \\
\hline
$f_{7}(x)$  & 0.7911 &	0.2238 &	0.0678 &	0.0466 &	0.02060 \\
\hline
$f_{8}(x)$  & 0.0563 &	0.0537 &	0.0351 &	0.0151 &	0.013 \\
\hline
$f_{9}(x)$  & 0.0715 &	0.0574 &	0.0385 &	0.0175 &	0.0151 \\
\hline
\end{tabular}}
\caption{RMSE results for $Beta(\frac{3}{2},\frac{3}{2})$ distribution with $\sigma^{2}=0.75$ using Coiflets 24 wavelet filter.} \label{tab:RMSECoifBetaSigma2}
}
\end{table}

\subsubsection{Remarks and comments}\label{SimResults}

\begin{enumerate}[(i)]
\item Practical choice of $J(n)$.  Since the optimal multiresolution index $J$ was obtained up to and unknown additive constant $\mathcal{K}_{1}$ (see Lemma 4), for implementation purposes it is possible to replace it with a predefined integer. However, a large value for this constant would cause an undesired inflation of the estimator variance and also, increase the computational complexity of the algorithm.
\item In the case of densities with exponentially decaying tails (i.e. largely deviated form uniformity), large samples are needed in order to obtain accurate estimates. In fact, during the simulation study we observed cases where abnormally large wavelet coefficients were obtained at the tails of the distribution (or regions with low density values). This was caused primarily due to possible violations of assumption (\textbf{A1}) and the lack of information available for a reasonable estimation of the coefficients in those regions. In this context, we suggest the following possible remedial actions:
    \begin{enumerate}[(a)]
      \item  Restricting the domain of estimation to the 95\% empirical quantiles along each of the dimensions of the predictors. This is a reasonable approach that can prevent the generation of large coefficient that induce error in the function estimation procedure. However, this reduces the effective sample size and also, restricts the possibility of estimation of unlikely or rare cases.
      \item Choosing parameter $\beta_{n}$ via cross-validation to minimize the RMSE. Abnormally large wavelet coefficients would lead (in general) to large function estimates. This can be prevented by truncating the final estimates using $\beta_{n}$ and the use of cross-validation would allow an evidence-based selection of this parameter.
   \end{enumerate}
\item Model without $\beta_{0}$. Because of the strang-fix condition, the estimation of a model with a constant $\beta_{0}$ turned out to be unstable. For this reason, we recommend a pre-processing stage in which the response is standardized so that it has zero mean and a standard deviation of 1. This approach is a natural result if we modify assumption (\textbf{A1}) to be instead $\mathbb{E}\left[f_{j}(X_{j}) \right]=0$ for $j=1,...,p$. Note that this does not alter at all the model structure, estimation procedure or statistical properties. In this case the natural estimator of the intercept would be given by $\hat{\beta}_{0}=\frac{1}{n}\sum_{i=1}^{n}y_{i}$.
\end{enumerate}

\section{Practical Application of Wavelet based Least Squares Method}\label{LSApplication}
In this section we consider the implementation of our proposed estimator using a dataset available at the machine learning repository of UCI\footnote{UCI Machine Learning Repository \url{http://archive.ics.uci.edu/ml}. Irvine, CA: University of California, School of Information and Computer Science.} concerning the study of hourly full load electrical output power (EP) of a combined cycle plant.

\medskip

This data set was extensively analized by Tufekci (2014)\cite{Tufekci2014} using different statistical models, with the goal of predicting EP based on 4 available features. That research utilized a variety of predictive methods including: Simple Linear Regression (SLR), Multilayer Perceptron (MLP), Radial Basis Function Neural Network (RBF), Additive Regression (AR, using back-fitting), KStar (instance-based classifier), Locally Weighted Learning, Bagging REP Tree (BREP, Bootstrap based tree methods), Model Tree rules, Model Tress Regression (M5P), REP Trees, Support Vector Regression, Least Median Square (LMS), etc. A total of 15 statistical models were used and compared using 2-fold Crossvalidation after randomly shuffling the data 5 times. Then, prediction accuracy was evaluated using RMSE as an error metric.

\medskip
\subsubsection*{Data set description}
The dataset contains 9568 data points collected from a Combined Cycle Power Plant\footnote{ A combined cycle power plant (CCPP) is composed of gas turbines (GT), steam turbines (ST) and heat recovery steam generators. In a CCPP, the electricity is generated by gas and steam turbines, which are combined in one cycle, and is transferred from one turbine to another. While the Vacuum is collected from and has effect on the Steam Turbine, he other three of the ambient variables effect the GT performance. } over 6 years (2006-2011), when the power plant was set to work with full load. The features are used to predict the net hourly electrical energy output (EP) of the plant and consist of :

\begin{enumerate}
  \item Temperature (AT) : This input variable is measured in degrees Celsius and it varies between 1.81C and 37.11C.
  \item Ambient Pressure (AP): This input variable is measured in millibar with an observed range from 992.89 to 1033.3 mbar.
  \item Relative Humidity (RH): This variable is measured as a percentage with an observed range from 25.56\% to 100.16\%.
  \item Exhaust Vacuum (V): This variable is measured in cm Hg with with an observed range from 25.36 to 81.56 cm Hg.
\end{enumerate} The characteristics of the data are the following:
Where the EP  is measured in mega watt with an observed range from 420.26 to 495.76 MW. Similarly, the details of the dataset can be summarized as follows:
\begin{table}[!htb]
\centering
\begin{tabular}{|c|c|}
\hline
Data Set characteristics & \verb"Multivariate" \\
\hline
Number of samples & \verb"9568" \\
\hline
Attribute characteristics & \verb"Real" \\
\hline
Number of Attributes & $5$ \\
\hline
\end{tabular}
\caption{Application Data Set characteristics, obtained from \cite{Tufekci2014}.} \label{tab:AppData}
\end{table}

More details about the data set and the problem in hand can be found in \cite{Tufekci2014}.
\medskip

\subsubsection*{Implementation settings and results}
\begin{figure}[!htb]
   \centering
   \begin{subfigure}[b]{0.25\textwidth}
       \includegraphics[width=\textwidth]{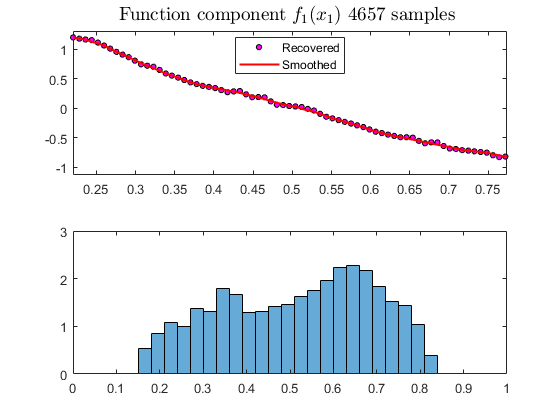}
       \caption{Estimated $f_{1}(x)$, corresponding to AT.}
       \label{fig:f1_power}
   \end{subfigure}\hfill
\begin{subfigure}[b]{0.25\textwidth}
       \includegraphics[width=\textwidth]{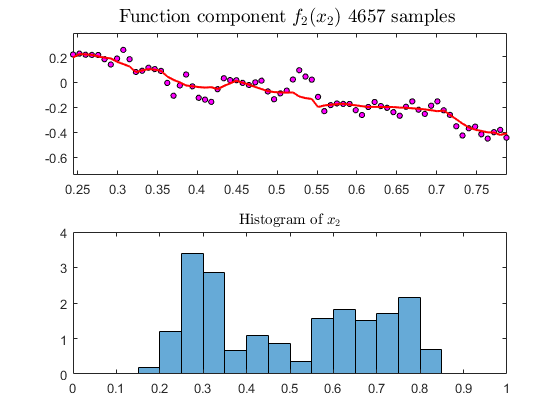}
       \caption{Estimated $f_{2}(x)$, corresponding to AP.}
       \label{fig:f2_power}
   \end{subfigure}\hfill
   \begin{subfigure}[b]{0.25\textwidth}
       \includegraphics[width=\textwidth]{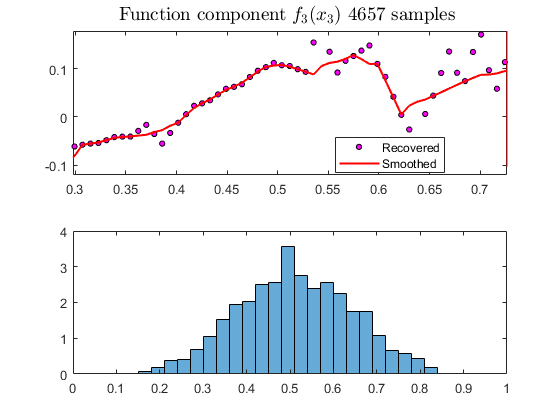}
       \caption{Estimated $f_{3}(x)$, corresponding to RH}
       \label{fig:f3_power}
    \end{subfigure}\hfill
\begin{subfigure}[b]{0.25\textwidth}
       \includegraphics[width=\textwidth]{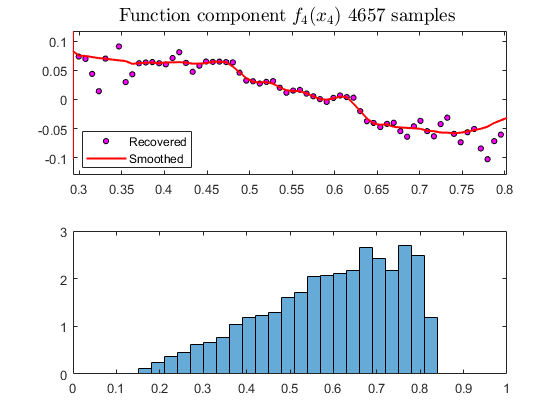}
       \caption{Estimated $f_{4}(x)$, corresponding V.}
       \label{fig:f4_power}
    \end{subfigure}
\caption{Estimated $f_{1}(x)$, $f_{2}(x)$, $f_{3}(x)$ and $f_{4}(x)$  over the 95\% empirical quantiles region. The bottom panel illustrates the sample histograms for
each considered feature, within the 95\% empirical quantiles region. }
\label{fig:f1f2_power}
\end{figure}

For this problem, we chose the following implementation settings:
\begin{enumerate}[(a)]
\item Daubechies 4 filter for the scaling functions.
\item $J(n)=1+\lfloor\log_{2}(n)-\log_{2}\left(\log(n)\,(\log(n)+1)\right)\rfloor$.
\item The response $y$ was centered and standardized and the predictors $\textbf{X}_{1},...,\textbf{X}_{n}$ where rescaled to $[0,1]^{p}$.
\item To prevent unstable estimates at the tails of the marginal distributions of the predictors, we restricted the estimation range to the 95\% empirical quantiles of the observed sample.
\item The data was randomly split into training and testing over the samples belonging to the hypercube defined by the 95\% empirical quantiles. 85\% of the data was selected for training and the remaining 15\% for testing purposes. The estimation process was repeated 100 times. The results for this procedure are illustrated in figures \ref{fig:f1f2_power}-\ref{fig:f4_power}.
\item For comparison purposes (with results presented in Table 10 \cite{Tufekci2014}), we also implemented the proposed method using 2-fold CV with Coiflets 24 filter. The process was replicated 10 times. In this case, the wavelet coefficients were obtained using the complete sample, without restricting the range of the estimation. Table \ref{tab:RMSEcomparison} illustrates the differences in accuracy for the wavelet-based estimator and the best regression techniques used in \cite{Tufekci2014}.
\end{enumerate}

\begin{figure}[!htb]
   \centering
\begin{subfigure}[b]{0.33\textwidth}
       \includegraphics[width=\textwidth]{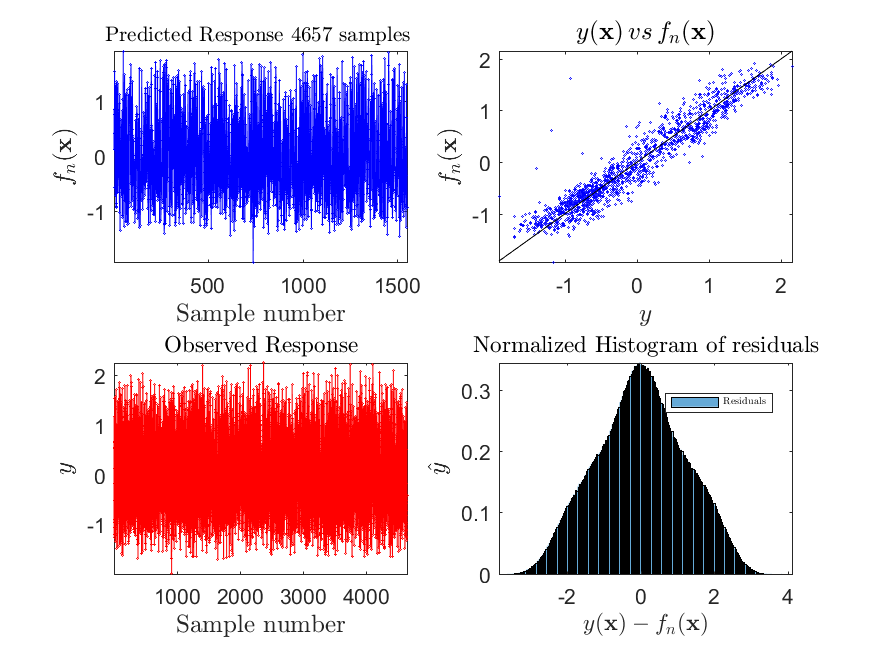}
       \caption{}
       \label{fig:Yhat_4657Application}
   \end{subfigure}
\begin{subfigure}[b]{0.33\textwidth}
       \includegraphics[width=\textwidth]{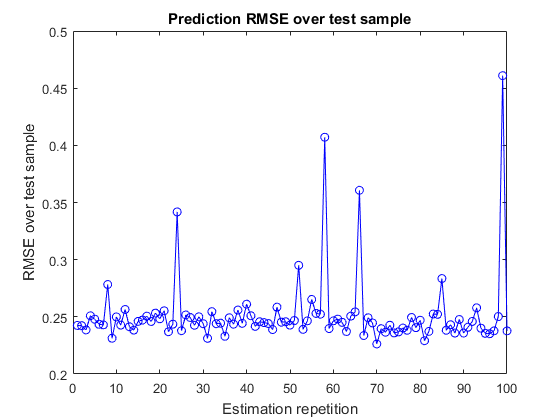}
       \caption{}
       \label{fig:RMSEs}
    \end{subfigure}
\caption{Estimaion result plots  over the 95\% empirical quantiles region and RMSE (computed using the standardized predictions) obtained over 100 replications.}
\label{fig:f3f4_power}
\end{figure}

\medskip
The obtained results are summarized in the following figures and tables:
\begin{enumerate}[(i)]
\item Figure \ref{fig:f1f2_power} shows the estimated unknown functions acting on each one of the problem features.
\item Figure \ref{fig:Yhat_4657Application} shows the estimated and actual standardized response, together with the $f_{n}(\textbf{x})$ vs $y$ plot and the residual plot $e_{i}=f_{n}(\textbf{x}_{i})-y_{i}$.
\item Table \ref{tab:RMSEcomparison} shows RMSE for best methods in \cite{Tufekci2014} and the Wavelet-based LS using 4 features.
\item Table \ref{tab:RMSEcomparison1} shows RMSE for best methods in \cite{Tufekci2014} and the Wavelet-based LS using 1 feature (AT).
\item Table \ref{tab:RMSEcomparison2} shows RMSE for best methods in \cite{Tufekci2014} and the Wavelet-based LS using 2 features (AT-V).
\item Table \ref{tab:RMSEcomparison3} shows RMSE for best methods in \cite{Tufekci2014} and the Wavelet-based LS using 3 features (AT-V-RH).
\end{enumerate}

\begin{table}[!htb]
\centering
\begin{tabular}{|c|c|c|c|c|c|c|c|c|c|c|}
\hline
Kstar & BREP & M5P &  MLP & RBF & LMS & SMOREg & M5R & REP & AR  & \ttgrc{Wavelet LS} \\
\hline
3.861 & 3.787 & 4.087 & 5.339 & 8.487 & 4.572  & 4.563  & 4.128  & 4.211 & 5.556 &  \ttgrc{4.325} \\
\hline
\end{tabular}
\caption{Comparison results for RMSE for best methods in \cite{Tufekci2014} and the Wavelet-based LS using 4 features.} \label{tab:RMSEcomparison}
\end{table}

\begin{table}[!htb]
\centering
\begin{tabular}{|c|c|c|c|c|c|c|c|}
\hline
Kstar & BREP & M5P &  LMS & SMOREg & M5R & REP &  \ttgrc{Wavelet LS} \\
\hline
5.381 & 5.208 & 5.086 & 5.433  & 5.433  & 5.085 & 5.229 &  \ttgrc{5.085} \\
\hline
\end{tabular}
\caption{Comparison results for RMSE for best methods in \cite{Tufekci2014} and the Wavelet-based LS using 1 feature (AT).} \label{tab:RMSEcomparison1}
\end{table}

\begin{table}[!htb]
\centering
\begin{tabular}{|c|c|c|c|c|c|c|c|}
\hline
Kstar & BREP & M5P &  LMS & SMOREg & M5R & REP &  \ttgrc{Wavelet LS} \\
\hline
4.634 & 4.026 & 4.359 & 4.968  & 4.968  & 4.419 & 4.339 &  \ttgrc{4.757} \\
\hline
\end{tabular}
\caption{Comparison results for RMSE for best methods in \cite{Tufekci2014} and the Wavelet-based LS using 2 features (AT-V).} \label{tab:RMSEcomparison2}
\end{table}

\begin{table}[!htb]
\centering
\begin{tabular}{|c|c|c|c|c|c|c|c|}
\hline
Kstar & BREP & M5P &  LMS & SMOREg & M5R & REP &  \ttgrc{Wavelet LS} \\
\hline
4.331 & 3.934 & 4.178 & 4.580  & 4.585  & 4.217 & 4.291 &  \ttgrc{4.776} \\
\hline
\end{tabular}
\caption{Comparison results for RMSE for best methods in \cite{Tufekci2014} and the Wavelet-based LS using 3 features (AT-V-RH).} \label{tab:RMSEcomparison3}
\end{table}

\subsubsection*{Remarks and Comments}
\begin{enumerate}[(i)]
\item From figures \ref{fig:f1_power}-\ref{fig:f4_power} it is possible to observe that the wavelet-based estimator is able to capture the non-linear influences of each of the features considered in the model. From the plots it is possible to assess the significance of each one of the uncovered functions in the model; in particular, \ref{fig:f1_power} shows an almost linear effect of the Temperature over EP with negative correlation. For the rest of the predictors, the effect on the response is almost negligible.
\item From figure \ref{fig:Yhat_4657Application}, we can conclude that the wavelet-based estimator is able to successfully predict the EP over the test sample. The predicted vs actual values lie in a straight line with no evident deviations apart from the noise in the data, showing a strong correlation between predicted and actual values.
\item In table \ref{tab:RMSEcomparison}, the average RMSE for the Wavelet-based LS method was 4.325 (non-standardized testing sample) which shows to be better than most of the results shown in Table 10 \cite{Tufekci2014}. In particular, the best regression methods studied in such reference (i.e. Bagging REP Tree, KStar, Model Trees Regression) achieve mean RMSE of 3.861, 3.787 and 4.087 respectively which shows how suitable the wavelet-based least squares estimator is for the non-linear additive model setting. Even though it could be argued that our comparison is based on results that were obtained under different settings than the baseline experiments, the obtained RMSE shows competitive results for the wavelet-based model. Moreover, the estimation experiments conducted using 85\% of the data for training and the remaining 15\% for testing suggest that the prediction RMSE could be even smaller than 4.17, which together with the simplicity of implementation positions the wavelet-based least squares method as a competitive for this kind of problems.
\end{enumerate}

\section{Conclusions and Discussion}\label{Conclusions}

This paper introduced a wavelet-based methodology for the non-parametric estimation and prediction of non-linear additive regression models with NESD. The proposed estimator is based on the projection of the unknown additive functions onto the space $V_{J}$ generated by an orthonormal wavelet basis. In this setting, the data driven wavelet coefficients that define the model are obtained using a thresholded least squares estimates.

\medskip
For the proposed estimator, we showed statistical properties of strong consistency and illustrated practical results using simulations with different exemplary baseline functions. Moreover, we provided convergence rates and optimal choices for the multiresolution index $J$ and the truncation parameter $\beta_{n}$.

\medskip
Our results show that our estimator doesn't suffer from the curse of dimensionality, and was observed to be robust with respect to sample size and noise variance in the model. In fact, our results show that the proposed method is able to successfully identify and predict the underlying model functions and response for relatively small sample sizes.

\medskip
As was seen in the sequel, the proposed estimators are completely data driven with only a few parameters of choice left to the user (multiresolution index $J$, wavelet filter and truncating parameter $\beta_{n}$). Also, the block-matrix based structure introduces computational speed and makes the estimators suitable for real-life applications. In our model, we used of Daubechies-Lagarias's algorithm for the evaluation of the scaling functions $\phi_{Jk}^{per}$ at the observed sample points $X_{ij}$.

\medskip
From a real data application viewpoint, in section \ref{LSApplication} we tested the proposed least squares method using a real data set that was extensively analyzed by Tufekci (2014) \cite{Tufekci2014}. The obtained results show that the proposed estimators are capable of uncover the existing non-linear relationships between the response and predictors, while achieving a high predictive accuracy. In particular, the wavelet-based least squares method showed to be more accurate than the additive model based on back-fitting used in \cite{Tufekci2014}.

\medskip

In terms of some of the drawbacks that were observed throughout this research for the proposed method, it is possible to obtain abnormally large wavelet coefficients in those design regions were the number of samples is small (this is highly likely to occur at the tails of the design distribution); Also, some problems may arise at the boundaries of the support due to the periodic wavelets extension. Nonetheless, it is possible to adjust the truncating parameter $\beta_{n}$ using cross-validation, which minimizes the effect of those large wavelet coefficients that induce errors in the prediction of the response and may contribute to reduce the effect predictors following exponentially decaying distributions.

\medskip

In summary, based on the theoretical properties and results obtained in this paper, we can argue that the proposed estimators posses interesting interpretations and results and add value to practical data analysis: it has good asymptotic properties, is able to identify models that might be hard to do using other methods and also, it is relatively easy to implement which increases its potential to reach a wide variety of users.

\newpage
\bibliography{Refpaper3}
\bibliographystyle{elsarticle-num}

\newpage

\appendix

\section{Previous Theorems and definitions}\label{previousThms}

In this section, we provide important definitions and results previously published that are used to derive the theoretical properties of the proposed estimators.

\subsection{Theorem P1 (Pollard 1984)}\label{ThPollard1}

Consider a class of functions $\mathcal{G}=\left\{g\,,\,g:\mathbb{R}^{p}\rightarrow[0,B]\right\}$, then for any $n\in\mathbb{N}$ and any $\epsilon>0$:

\begin{equation}\label{eqLSTh1:1}
\mathbb{P}\left\{\mathop{\sup}\limits_{g\in\mathcal{G}}\left|\frac{1}{n}\sum_{i=1}^{n}g(\textbf{Z}_{i})-\mathbb{E}[g(\textbf{Z})] \right|>\epsilon \right\} \leq 8\cdot \mathbb{E}\left[ \mathcal{N}_{1}\left(\frac{\epsilon}{8},\mathcal{G},\textbf{z}_{1}^{n}\right)\right]\cdot e^{-\frac{n\cdot\epsilon^{2}}{128B^{2}}}\,,
\end{equation}
where $B<\infty$ (i.e. the functions $g$ are uniformly bounded over the class $\mathcal{G}$), $\left\{\textbf{Z},\textbf{Z}_{i}\right\}_{i=1}^{n}$ is an iid sample of random variables in $\mathbb{R}^{p}$, $\mathcal{N}_{1}\left(\frac{\epsilon}{8},\mathcal{G},\textbf{z}_{1}^{n}\right)$ is the $\mathbb{L}_{1}$ $\frac{\epsilon}{8}$-covering number of $\mathcal{G}$ on $\textbf{z}_{1}^{n}=\left\{\textbf{Z}_{i}\right\}_{i=1}^{n}$.
\medskip
This is the smallest $N\in\mathbb{N}$ such that for every function $g\in\mathcal{G}$ and a given probability measure $\mu$ on $\mathbb{R}^{p}$ and $s\geq1$ there exists a $j=j(g)\in \left\{1,...,N\right\}$ for which $||g-g_{j}||_{\mathbb{L}_{1}(\mu)} < \epsilon$, for $||g||_{\mathbb{L}_{1}(\mu)}:=\left(\int |f(z)|d\mu_{n}\right)=\left(\frac{1}{n}\sum_{i=1}^{n}|g(\textbf{z}_{i})-g_{j}(\textbf{z}_{i}) |^{s} \right)^{\frac{1}{s}}$.

\medskip

A detailed proof of this theorem and a illustrative discussion about covering numbers can be found in \cite{Pollard1984} and \cite{Gyorfi2002}.

\subsection{Lemma G1 (Gyorfi et al. 2002)}\label{LemmaG1}

Consider a probability measure $\mu$ on $\mathbb{R}^{p}$, $s\geq1$, $\epsilon>0$ and a class of functions $\mathcal{G}$ on $\mathbb{R}^{p}$. Then:

\begin{equation}\label{eqLemma1:1}
\mathcal{M}\left(2\epsilon,\mathcal{G},||\cdot||_{\mathbb{L}_{s}(\mu)}\right)\leq \mathcal{N}\left(\epsilon,\mathcal{G},||\cdot||_{\mathbb{L}_{s}(\mu)}\right)\leq \mathcal{M}\left(\epsilon,\mathcal{G},||\cdot||_{\mathbb{L}_{s}(\mu)}\right)\,.
\end{equation}
Here, $\mathcal{M}\left(\epsilon,\mathcal{G},||\cdot||_{\mathbb{L}_{s}(\mu)}\right)$ represents the size of the largest $\epsilon$-packing of $\mathcal{G}$ with respect to $||\cdot||_{\mathbb{L}_{s}(\mu)}$. This is the largest $N\in\mathbb{N}$ such that the collection of functions $\left\{g_{1},...,g_{N} \right\}\,\,\in \mathcal{G}$ satisfy $||g_{j}-g_{l}||_{\mathbb{L}_{s}(\mu)}\geq \epsilon$, for $||g||_{\mathbb{L}_{s}(\mu)}:=\left(\int |f(z)|^{s}d\mu\right)^{\frac{1}{s}}$.

\medskip

A detailed proof of this Lemma, together with definitions and details about covering and packing numbers can be found in section 9 of \cite{Gyorfi2002}.

\subsection{Theorem G2 (Gyorfi et al. 2002)}\label{ThG2}

Before stating this theorem, consider the following definitions:

\paragraph*{Definitions G2.1}

Consider a class of subsets of $\mathbb{R}^{p}$ denoted by $\mathcal{A}$. Let $n\in\mathbb{N}$. Then,

\begin{enumerate}[(i)]
\item For a sample $\textbf{z}_{1},...,\textbf{z}_{n}$ $\in$ $\mathbb{R}^{p}$, define $s\left(\mathcal{A},\left\{\textbf{z}_{1},...,\textbf{z}_{n} \right\} \right)$ as the number of different subsets of $\left\{\textbf{z}_{1},...,\textbf{z}_{n} \right\}$ that can be expressed as sets of the form $A\cap\left\{\textbf{z}_{1},...,\textbf{z}_{n} \right\} $ for $A\in\mathcal{A}$. This is $s\left(\mathcal{A},\left\{\textbf{z}_{1},...,\textbf{z}_{n} \right\} \right)=\left|A\cap\left\{\textbf{z}_{1},...,\textbf{z}_{n} \right\}  : A\in\mathcal{A} \right|$.
\item If for a set $H \subseteq \mathbb{R}^{p}$ $s\left(\mathcal{A},H\right)=2^{n}$ (i.e. every subset of $H$ can be represented as $A\cap H$ for $A\in\mathcal{A}$), then we say that $\mathcal{A}$ shatters $H$.
\item The $n$-th shatter coefficient of $\mathcal{A}$ given a sample containing $n$ points is the maximal number of different subsets of the $n$ points that are contained by sets in $\mathcal{A}$, therefore, they can be represented as $A\cap H$ for $A\in\mathcal{A}$. We denote the $n$-th shatter coefficient of $\mathcal{A}$ as $S(\mathcal{A},n)$. Note that for all $n>k$ we have that $S(\mathcal{A},k)<2^{k}$ implies $S(\mathcal{A},n)<2^{n}$.
\item Suppose that $\mathcal{A} \subseteq \mathbb{R}^{p} \neq \emptyset$, the VC dimension (Vapnis-Chervonenkis dimension) $V_{\mathcal{A}}$ of $\mathcal{A}$ corresponds to the largest integer $n$ such that there exists a set of $n$ points in $\mathbb{R}^{p}$ that can be shattered by $\mathcal{A}$. This is $V_{\mathcal{A}}=\sup\left\{n\in\mathbb{N}:\, S(\mathcal{A},n)=2^{n}\right\}$.
\item Suppose $\mathcal{G}$ is a class of functions in $\mathbb{R}^{p}$ such that $\forall g\in\mathcal{G}\,,g:\mathbb{R}^{p}\rightarrow[0,B]$. Let's define the set $\mathcal{G}^{+}:=\left\{{(\textbf{z},t)\in \mathbb{R}^{p}\times\mathbb{R}\,;t\leq(\textbf{z})}\,;g\in\mathcal{G} \right\}$. This set corresponds to the set of all subgraphs of the functions contained in the set $\mathcal{G}$.
\end{enumerate}

Now, consider a class of functions $\mathcal{G}$ in $\mathbb{R}^{p}$ such that $\forall g\in\mathcal{G}\,,g:\mathbb{R}^{p}\rightarrow[0,B]$ with $V_{\mathcal{G}^{+}}\geq2$. Let $s\geq1$ and $\mu$ a probability measure on $\mathbb{R}^{p}$ and let $0<\epsilon<\frac{B}{4}$; then:

\begin{equation}\label{eqTH2:1}
\mathcal{M}\left(\epsilon,\mathcal{G},||\cdot||_{\mathbb{L}_{s}(\mu)}\right) \leq 3\left(\frac{2eB^{s}}{\epsilon^{s}}\log\left(\frac{3eB^{s}}{\epsilon^{s}}\right) \right)^{V_{\mathcal{G}^{+}}}\,.
\end{equation}
A detailed proof of this Theorem, together with definitions and details about shattering numbers and VC dimension can be found in section 9 of \cite{Gyorfi2002}.

\subsection{Theorem G3 (Gyorfi et al. 2002)}\label{ThG3}

This theorem provides an upper bound on the VC dimension for $r$-dimensional vector spaces. Consider $\mathcal{G}$ to be a $r$-dimensional vector space of real functions defined on $\mathbb{R}^{p}$. Let $\mathcal{A}=\left\{{\textbf{z}:g(\textbf{x})\geq0}:g\in\mathcal{G} \right\}$. Then:

\begin{equation}\label{eqTH3:1}
V_{\mathcal{A}}\leq r\,.
\end{equation}
A detailed proof of this Theorem can be found in section 9.4 of \cite{Gyorfi2002}.
\medskip

\subsection{Theorem G4 (Gyorfi et al. 2002)}\label{ThG4}

This theorem provides necessary and sufficient conditions for the consistency of least squares estimators. Consider $\mathcal{F}_{n}=\mathcal{F}_{n}\left(\left\{(Y_{i},\textbf{X}_{i}\right\}_{i=1}^{n} \right)$ a class of functions $f:\mathbb{R}^{p}\rightarrow\mathbb{R}$. Let $\beta_{n}$ be a parameter depending on the sample size $n$ such that $\beta_{n}\rightarrow\infty$ as $n\rightarrow\infty$. Let $\hat{f}_{J(n)}$ be defined as in (\ref{eqLS:3}) and $f_{J(n)}=T_{\beta_{n}}\hat{f}_{J(n)}$ (i.e. the truncated version of $\hat{f}_{J(n)}$) and $\mu$ be a lebesgue measure in $\mathbb{R}^{p}$; Then :

\begin{enumerate}[(i)]
\item If for all $L>0$ the following conditions hold: \begin{eqnarray}
            \label{eqTh4:1}
            \mathop{\lim}\limits_{n\rightarrow\infty} \mathop{\inf}\limits_{f\in\mathcal{F}_{n}:||f||_{\infty}\leq\beta_{n}}\int \left|f(\textbf{x})-f_{A}(\textbf{x}) \right|^{2}\mu(d\textbf{x})&=&0\,\,(a.s.) \,,\\
            \label{eqTh4:2}
            \mathop{\lim}\limits_{n\rightarrow\infty} \mathop{\sup}\limits_{f\in T_{\beta_{n}}\mathcal{F}_{n}}\left|\frac{1}{n}\sum_{i=1}^{n}|f(\textbf{X}_{i})-Y_{i,L}|^{2}-\mathbb{E}\left[(f(\textbf{X})-Y_{L} )^{2} \right] \right|&=&0\,\,(a.s.)\,,
         \end{eqnarray}%

then: \begin{center} $\mathop{\lim}\limits_{n\rightarrow\infty} \int \left|f_{J(n)}(\textbf{x})-f_{A}(\textbf{x}) \right|^{2}\mu(d\textbf{x})=0$, almost surely (a.s.).\end{center}

\medskip

Here, $Y_{L}=T_{L}Y=\left\{
                      \begin{array}{lr}
                        Y & |Y| \leq \beta_{n} \\
                        \beta_{n}\cdot \text{sign}(Y) & |Y|>\beta_{n}
                      \end{array}
                    \right\}$.

\item If for all $L>0$ the following conditions hold: \begin{eqnarray}
            \label{eqTh4:3}
            \mathop{\lim}\limits_{n\rightarrow\infty}\mathbb{E}\left\{ \mathop{\inf}\limits_{f\in\mathcal{F}_{n}:||f||_{\infty}\leq\beta_{n}}\int \left|f(\textbf{x})-f_{A}(\textbf{x}) \right|^{2}\mu(d\textbf{x})\right\}&=&0\,, \\
            \label{eqTh4:4}
            \mathop{\lim}\limits_{n\rightarrow\infty}\mathbb{E}\left\{ \mathop{\sup}\limits_{f\in T_{\beta_{n}}\mathcal{F}_{n}}\left|\frac{1}{n}\sum_{i=1}^{n}|f(\textbf{X}_{i})-Y_{i,L}|^{2}-\mathbb{E}\left[(f(\textbf{X})-Y_{L} )^{2} \right] \right|\right\}&=&0\,,
         \end{eqnarray}%

then: \begin{center} $\mathop{\lim}\limits_{n\rightarrow\infty} \mathbb{E}\left\{\int \left|f_{J(n)}(\textbf{x})-f_{A}(\textbf{x}) \right|^{2}\mu(d\textbf{x})\right\}=0$.\end{center}
\end{enumerate}
A detailed proof of this Theorem can be found in section 10.1 of \cite{Gyorfi2002}.

\medskip

This theorem shows that strong consistency is achieved for any least squares estimator obtained over a data-dependent class of functions $\mathcal{F}_{n}$, truncated by a suitable parameter $\beta_{n}$ that depends on the sample size and converges to $\infty$,  and provided that the approximation error (\ref{eqTh4:1}) converges to zero a.s. (i.e. for every $\omega\in\Omega$ such that $\mathbb{P}(\omega)\neq 0$, $f_{n}(\omega)\rightarrow f_{A}$ with probability 1), and that the empirical $\mathbb{L}_{2}$ norm uniformly converges to the $\mathbb{L}_{2}(\mu)$ norm over the set of functions $T_{\beta_{n}}\mathcal{F}_{n}$.

\subsection{Theorem P2 (Pollard 1984)}\label{ThPollard2}

Suppose $\mathcal{F}$ is a class of functions $f\,:\,\mathbb{R}^{p}\rightarrow\mathbb{R}$ such that $\forall\,\textbf{x}\in\mathbb{R}^{p}$, $|f(\textbf{x})|<B$, for $0<B<\infty$. Then, for $\epsilon>0$ (arbitrary) it follows:
\begin{equation}\label{eqLSThP2:1}
\mathbb{P}\left\{\exists\,f\in\mathcal{F}:\,||f||-2||f||_{n}>\epsilon \right\} \leq 3\cdot\mathbb{E}\left[\mathcal{N}_{2}\left(\frac{\sqrt{2}}{24}\epsilon,\,\mathcal{F},\textbf{X}_{1}^{2n}\right)\right]e^{-\frac{n\epsilon^{2}}{288\,B^{2}}}\,,
\end{equation}
where $||g||^{2}=\int_{\mathbb{R}^{p}}|g(\textbf{x})|^{2}d\textbf{x}$ and $||g||^{2}_{n}=\frac{1}{n}\sum_{i=1}^{n}|g(\textbf{x}_{i})|^{2}$. A detailed proof of this Lemma, together with definitions and details about covering and packing numbers can be found in section 11 of \cite{Gyorfi2002}.

\section{Proof of Theorem 1.} \label{proof:Theorem1}
Suppose an orthonormal set of functions $\left\{ \phi^{per}_{J,k}(x), k =0,...,2^{J}-1\right\}$ which is dense in $\mathbb{L}_{2}(\nu([0,1]))$ for $\nu\in \Upsilon$, and let $\Upsilon$ be the set of bounded lebesgue measures in $[0,1]$. Suppose $\mu$ is a bounded lebesgue measure in $[0,1]^{p}$, and the following conditions are satisfied for the scaling function $\phi$:

\begin{enumerate}
  \item $\exists\,\Phi$, bounded and non-increasing function in $\mathbb{R}$ such that $\int\Phi(|u|)du<\infty$ and $|\phi(u)|\leq \Phi(|u|)$ almost everywhere (a.e.).
  \item In addition, $\int_{\mathbb{R}}|u|^{N+1}\Phi(|u|)du<\infty$ for some $N\geq0$.
  \item $\exists\,F$, integrable, such that $|K(x,y)|\leq F(x-y)$, $\forall x,y \in \mathbb{R}$, for $K(x,y)=\sum_{k}\phi(x-k)\phi(y-k) $.
  \item Suppose $\phi$ satisfies:
  \begin{enumerate}
    \item $\sum_{k}|\hat{\phi}(\xi+2k\pi)|^{2}=1$, a.e., where $\hat{\phi}$ denotes the Fourier transform of the scaling function $\phi$. \label{ass:w1}
    \item $\hat{\phi}(\xi)=\hat{\phi}(\frac{\xi}{2})m_{0}(\frac{\xi}{2})$, where $m_{0}(\xi)$ is a $2\pi$-periodic function and $m_{0}\,\in\,\mathbb{L}_{2}(0,2\pi)$.\label{ass:w2}
  \end{enumerate}
  \item $\int_{\mathbb{R}}x^{k}\psi(x)dx=0$, for $k=0,1,...,N$, $N\geq1$ where $\psi$ is the mother wavelet corresponding to $\phi$.
  \item The functions $\left\{f_{l}\right\}_{l=1}^{p}$, are such that $f_{l}\,\in\,L_{\infty}([0,1])$ and $f_{l}\,\in\,W_{\infty}^{m+1}([0,1])\,,\,m \geq N$, where $W_{\infty}^{m}([0,1])$ denotes the space of functions that are $m$-times weakly-differentiable and $f_{l}^{(k)}\,\in\,L_{\infty}([0,1])\,,\, k=1,...,m$.
  \item $\theta_{\phi}(x):=\sum_{k}|\phi(x-k)|$ such that $||\theta_{\phi}||_{\infty}<\infty$.
\end{enumerate}
Under Corollary 8.2 \cite{Hardle1998}, if $f\in W_{\infty}^{N+1}([0,1])$ then $||K_{J}f-f||_{\infty}^{p}=\mathcal{O}\left(2^{-pJ(N+1)}\right)\,,\,p\geq1$. Furthermore, assume condition (\textbf{A3}) is satisfied. Define the set of functions:
\begin{equation} \mathcal{F}_{n}=\left\{f:[0,1]^{p}\rightarrow\mathbb{R}\,|\,f(\textbf{x})=\sum_{j=1}^{p}\sum_{k=0}^{2^{J}-1}c_{Jk}^{(j)}\phi_{Jk}^{per}(x_{j})\,;\,J=J(n)\right\}\,, \label{eqTh1:3} \end{equation} %
where $x_{j}\,,j=1,...,p$ corresponds to the $j$-th component of the vector $\textbf{x}\in[0,1]^{p}$. Also, let $\beta_{n}>0$ be a parameter depending on the sample and assume $\mathbb{E}\left[Y^{2}\right]<\infty$. Define $\hat{f}_{J(n)}$ as in (\ref{eqLS:3}) and let $f_{J(n)}=T_{\beta_{n}}\hat{f}_{J(n)}:=\hat{f}_{J(n)}\mathbbm{1}_{\left\{|\hat{f}_{J(n)}|\leq \beta_{n}\right\}}+\text{sign}(\hat{f}_{J(n)})\beta_{n}\mathbbm{1}_{\left\{|\hat{f}_{J(n)}|> \beta_{n}\right\}}$, $\mathcal{K}_{n}=2^{J(n)}$. Assume the following conditions hold:
\begin{enumerate}[(i)]
\item $\beta_{n}\rightarrow\infty$ as $n\rightarrow\infty$.
\item $\frac{\mathcal{K}_{n}\beta_{n}^{4}\log\left(\beta_{n}\right)}{n}\rightarrow0$ as $n\rightarrow\infty$.
\item For some $\delta>0$ as $n\rightarrow\infty$ $\frac{n^{1-\delta}}{\beta_{n}^{4}}\rightarrow\infty$.
\end{enumerate}
Then:
\begin{eqnarray}
\label{eqproofTh1:1}
\mathop{\lim}\limits_{n\rightarrow\infty} \int \left|f_{J(n)}(\textbf{x})-f_{A}(\textbf{x}) \right|^{2}\mu(d\textbf{x})=0\,\,\,\, \text{(a.s.)}\,, \\
\label{eqproofTh1:2}
\mathop{\lim}\limits_{n\rightarrow\infty} \mathbb{E}\left\{\int \left|f_{J(n)}(\textbf{x})-f_{A}(\textbf{x}) \right|^{2}\mu(d\textbf{x})\right\}=0\,.
\end{eqnarray}
\subsection*{Proof}
The proof for this theorem is based on the application of Theorem G4 (Gyorfi et al. 2002) described in \ref{ThG4}, checking conditions (\ref{eqTh4:1})-(\ref{eqTh4:4}) are satisfied.
\medskip

This proof is composed of 2 parts: the first shows that conditions (\ref{eqTh4:1}) and (\ref{eqTh4:3}) are implied by assumption (i). The second part shows that assumptions (ii) and (iii) imply conditions (\ref{eqTh4:2}) and (\ref{eqTh4:4}) of Theorem \ref{ThG4}.

\subsubsection*{Part 1}
Consider an arbitrary $\epsilon>0$. Then for $f\in\mathcal{F}_{n}$, it follows:

\begin{eqnarray}
\nonumber
\int_{[0,1]^{p}}\left|f(\textbf{x})-f_{A}(\textbf{x})\right|^{2}\mu(d\textbf{x}) &=& \int_{[0,1]^{p}}\left|\sum_{j=1}^{p}\left(\sum_{k=0}^{2^{J}-1}c_{Jk}^{(j)}\phi_{Jk}^{per}(x_{j})-f_{j}(x_{j})\right)\right|^{2}\mu(d\textbf{x}) \\
\nonumber
&\leq & p\cdot \sum_{j=1}^{p}\int_{[0,1]^{p}}\left(\sum_{k=0}^{2^{J}-1}c_{Jk}^{(j)}\phi_{Jk}^{per}(x_{j})-f_{j}(x_{j})\right)^{2}\mu(d\textbf{x}) \\
\label{eqProofTh1P1:1}
&\leq & p\cdot \sum_{j=1}^{p}\int_{0}^{1}\left(\sum_{k=0}^{2^{J}-1}c_{Jk}^{(j)}\phi_{Jk}^{per}(x_{j})-f_{j}(x_{j})\right)^{2}\nu_{j}(dx_{j})\,,
\end{eqnarray}
where $\nu_{1},...,\nu_{p}$ are bounded lebesgue measures on $[0,1]$ (since $\mu$ is a bounded lebesgue measure in $[0,1]^{p}$). Since $\left\{ \phi^{per}_{j,k}(x), k =0,...,2^{j}-1,\,j\geq0\right\}$ is dense in $\mathbb{L}_{2}(\nu([0,1]))$, by Proposition 1 in \ref{prop8}:
\begin{center} $\,\,\,\exists\,\,\left\{c_{J,0}^{(1)*},...,c_{J,2^{J}-1}^{(1)*},...,c_{J,0}^{(p)*},...,c_{J,2^{J}-1}^{(p)*} \right\}$, \end{center}
for which $J=J^{*}(n_{0}(\epsilon))$ such that:

\begin{equation} \label{eqProofTh1P1:2}
\int_{[0,1]^{p}}\left|\sum_{j=1}^{p}\left(\sum_{k=0}^{2^{J}-1}c_{J,k}^{(j)}\phi_{J,k}^{per}(x_{j})-f_{j}(x_{j})\right) \right|^{2}\mu(d\textbf{x})\leq \epsilon\,.
\end{equation}
Therefore, for a given $\epsilon>0$, it is possible to find $n_{0}(\epsilon)$ such that for $J^{*}=J(n_{0}(\epsilon))$ (\ref{eqProofTh1P1:2}) holds.
\medskip

Now for a fixed $n=n_{0}(\epsilon)$ the set $\mathcal{F}_{n}$ is composed of functions that are uniformly bounded by a parameter depending on the sample size. In fact, it is possible to show that $||f||_{\infty}\leq ||\theta_{\phi}||_{\infty}||f_{j}^{*}||_{\infty}\cdot 2^{\frac{J(n_{0}(\epsilon))}{2}}$, where $||f_{j}^{*}||_{\infty}=\mathop{\max}\limits_{j=1,...,p}||f_{j}||_{\infty}$. Therefore, for an arbitrary $\epsilon>0$, and for all $n\leq n_{0}(\epsilon)$, $\,\exists\,\beta_{n}>0$ such that:

\begin{equation}
\nonumber
\sum_{j=1}^{p}\sum_{k=0}^{2^{J(n)}-1}c_{Jk}^{(j)*}\phi_{Jk}^{per}(x_{j})\,\,\in \,\,\left\{f\in\mathcal{F}_{n}\, | \,||f||_{\infty} \leq \beta_{n_{0}(\epsilon)} \right\}\,.
\end{equation}
From this last result and (\ref{eqProofTh1P1:1}),(\ref{eqProofTh1P1:2}), for $n\geq n_{0}(\epsilon)$ it follows:

\begin{equation}\label{eqProofTh1P1:3}
\mathop{\inf}\limits_{\left\{f\in\mathcal{F}_{n}\, | \,||f||_{\infty} \leq \beta_{n} \right\}}\int_{[0,1]^{p}}\left|f(\textbf{x})-f_{A}(\textbf{x})\right|^{2}\mu(d\textbf{x})\leq \epsilon\,.
\end{equation}
Since $\epsilon>0$ is arbitrary, (\ref{eqProofTh1P1:3}) implies:

\begin{equation}\label{eqProofTh1P1:4}
\mathop{\lim}\limits_{n\rightarrow\infty}\left\{\mathop{\inf}\limits_{\left\{f\in\mathcal{F}_{n}\, | \,||f||_{\infty} \leq \beta_{n} \right\}}\int_{[0,1]^{p}}\left|f(\textbf{x})-f_{A}(\textbf{x})\right|^{2}\mu(d\textbf{x})\right\} = 0\,,
\end{equation}
which shows that as $J=J(n)\rightarrow\infty$ ($n\rightarrow\infty$) and $\beta_{n}\rightarrow\infty$ ($n\rightarrow\infty$),  (\ref{eqTh4:1}) is satisfied.
\medskip

From (\ref{eqProofTh1P1:3}) and the last result, the dominated convergence theorem implies:

\begin{equation}\label{eqProofTh1P1:5}
\mathop{\lim}\limits_{n\rightarrow\infty}\mathbb{E}\left\{\mathop{\inf}\limits_{\left\{f\in\mathcal{F}_{n}\, | \,||f||_{\infty} \leq \beta_{n} \right\}}\int_{[0,1]^{p}}\left|f(\textbf{x})-f_{A}(\textbf{x})\right|^{2}\mu(d\textbf{x})\right\} = 0\,,
\end{equation}
therefore, (\ref{eqTh4:3}) is also implied, provided $J=J(n)\rightarrow\infty$ ($n\rightarrow\infty$) and $\beta_{n}\rightarrow\infty$ ($n\rightarrow\infty$).

\subsubsection*{Part 2}

In this part, we use results provided in section \ref{ThG4} of the appendix. Consider $L>0$ arbitrary and assume (wlog) that $L<\beta_{n}$. Define $\textbf{Z}=(\textbf{X},Y)$ and $\textbf{Z}_{i}=(\textbf{X}_{i},Y_{i})$ for $i=1,...,n$. Also, define the set of functions:
\begin{equation}
\nonumber
\mathcal{G}_{n}=\left\{g,:\,[0,1]^{p}\times\mathbb{R}\rightarrow\mathbb{R}\,:\,\exists\,f\in T_{\beta_{n}}\mathcal{F}_{n}\,\,\text{s.t.}\,\,g(\textbf{X},y)=|f(\textbf{X})-T_{L}Y|^{2}\right\}\,.
\end{equation}
Note that the last definition implies that $\mathop{\sup}\limits_{f\in T_{\beta_{n}}\mathcal{F}_{n}}\left|\frac{1}{n}\sum_{i=1}^{n}|f(\textbf{X}_{i})-Y_{i,L}|^{2}-\mathbb{E}\left[(f(\textbf{X})-Y_{L} )^{2} \right] \right|$ is equivalent to:
\begin{equation}
\nonumber
\mathop{\sup}\limits_{g\in \mathcal{G}_{n}}\left|\frac{1}{n}\sum_{i=1}^{n}g(\textbf{Z}_{i})-\mathbb{E}\left[g(\textbf{Z})\right] \right|\,.
\end{equation}
Moreover, since it is assumed that $L<\beta_{n}$, every function $g\in\mathcal{G}_{n}$ satisfies $0\leq g(\textbf{Z})\leq 4\beta_{n}^{2}$. This allows the application of Theorem P1 (Pollard 1984) as follows:

\medskip

For an arbitrary $\epsilon>0$, it follows:

\begin{equation}\label{eqProofTh1P2:1}
\mathbb{P}\left\{\mathop{\sup}\limits_{g\in \mathcal{G}_{n}}\left|\frac{1}{n}\sum_{i=1}^{n}g(\textbf{Z}_{i})-\mathbb{E}\left[g(\textbf{Z})\right] \right|>\epsilon \right\} \leq 8\cdot\mathbb{E}\left[\mathcal{N}_{1}\left(\frac{\epsilon}{8},\mathcal{G}_{n},\textbf{z}_{1}^{n} \right) \right]e^{-\frac{n\epsilon^{2}}{2048\beta_{n}^{4}}}\,.
\end{equation}
Lemma G1 shows that $\mathcal{N}_{1}\left(\frac{\epsilon}{8},\mathcal{G}_{n},\textbf{z}_{1}^{n} \right)\leq \mathcal{M}_{1}\left(\frac{\epsilon}{8},\mathcal{G}_{n},\textbf{z}_{1}^{n} \right)$. Therefore, a relation between $\mathcal{M}_{1}\left(\frac{\epsilon}{8},\mathcal{G}_{n},\textbf{z}_{1}^{n} \right)$ and $\mathcal{M}_{1}\left(\lambda,T_{\beta_{n}}\mathcal{F}_{n},\textbf{X}_{1}^{n} \right)$ needs to be established for some $\lambda=\lambda(\epsilon)>0$.

\medskip

Consider $g_{1},g_{2}$ $\in \mathcal{G}_{n}$ (i.e. $\exists \, f_{1},f_{2}\,\in T_{\beta_{n}}\mathcal{F}_{n}\,\text{s.t.}\,g(\textbf{X},y)=|f(\textbf{X})-T_{L}Y|^{2}$), then if $\left\{g_{1},...,g_{M}\right\}$ is an $\mathbb{L}_{1}$-$\frac{\epsilon}{8}$ packing of $\mathcal{G}_{n}$ on $\textbf{z}_{1}^{n}$, $\forall 1\leq j < m \leq M$ it holds:
\begin{eqnarray}
\nonumber
  \frac{1}{n}\sum_{i=1}^{n}\left|g_{j}(\textbf{z}_{i})-g_{m}(\textbf{z}_{i}) \right| &\geq& \frac{\epsilon}{8}\,.
\end{eqnarray}
Using the definition of $\mathcal{G}_{n}$, it follows:
\begin{eqnarray}
\nonumber 
  \frac{1}{n}\sum_{i=1}^{n}\left|g_{1}(\textbf{z}_{i})-g_{2}(\textbf{z}_{i}) \right| &=& \frac{1}{n}\sum_{i=1}^{n}\left||f_{1}(\textbf{X}_{i})-T_{L}Y_{i}|^{2}-|f_{2}(\textbf{X}_{i})-T_{L}Y_{i}|^{2} \right| \\
\nonumber
  &=& \frac{1}{n}\sum_{i=1}^{n}\left(\left|f_{1}(\textbf{X}_{i})-f_{2}(\textbf{X}_{i}) \right|\left|f_{1}(\textbf{X}_{i})+f_{2}(\textbf{X}_{i})-2T_{L}Y_{i} \right| \right) \\
\nonumber
 &\leq & \frac{1}{n}\sum_{i=1}^{n}\left|f_{1}(\textbf{X}_{i})-f_{2}(\textbf{X}_{i}) \right|\cdot 4\beta_{n} \\
\nonumber
\frac{\epsilon}{32\beta_{n}} &\leq& \sum_{i=1}^{n}\left|f_{1}(\textbf{X}_{i})-f_{2}(\textbf{X}_{i}) \right|\,.
\end{eqnarray}
Therefore, if $\left\{g_{1},...,g_{M}\right\}$ is an $\mathbb{L}_{1}$-$\frac{\epsilon}{8}$ packing of $\mathcal{G}_{n}$ on $\textbf{z}_{1}^{n}$, then $\left\{f_{1},...,f_{M}\right\}$ is an $\mathbb{L}_{1}$-$\frac{\epsilon}{32\beta_{n}}$ packing of $T_{\beta_{n}}\mathcal{F}_{n}$ on $\textbf{X}_{1}^{n}$. Thus this result implies:
\begin{equation}
\mathcal{M}_{1}\left(\frac{\epsilon}{8},\mathcal{G}_{n},\textbf{z}_{1}^{n} \right) \leq \mathcal{M}_{1}\left(\frac{\epsilon}{32\beta_{n}},T_{\beta_{n}}\mathcal{F}_{n},\textbf{X}_{1}^{n} \right)
\end{equation}
Substituting the last result in (\ref{eqProofTh1P2:1}), leads to:

\begin{equation}\label{eqProofTh1P2:2}
\mathbb{P}\left\{\mathop{\sup}\limits_{g\in \mathcal{G}_{n}}\left|\frac{1}{n}\sum_{i=1}^{n}g(\textbf{Z}_{i})-\mathbb{E}\left[g(\textbf{Z})\right] \right|>\epsilon \right\} \leq 8\cdot\mathbb{E}\left[\mathcal{M}_{1}\left(\frac{\epsilon}{32\beta_{n}},T_{\beta_{n}}\mathcal{F}_{n},\textbf{X}_{1}^{n} \right) \right]e^{-\frac{n\epsilon^{2}}{2048\beta_{n}^{4}}}\,.
\end{equation}
Now, applying Theorem G2, for $0<\epsilon<\frac{\beta_{n}}{4}$ it follows:
\begin{equation}\label{eqProofTh1P2:3}
\mathcal{M}_{1}\left(\frac{\epsilon}{32\beta_{n}},T_{\beta_{n}}\mathcal{F}_{n},\textbf{X}_{1}^{n} \right) \leq 3\left(\frac{128\,e\,\beta_{n}^{2}}{\epsilon}\log\left(\frac{192\,e\,\beta_{n}^{2}}{\epsilon} \right) \right)^{V_{T_{\beta_{n}}\mathcal{F}_{n}^{+}}}\,.
\end{equation}
Since $T_{\beta_{n}}\mathcal{F}_{n}^{+}=\left\{(\textbf{x},t)\in[0,1]^{p}\times\mathbb{R}\,:\,t\leq f(\textbf{x})\,,\,f\in T_{\beta_{n}}\mathcal{F}_{n} \right\} $, for $t>\beta_{n}$ the pair $(\textbf{x},t)\notin T_{\beta_{n}}\mathcal{F}_{n}^{+}$. On the contrary, when $t\leq\beta_{n}$ since $\forall f\in T_{\beta_{n}}\mathcal{F}_{n}$ $\beta_{n}\leq f \leq \beta_{n}$, every pair $(\textbf{x},t)\in T_{\beta_{n}}\mathcal{F}_{n}^{+}$. This implies:
\begin{equation}\label{eqProofTh1P2:4}
V_{T_{\beta_{n}}\mathcal{F}_{n}^{+}}\leq V_{\mathcal{F}_{n}^{+}}\,.
\end{equation}
Similarly, since $dim(\mathcal{F}_{n})=p\cdot2^{J}$, Theorem G3 implies:
\begin{equation}\label{eqProofTh1P2:5}
V_{\mathcal{F}_{n}^{+}}\leq p\cdot2^{J}+1\,.
\end{equation}
Combining (\ref{eqProofTh1P2:3}), (\ref{eqProofTh1P2:4}), and (\ref{eqProofTh1P2:5}), it is possible to express (\ref{eqProofTh1P2:2}) as:

\begin{eqnarray}
\nonumber
\mathbb{P}\left\{\mathop{\sup}\limits_{g\in \mathcal{G}_{n}}\left|\frac{1}{n}\sum_{i=1}^{n}g(\textbf{Z}_{i})-\mathbb{E}\left[g(\textbf{Z})\right] \right|>\epsilon \right\} &\leq & 24\cdot\left(\left(\frac{128\,e\,\beta_{n}^{2}}{\epsilon}\log\left(\frac{192\,e\,\beta_{n}^{2}}{\epsilon} \right) \right)^{(p\cdot2^{J}+1)} \right) e^{-\frac{n\epsilon^{2}}{2048\beta_{n}^{4}}}\\
\nonumber
&\leq & 24\cdot e^{2(p\cdot2^{J}+1)\log\left(\frac{192\,e\,\beta_{n}^{2}}{\epsilon}\right)-\frac{n\epsilon^{2}}{2048\beta_{n}^{4}}}\,.
\end{eqnarray}
Finally, it follows:
\begin{eqnarray}
\nonumber 
\mathbb{P}\left\{\mathop{\sup}\limits_{g\in \mathcal{G}_{n}}\left|\frac{1}{n}\sum_{i=1}^{n}g(\textbf{Z}_{i})-\mathbb{E}\left[g(\textbf{Z})\right] \right|>\epsilon \right\} &\leq & \sum_{n=1}^{\infty}\mathbb{P}\left\{\mathop{\sup}\limits_{g\in \mathcal{G}_{n}}\left|\frac{1}{n}\sum_{i=1}^{n}g(\textbf{Z}_{i})-\mathbb{E}\left[g(\textbf{Z})\right] \right|>\epsilon \right\} \\
\nonumber
  &\leq & \sum_{n=1}^{\infty}24\cdot e^{2(p\cdot2^{J}+1)\log\left(\frac{192\,e\,\beta_{n}^{2}}{\epsilon}\right)-\frac{n\epsilon^{2}}{2048\beta_{n}^{4}}} \\
\label{eqProofTh1P2:6}
 &\leq & \sum_{n=1}^{\infty}24\cdot e^{\left\{-n^{\delta}\frac{n^{1-\delta}}{\beta_{n}^{4}}\left(\frac{\epsilon^{2}}{2048}-\frac{2(p\cdot2^{J}+1)\beta_{n}^{4}}{n} \log\left(\frac{192\,e\,\beta_{n}^{2}}{\epsilon}\right)\right) \right\}}\,.
\end{eqnarray}
Notice that if for some $\delta>0$ the following conditions hold:
\begin{enumerate}[(a)]
  \item $\frac{n^{1-\delta}}{\beta_{n}^{4}}\longrightarrow\infty$ as $n\rightarrow\infty$\,,
  \item $\frac{2(p\cdot2^{J}+1)\beta_{n}^{4}}{n} \log\left(\frac{192\,e\,\beta_{n}^{2}}{\epsilon}\right)\longrightarrow\infty$ as $n\rightarrow\infty$,
\end{enumerate}
then the series (\ref{eqProofTh1P2:6}) is absolutely convergent. Denote $\mathcal{K}_{n}=p\cdot2^{J}$ and observe that condition (b) can be bounded as:

\begin{eqnarray}
\nonumber 
  \frac{2(\mathcal{K}_{n}+1)\beta_{n}^{4}}{n} \log\left(\frac{192\,e\,\beta_{n}^{2}}{\epsilon}\right) &\leq & \frac{4(\mathcal{K}_{n}+1)\beta_{n}^{4}\log(\beta_{n})}{n}+\frac{C_{1}(\mathcal{K}_{n}+1)\beta_{n}^{4}}{n} \\
   &\leq & C_{2}\frac{\mathcal{K}_{n}\beta_{n}^{4}\log(\beta_{n})}{n}\,,
\end{eqnarray}
for a constant $C_{2}>0$ independent of $n$.
\medskip

Therefore, if $\frac{\mathcal{K}_{n}\beta_{n}^{4}\log(\beta_{n})}{n}\longrightarrow\infty$ as $n\rightarrow\infty$, then we get condition (b) satisfied by assumption (ii). This implies that the terms in the series (\ref{eqProofTh1P2:6}) go to zero. Therefore:
\begin{equation}
\nonumber
\sum_{n=1}^{\infty}24\cdot e^{\left\{-n^{\delta}\frac{n^{1-\delta}}{\beta_{n}^{4}}\left(\frac{\epsilon^{2}}{2048}-\frac{2(p\cdot2^{J}+1)\beta_{n}^{4}}{n} \log\left(\frac{192\,e\,\beta_{n}^{2}}{\epsilon}\right)\right) \right\}}<\infty\,.
\end{equation}
This result implies that $\exists\,n_{0}(\epsilon)$ such that for $n>n_{0}(\epsilon)$, it follows:

\begin{equation}
\label{eqProofTh1P2:7}
\mathbb{P}\left\{\mathop{\sup}\limits_{g\in \mathcal{G}_{n}}\left|\frac{1}{n}\sum_{i=1}^{n}g(\textbf{Z}_{i})-\mathbb{E}\left[g(\textbf{Z})\right] \right|>\epsilon \right\}\longrightarrow0\,\,\,(n\rightarrow\infty)\,.
\end{equation}
Similarly, for $\epsilon>0$ it follows:

\begin{eqnarray}
\nonumber
\mathbb{E}\left\{\mathop{\sup}\limits_{g\in \mathcal{G}_{n}}\left|\frac{1}{n}\sum_{i=1}^{n}g(\textbf{Z}_{i})-\mathbb{E}\left[g(\textbf{Z})\right] \right|\right\}&=&\int_{0}^{\infty}\mathbb{P}\left\{\mathop{\sup}\limits_{g\in \mathcal{G}_{n}}\left|\frac{1}{n}\sum_{i=1}^{n}g(\textbf{Z}_{i})-\mathbb{E}\left[g(\textbf{Z})\right] \right|>t \right\}dt \\
\nonumber
& \leq & \epsilon + \int_{\epsilon}^{\infty}\mathbb{P}\left\{\mathop{\sup}\limits_{g\in \mathcal{G}_{n}}\left|\frac{1}{n}\sum_{i=1}^{n}g(\textbf{Z}_{i})-\mathbb{E}\left[g(\textbf{Z})\right] \right|>t \right\}dt \\
\nonumber
& \leq & \epsilon + \int_{\epsilon}^{\infty}24\cdot\left(\left(\frac{192\,e\,\beta_{n}^{2}}{t} \right)^{2(\mathcal{K}_{n}+1)} \right) e^{-\frac{n\,t^{2}}{2048\beta_{n}^{4}}}dt \\
\nonumber
& \leq & \epsilon + 24\frac{2048\beta_{n}^{4}}{n\epsilon}e^{2(\mathcal{K}_{n}+1)\log\left(\frac{192\,e\,\beta_{n}^{2}}{\epsilon}\right)-\frac{n\epsilon^{2}}{2048\beta_{n}^{4}}} \\
\nonumber
& \leq & \epsilon + 24\cdot2048\frac{1}{n^{\delta}}\frac{\beta_{n}^{4}}{n^{1-\delta}}e^{-n^{\delta}\frac{n^{1-\delta}}{\beta_{n}^{4}}\left(\frac{\epsilon^{2}}{2048}-\frac{2(\mathcal{K}_{n}+1)\beta_{n}^{4}}{n} \log\left(\frac{192\,e\,\beta_{n}^{2}}{\epsilon}\right)\right)}\,.
\end{eqnarray}
Clearly, since condition (a) and (b) are satisfied by assumptions (ii) and (iii), the second term of the above equation goes to zero as $n\rightarrow\infty$. Since $\epsilon$ is arbitrary, this implies:

\begin{equation}\label{eqProofTh1P2:8}
\mathbb{E}\left\{\mathop{\sup}\limits_{g\in \mathcal{G}_{n}}\left|\frac{1}{n}\sum_{i=1}^{n}g(\textbf{Z}_{i})-\mathbb{E}\left[g(\textbf{Z})\right] \right|\right\}\longrightarrow0\,\,\,(n\rightarrow\infty)\,.
\end{equation}
By the Borel-Cantelli Lemma, (\ref{eqProofTh1P2:8}) and (\ref{eqProofTh1P2:7}) show assumptions (ii) and (iii) imply conditions (\ref{eqTh4:2}) and (\ref{eqTh4:4}) of Theorem \ref{ThG4}. This, together with results from Part 1 show that (\ref{eqproofTh1:1}) and (\ref{eqproofTh1:2}) hold, and the Theorem is proved.

\newpage

\section{Proof of Lemma 1.} \label{proof:Lemma1}

Suppose an orthonormal set of functions $\left\{ \phi^{per}_{j,k}(x), k =0,...,2^{j}-1,\,j\geq0\right\}$ which is dense in $\mathbb{L}_{2}(\nu([0,1]))$ for $\nu\in \Upsilon$, which represents the set of bounded lebesgue measures in $[0,1]$. Suppose $\mu$ is a bounded lebesgue measure in $[0,1]^{p}$ and that conditions stated in Theorem 1 for the scaling function $\phi$, and assumptions (\textbf{A1})-(\textbf{A4}) presented in \ref{ARM} hold.
\medskip

Define the set of functions $\mathcal{F}_{n}$ as in (\ref{eqTh1:3}). Also, let $\beta_{n}>0$ be a parameter depending on the sample and assume $\mathbb{E}\left[Y^{2}\right]<\infty$. Define $\hat{f}_{J(n)}$ as in (\ref{eqLS:3}) and let $f_{J(n)}=T_{\beta_{n}}\hat{f}_{J(n)}$, let $\mathcal{K}_{n}=p\,2^{J(n)}$.
\medskip

Furthermore, assume the following condition holds:
\begin{enumerate}[(i)]
  \item $\sum_{j=1}^{p}||f_{j}||_{\infty}<L$, for some $L<\beta_{n}$.
\end{enumerate}
Then:
\begin{equation}\label{eqProoflemma1:1}
\mathbb{E}\left[\frac{1}{n}\sum_{i=1}^{n}\left|f_{J(n)}(\textbf{x}_{i})-f_{A}(\textbf{x}_{i}) \right|^{2} \mid \textbf{X}_{1}^{n}\right]\leq \mathop{\min}\limits_{f\in\mathcal{F}_{n}}\left\{||f-f_{A} ||_{n}^{2} \right\}+\frac{\sigma^{2}}{n}\mathcal{K}_{n}
\end{equation}
\subsubsection*{Proof}
First, note that $||f_{A}||_{\infty}<\beta_{n}$ (from condition (i)), implies that $||f_{J(n)}-f_{A}||_{n}^{2}\leq ||\hat{f}_{J(n)}-f_{A}||_{n}^{2}$. Therefore, this further implies:
\begin{eqnarray}
\nonumber
\mathbb{E}\left[||f_{J(n)}-f_{A}||_{n}^{2} \mid \textbf{X}_{1}^{n}\right] &\leq& \mathbb{E}\left[||\hat{f}_{J(n)}-f_{A}||_{n}^{2} \mid \textbf{X}_{1}^{n}\right] \\
\nonumber
& \leq & \mathbb{E}\left[\left\|\hat{f}_{J(n)}-\mathbb{E}\left[\hat{f}_{J(n)}\mid \textbf{X}_{1}^{n} \right]+\mathbb{E}\left[\hat{f}_{J(n)}\mid \textbf{X}_{1}^{n} \right]-f_{A}\right\|_{n}^{2} \mid \textbf{X}_{1}^{n}\right] \\
\nonumber
& \leq & \mathbb{E}\left[\left\|\hat{f}_{J(n)}-\mathbb{E}\left[\hat{f}_{J(n)}\mid \textbf{X}_{1}^{n} \right]\right\|_{n}^{2} \mid \textbf{X}_{1}^{n}\right]+\mathbb{E}\left[\left\|\mathbb{E}\left[\hat{f}_{J(n)}\mid \textbf{X}_{1}^{n} \right]-f_{A}\right\|_{n}^{2} \mid \textbf{X}_{1}^{n}\right] \\
\nonumber
& & + 2\mathbb{E}\left\{ \frac{1}{n}\sum_{i=1}^{n}\left(\hat{f}_{J(n)}(\textbf{X}_{i})-\mathbb{E}\left[\hat{f}_{J(n)}\mid \textbf{X}_{1}^{n} \right] \right)\,\left(\mathbb{E}\left[\hat{f}_{J(n)}\mid \textbf{X}_{1}^{n} \right]-f_{A}(\textbf{X}_{i})\right)\,\mid \textbf{X}_{1}^{n} \right\} \\
\label{eqProoflemma1:2}
& \leq & \mathbb{E}\left[\left\|\hat{f}_{J(n)}-\mathbb{E}\left[\hat{f}_{J(n)}\mid \textbf{X}_{1}^{n} \right]\right\|_{n}^{2} \mid \textbf{X}_{1}^{n}\right]+\left\|\mathbb{E}\left[\hat{f}_{J(n)}\mid \textbf{X}_{1}^{n} \right]-f_{A}\right\|_{n}^{2}\,,
\end{eqnarray}
where the last result follows since the last term in the third inequality is zero. From definitions (\ref{eqLS:2}), (\ref{eqLS:4}), and (\ref{eqLS:5}), for any $i\in\left\{1,...,n \right\}$ it follows:
\begin{eqnarray}
\nonumber
\mathbb{E}\left[\hat{f}_{J(n)}(\textbf{X}_{i}) \mid \textbf{X}_{1}^{n} \right] & = & \mathbb{E}\left[\textbf{B}^{T}(\textbf{X}_{i})\,\textbf{c}^{*} \mid \textbf{X}_{1}^{n}\right] \\
\nonumber
& = & \textbf{B}^{T}(\textbf{X}_{i})\,\left(\textbf{B}^{T}\textbf{B}\right)^{-1}\textbf{B}^{T}\, \mathbb{E}\left[\textbf{Y} \mid \textbf{X}_{1}^{n}\right] \\
\nonumber
& = & \textbf{B}^{T}(\textbf{X}_{i})\,\left(\textbf{B}^{T}\textbf{B}\right)^{-1}\textbf{B}^{T}\,\begin{bmatrix}
           f_{A}(\textbf{X}_{1}) \\
           \vdots \\
            f_{A}(\textbf{X}_{n})
           \end{bmatrix} \\
\label{eqProoflemma1:3}
& = & \textbf{B}^{T}(\textbf{X}_{i})\,\left(\textbf{B}^{T}\textbf{B}\right)^{-1}\textbf{B}^{T}\,\textbf{f}\,.
\end{eqnarray}
Now, from the last set of equations, it follows that $\mathbb{E}\left[\textbf{c}^{*} \mid \textbf{X}_{1}^{n}\right]=\left(\textbf{B}^{T}\textbf{B}\right)^{-1}\textbf{B}^{T}\,\textbf{f}$, which implies:
\begin{equation}
\nonumber
\frac{1}{n}\left(\textbf{B}^{T}\textbf{B}\right)\mathbb{E}\left[\textbf{c}^{*} \mid \textbf{X}_{1}^{n}\right]=\frac{1}{n}\textbf{B}^{T}\,\textbf{f}\,.
\end{equation}
Therefore, $\mathbb{E}\left[\textbf{c}^{*} \mid \textbf{X}_{1}^{n}\right]$ is the least squares solution for the problem: $\mathop{\min}\limits_{\textbf{a}\in\mathbb{R}^{\mathcal{K}_{n}}}\left\{\left\|\textbf{B}\,\textbf{a}-\textbf{f} \right\|^{2}_{n} \right\}$. This implies that $\left\|\mathbb{E}\left[\hat{f}_{J(n)}\mid \textbf{X}_{1}^{n} \right]-f_{A}\right\|_{n}^{2}=\mathop{\min}\limits_{f\in\mathcal{F}_{n}}||f-f_{A}||_{n}^{2}$.
\medskip

Therefore, this result and (\ref{eqProoflemma1:2}), imply:
\begin{equation}
\nonumber
\mathbb{E}\left[||f_{J(n)}-f_{A}||_{n}^{2} \mid \textbf{X}_{1}^{n}\right] \leq \mathbb{E}\left[\left\|\hat{f}_{J(n)}-\mathbb{E}\left[\hat{f}_{J(n)}\mid \textbf{X}_{1}^{n} \right]\right\|_{n}^{2} \mid \textbf{X}_{1}^{n}\right] + \mathop{\min}\limits_{f\in\mathcal{F}_{n}}||f-f_{A}||_{n}^{2}\,.
\end{equation}
For a fixed $\textbf{x}$, from definitions (\ref{eqLS:2}), (\ref{eqLS:4}) and (\ref{eqLS:5}), it follows:

\begin{eqnarray}
\nonumber
\mathbb{E}\left[\left|\hat{f}_{J(n)}(\textbf{x})-\mathbb{E}\left[\hat{f}_{J(n)}(\textbf{x})\mid \textbf{X}_{1}^{n} \right]\right|^{2} \mid \textbf{X}_{1}^{n}\right] &=&
\mathbb{E}\left[\left|\textbf{B}(\textbf{x})^{T}\,\textbf{c}^{*}-\textbf{B}(\textbf{x})^{T}\mathbb{E}\left[\textbf{c}^{*} \mid \textbf{X}_{1}^{n}\right] \right|^{2} \mid \textbf{X}_{1}^{n}\right]\\
\nonumber
&=& \mathbb{E}\left[\left|\textbf{B}(\textbf{x})^{T}\,\left(\textbf{B}^{T}\textbf{B}\right)^{-1}\textbf{B}^{T}\left(\textbf{Y}-\textbf{f} \right)\right|^{2} \mid \textbf{X}_{1}^{n}\right] \\
&=& \textbf{B}(\textbf{x})^{T}\,\textbf{H}\,\mathbb{E}\left[\left(\textbf{Y}-\textbf{f} \right)\left(\textbf{Y}-\textbf{f} \right)^{T} \right]\textbf{H}^{T}\textbf{B}(\textbf{x})\,, \label{eqProoflemma1:2.1}
\end{eqnarray}
where $\textbf{H}=\left(\textbf{B}^{T}\textbf{B}\right)^{-1}\textbf{B}^{T}$. By the assumptions of model (\ref{eq:3.1}), it follows that $\mathbb{E}\left[\left(\textbf{Y}-\textbf{f} \right)\left(\textbf{Y}-\textbf{f} \right)^{T} \right]=\sigma^{2}\textbf{I}_{\mathcal{K}_{n}}$. Therefore, (\ref{eqProoflemma1:2.1}) can be expressed as:
\begin{eqnarray}
\nonumber
\mathbb{E}\left[\left|\hat{f}_{J(n)}(\textbf{x})-\mathbb{E}\left[\hat{f}_{J(n)}(\textbf{x})\mid \textbf{X}_{1}^{n} \right]\right|^{2} \mid \textbf{X}_{1}^{n}\right] &=&
\sigma^{2}\textbf{B}(\textbf{x})^{T}\left(\textbf{B}^{T}\textbf{B}\right)^{-1}\textbf{B}(\textbf{x})\,.
\end{eqnarray}
By substituting this result in $\mathbb{E}\left[\left\|\hat{f}_{J(n)}-\mathbb{E}\left[\hat{f}_{J(n)}\mid \textbf{X}_{1}^{n} \right]\right\|_{n}^{2} \mid \textbf{X}_{1}^{n}\right]$, it follows:

\begin{equation}
\label{eqProoflemma1:4}
\mathbb{E}\left[||f_{J(n)}-f_{A}||_{n}^{2} \mid \textbf{X}_{1}^{n}\right] \leq \mathop{\min}\limits_{f\in\mathcal{F}_{n}}||f-f_{A}||_{n}^{2} + \frac{\sigma^{2}}{n}\sum_{i=1}^{n}\textbf{B}(\textbf{x}_{i})^{T}\left(\textbf{B}^{T}\textbf{B}\right)^{-1}\textbf{B}(\textbf{x}_{i})\,.
\end{equation}
Notice that:
\begin{eqnarray}
\nonumber
\sum_{i=1}^{n}\textbf{B}(\textbf{x}_{i})^{T}\left(\textbf{B}^{T}\textbf{B}\right)^{-1}\textbf{B}(\textbf{x}_{i}) &=& \text{trace}\left\{\sum_{i=1}^{n}\textbf{B}(\textbf{x}_{i})^{T}\left(\textbf{B}^{T}\textbf{B}\right)^{-1}\textbf{B}(\textbf{x}_{i}) \right\} \\
\nonumber
&=& \sum_{i=1}^{n}\text{trace}\left\{\textbf{B}(\textbf{x}_{i})^{T}\left(\textbf{B}^{T}\textbf{B}\right)^{-1}\textbf{B}(\textbf{x}_{i})\right\} \\
\nonumber
&=& \sum_{i=1}^{n}\text{trace}\left\{\left(\textbf{B}^{T}\textbf{B}\right)^{-1}\textbf{B}(\textbf{x}_{i})\textbf{B}(\textbf{x}_{i})^{T}\right\} \\
\nonumber
&=& \text{trace}\left\{\left(\textbf{B}^{T}\textbf{B}\right)^{-1}\sum_{i=1}^{n}\textbf{B}(\textbf{x}_{i})\textbf{B}(\textbf{x}_{i})^{T} \right\} \\
&=& \text{trace}\left\{\textbf{I}_{\mathcal{K}_{n}}\right\} = \mathcal{K}_{n}\,,
\end{eqnarray}
where the last 2 equalities follow from definitions (\ref{eqLS:2}) and (\ref{eqLS:4}). In fact, it is possible to observe that $\sum_{i=1}^{n}\textbf{B}(\textbf{x}_{i})\textbf{B}(\textbf{x}_{i})^{T}=\textbf{B}^{T}\textbf{B}$. Therefore, this result applied to (\ref{eqProoflemma1:4}) implies:

\begin{equation}
\nonumber
\mathbb{E}\left[\frac{1}{n}\sum_{i=1}^{n}\left|f_{J(n)}(\textbf{x}_{i})-f_{A}(\textbf{x}_{i}) \right|^{2} \mid \textbf{X}_{1}^{n}\right]\leq \mathop{\min}\limits_{f\in\mathcal{F}_{n}}\left\{||f-f_{A} ||_{n}^{2} \right\}+\frac{\sigma^{2}}{n}\mathcal{K}_{n}\,.
\end{equation}
which proves assertion (\ref{eqProoflemma1:1}).

\newpage
\section{Proof of Lemma 2.} \label{proof:Lemma2}
Suppose an orthonormal basis $\left\{ \phi^{per}_{j,k}(x), k =0,...,2^{j}-1,\,j\geq0\right\}$ which is dense in $\mathbb{L}_{2}(\nu([0,1]))$ for $\nu\in \Upsilon$, which represents the set of bounded lebesgue measures in $[0,1]$. Suppose assumptions stated in Theorem 1 for the scaling function $\phi$, and conditions (\textbf{A1})-(\textbf{A4}) defined in \ref{ARM} hold. Let the set of functions $\mathcal{F}_{n}$ to be defined as in (\ref{eqTh1:3}).

\medskip
Then it follows:
\begin{equation}\label{eqProoflemma2:1}
\mathop{\inf}\limits_{f\in\mathcal{F}_{n}}\int_{[0,1]^{p}}\left|f(\textbf{x})-f_{A}(\textbf{x}) \right|^{2}\mu(d\textbf{x}) \leq p^{2}\,C_{2}^{2}\,2^{-2(N+1)\,J(n)}\,.
\end{equation}
\subsubsection*{Proof}
Denote $f_{j}^{J}=\sum_{k=0}^{2^{J}-1}c_{Jk}^{(j)}\phi_{Jk}^{per}$. Consider:
\begin{eqnarray}
\nonumber
\mathop{\inf}\limits_{f\in\mathcal{F}_{n}}\int_{[0,1]^{p}}\left|f(\textbf{x})-f_{A}(\textbf{x}) \right|^{2}\mu(d\textbf{x}) &=& \mathop{\inf}\limits_{f\in\mathcal{F}_{n}}\int_{[0,1]^{p}}\left| \sum_{j=1}^{p}\left(f_{j}^{J}(x_{j})-f_{j}(x_{j}) \right) \right|^{2}\mu(d\textbf{x}) \\
\nonumber
& \leq & p\,\mathop{\inf}\limits_{f\in\mathcal{F}_{n}}\int_{[0,1]^{p}}\sum_{j=1}^{p}\left| f_{j}^{J}(x_{j})-f_{j}(x_{j}) \right|^{2}\mu(d\textbf{x}) \\
\nonumber
& \leq & p\,\mathop{\inf}\limits_{f\in\mathcal{F}_{n}}\sum_{j=1}^{p}\mathop{\sup}\limits_{x_{j}\in[0,1]}\left| f_{j}^{J}(x_{j})-f_{j}(x_{j}) \right|^{2} \\
\nonumber
& \leq & p\,\mathop{\inf}\limits_{f\in\mathcal{F}_{n}}\sum_{j=1}^{p}\left(\mathop{\sup}\limits_{x_{j}\in[0,1]}\left| f_{j}^{J}(x_{j})-f_{j}(x_{j}) \right|\right)^{2}\,.
\end{eqnarray}
By corollary 8.2 of \cite{Hardle1998}, it follows that $\mathop{\sup}\limits_{x_{j}\in[0,1]}\left| f_{j}^{J}(x_{j})-f_{j}(x_{j}) \right|=\mathcal{O}\left(2^{-J\,(N+1)} \right)$; therefore, $\exists\,C_{2}$ independent of $n,\, \text{and}\,\,J$ such that $\mathop{\sup}\limits_{x_{j}\in[0,1]}\left| f_{j}^{J}(x_{j})-f_{j}(x_{j}) \right|\leq C_{2}\,2^{-J\,(N+1)}$. Thus:
\begin{eqnarray}
\nonumber
\mathop{\inf}\limits_{f\in\mathcal{F}_{n}}\int_{[0,1]^{p}}\left|f(\textbf{x})-f_{A}(\textbf{x}) \right|^{2}\mu(d\textbf{x}) &\leq& p^{2}\,C_{2}^{2}\,2^{-2(N+1)\,J(n)}\,,
\end{eqnarray}
as desired.

\newpage

\section{Proof of Theorem 2.} \label{proof:Theorem2}
This proof follows the same methodology as in section 10 of \cite{Gyorfi2002}. Consider assumptions stated for Lemma 1 and conditions (i)-(iii) for Theorem 1 hold . Then:
\begin{equation}\label{eqProofTheorem2:1}
\mathbb{E}\left[\int_{[0,1]^{p}}\left|f_{J(n)}(\textbf{x})-f_{A}(\textbf{x})\right|^{2}\mu(d\textbf{x})\right] \leq \tilde{C}\max\left\{\beta_{n}^{2},\sigma^{2}\right\}\frac{p\,2^{J(n)}}{n}\left(\log(n)+1\right)\,,
+ 8\,C_{2}^{2}\,p^{2}\,2^{-2(N+1)J(n)}\,,
\end{equation}
for proper constants $\tilde{C}>0$ and $C_{2}>0$ independent of $n,N,p$.

\subsubsection*{Proof}

Note that $||f_{J(n)}-f_{A}||^{2}=\int_{[0,1]^{p}}\left|f_{J(n)}(\textbf{x})-f_{A}(\textbf{x}) \right|^{2}\mu(d\textbf{x})$ can be expressed as follows:

\begin{eqnarray}
\nonumber
||f_{J(n)}-f_{A}||^{2}&=&\left(||f_{J(n)}-f_{A}||-2\,||f_{J(n)}-f_{A}||_{n}+2\,||f_{J(n)}-f_{A}||_{n} \right)^{2} \\
\nonumber
& \leq & \left(\max\left\{0\,,\,||f_{J(n)}-f_{A}||-2\,||f_{J(n)}-f_{A}||_{n} \right\}+ 2\,||f_{J(n)}-f_{A}||_{n}\right)^{2} \\
\nonumber
& \leq & 2\,\left(\max\left\{0\,,\,||f_{J(n)}-f_{A}||-2\,||f_{J(n)}-f_{A}||_{n} \right\}\right)^{2}+8\,||f_{J(n)}-f_{A}||_{n}^{2}\,, \\
\nonumber
& \leq & \,S_{1,n}+8\,S_{2,n}\,.
\end{eqnarray}
Observe that $\mathbb{E}\left[S_{2,n}\right]=\mathbb{E}_{\textbf{X}_{1}^{n}}\left[\mathbb{E}\left(||f_{J(n)}-f_{A}||_{n}^{2}\,\mid \textbf{X}_{1}^{n} \right) \right]$. Similarly, from the definition of $f_{J(n)}$ and condition (i) of Lemma 1, it follows that $||f_{J(n)}-f_{A}||_{n}^{2}\leq||\hat{f}_{J(n)}-f_{A}||_{n}^{2}$. These 2 results and Lemma 1 imply:
\begin{eqnarray}
\nonumber
\mathbb{E}\left[S_{2,n}\right] &\leq& \mathbb{E}_{\textbf{X}_{1}^{n}}\left[\mathop{\min}\limits_{f\in\mathcal{F}_{n}}\left\{||f-f_{A} ||_{n}^{2} \right\} \right]+\frac{\sigma^{2}}{n}\mathcal{K}_{n} \\
\nonumber
&\leq& \mathbb{E}_{\textbf{X}_{1}^{n}}\left[\mathop{\min}\limits_{f\in\mathcal{F}_{n}}\left\{\frac{1}{n}\sum_{i=1}^{n}\left|f(\textbf{x}_{i})-f_{A}(\textbf{x}_{i})\right|^{2} \right\} \right]+\frac{\sigma^{2}}{n}\mathcal{K}_{n} \\
\label{eqProofTheorem2:2}
&\leq& \mathop{\inf}\limits_{f\in\mathcal{F}_{n}}\left\{\int_{[0,1]^{p}}\left|f(\textbf{x})-f_{A}(\textbf{x})\right|^{2}\mu(d\textbf{x})\right\}+\frac{\sigma^{2}}{n}\mathcal{K}_{n}\,,
\end{eqnarray}
where the last inequality follows from the properties of the expected value and the iid condition of the sample $\textbf{X}_{1}^{n}=\left(\textbf{X}_{1},...,\textbf{X}_{n}\right)$. Now, for $S_{1,n}$, define:
\begin{equation}
\nonumber
\mathcal{G}_{n}=\left\{g_{n}\,:\,[0,1]^{p}\rightarrow\mathbb{R}\,;\,g_{n}=f_{J(n)}-f_{A}\,\mid\,f_{J(n)}\in\,T_{\beta_{n}}\mathcal{F}_{n}\right\}\,.
\end{equation}
Observe that $\forall g\in\mathcal{G}_{n}\,\,|g_{n}|\leq2\,\beta_{n}$. Consider $u>0$ (arbitrary) and:
\begin{eqnarray}
\nonumber
\mathbb{P}\left\{S_{1,n}>u\right\} & = &\mathbb{P}\left\{2\,\left(\max\left\{0\,,\,||f_{J(n)}-f_{A}||-2\,||f_{J(n)}-f_{A}||_{n} \right\}\right)^{2}>u \right\} \\
\nonumber
& = & \mathbb{P}\left\{\max\left\{0\,,\,||f_{J(n)}-f_{A}||-2\,||f_{J(n)}-f_{A}||_{n} \right\}>\sqrt{\frac{u}{2}} \right\} \\
\nonumber
& \leq & \mathbb{P}\left\{\max\left\{0\,,\,||f_{J(n)}-f_{A}||-2\,||f_{J(n)}-f_{A}||_{n} \right\}>\sqrt{\frac{u}{2}} \right\}\,.
\end{eqnarray}
From Theorem P2, it follows:
\begin{eqnarray}
\nonumber
\mathbb{P}\left\{S_{1,n}>u\right\} &\leq& 3\,\mathbb{E}\left[\mathcal{N}_{2}\left(\frac{\sqrt{2}}{24}\sqrt{\frac{u}{2}}\,,\mathcal{G}_{n},\textbf{X}_{1}^{2n}\right) \right]\,e^{-\frac{n\,\frac{u}{2}}{288\,(2\beta_{n})^{2}}} \\
\nonumber
&\leq& 3\,\mathbb{E}\left[\mathcal{N}_{2}\left(\frac{\sqrt{u}}{24}\,,\mathcal{G}_{n},\textbf{X}_{1}^{2n}\right) \right]\,e^{-\frac{n\,u}{576\cdot4\beta_{n}^{2}}}\,.
\end{eqnarray}
Lemma G1 implies that $\mathcal{N}_{2}\left(\frac{\sqrt{u}}{24}\,,\mathcal{G}_{n},\textbf{X}_{1}^{2n}\right)\leq \mathcal{M}_{2}\left(\frac{\sqrt{u}}{24}\,,\mathcal{G}_{n},\textbf{X}_{1}^{2n}\right)$. Similarly, from Theorem G2, it follows that $\mathcal{M}_{2}\left(\frac{\sqrt{u}}{24}\,,\mathcal{G}_{n},\textbf{X}_{1}^{2n}\right)\leq 3\,\left(\frac{2\,e\,4\beta_{n}^{2}}{\left(\frac{\sqrt{u}}{24} \right)^{2}}\log\left(\frac{3\,e\,4\beta_{n}^{2}}{\left(\frac{\sqrt{u}}{24} \right)^{2}} \right) \right)^{V_{\mathcal{G}_{n}^{+}}}$. Using the same argument as in the proof of Theorem 1, Theorem G3 implies that $V_{\mathcal{G}_{n}^{+}}\leq \mathcal{K}_{n}+1$.
\medskip
Therefore:
\begin{eqnarray}
\nonumber
\mathbb{P}\left\{S_{1,n}>u\right\} &\leq& 3\,\left(\frac{24^{2}\,12\,e\,\beta_{n}^{2}}{u} \right)^{2(\mathcal{K}_{n}+1)}\,e^{-\frac{n\,u}{576\cdot4\beta_{n}^{2}}}\,.
\end{eqnarray}
Note that for $u>\frac{576\,\beta_{n}^{2}}{n}$, $\frac{24^{2}\,12\,e\,\beta_{n}^{2}}{u} \leq 12\,e\,n$; Therefore, it follows:
\begin{eqnarray}
\nonumber
\mathbb{P}\left\{S_{1,n}>u\right\} &\leq& 3\,\left(12\,e\,n\right)^{2(\mathcal{K}_{n}+1)}\,e^{-\frac{n\,u}{576\cdot4\beta_{n}^{2}}}\,.
\end{eqnarray}
Now, consider $\delta>0$. For $u>\frac{576\,\beta_{n}^{2}}{n}$, $\mathbb{E}\left[S_{1,n}\right]$ can be bounded as follows:

\begin{eqnarray}
\nonumber
\mathbb{E}\left[S_{1,n}\right] &\leq& \int_{0}^{\infty}\mathbb{P}\left\{S_{1,n}>t\right\}dt \\
\nonumber
& \leq & \delta + \int_{\delta}^{\infty}\mathbb{P}\left\{S_{1,n}>t\right\}dt \\
\nonumber
& \leq & \delta + 3\,\left(12\,e\,n\right)^{2(\mathcal{K}_{n}+1)}\int_{\delta}^{\infty}e^{-\frac{n\,t}{576\cdot4\beta_{n}^{2}}}dt \\
\label{eqProofTheorem2:2.2}
& \leq & \delta + 3\,\left(12\,e\,n\right)^{2(\mathcal{K}_{n}+1)}\left(\frac{2304\,\beta_{n}^{2}}{n}\right) e^{-\frac{n\,\delta}{576\cdot4\beta_{n}^{2}}}\,.
\end{eqnarray}
Observe that the rhs of (\ref{eqProofTheorem2:2.2}) is continuous for $\delta>0$. Therefore, it is possible to obtain a value of $\delta$ that minimizes the upper bound. In this context, it is possible to show that $\delta^{*}=\frac{2304\,\beta_{n}^{2}}{n}\log\left(9\cdot\left(12\,e\,n \right)^{2(\mathcal{K}_{n}+1)} \right)$ is the aforementioned minimizer. Using this result, it follows:
\begin{eqnarray}
\nonumber
\mathbb{E}\left[S_{1,n}\right] &\leq& \frac{2304\,\beta_{n}^{2}}{n}\log\left(9\cdot\left(12\,e\,n \right)^{2(\mathcal{K}_{n}+1)} \right)+ \frac{2304\,\beta_{n}^{2}}{n}\,.
\end{eqnarray}
After some algebra, the last expression takes the form:
\begin{eqnarray}
\label{eqProofTheorem2:3}
\mathbb{E}\left[S_{1,n}\right] &\leq& \frac{\tilde{C}\,\beta_{n}^{2}\,\mathcal{K}_{n}\left(\log(n)+1 \right)}{n}\,,
\end{eqnarray}
for $\tilde{C}=4608\log(12)$. This, together with (\ref{eqProofTheorem2:2}) imply:
\begin{eqnarray}
\nonumber
\mathbb{E}\left\{||f_{J(n)}-f_{A}||^{2}\right\} &\leq & \frac{\tilde{c}\,\beta_{n}^{2}\,\mathcal{K}_{n}\left(\log(n)+1 \right)}{n} + 8\,\mathop{\inf}\limits_{f\in\mathcal{F}_{n}}\left\{\int_{[0,1]^{p}}\left|f(\textbf{x})-f_{A}(\textbf{x})\right|^{2}\mu(d\textbf{x})\right\}+\frac{8\sigma^{2}}{n}\mathcal{K}_{n}\,.
\end{eqnarray}
Finally, from Lemma 2 it follows:
\begin{eqnarray}
\label{eqProofTheorem2:4}
\mathbb{E}\left\{||f_{J(n)}-f_{A}||^{2}\right\} &\leq & \tilde{C}\max\left\{\beta_{n}^{2},\sigma^{2}\right\}\frac{p\,2^{J(n)}}{n}\left(\log(n)+1\right)\,,
+ 8\,C_{2}^{2}\,p^{2}\,2^{-2(N+1)J(n)}
\end{eqnarray}
which proves the desired result.

\newpage

\section{Proof of Lemma 3.} \label{proof:Lemma3}

Suppose a model of the form (\ref{eq:LSa1}). Assume $\epsilon$ is a sub-gaussian random variable independent of $\textbf{X}$ such that $\mathbb{E}[\epsilon]=0$, $\mathbb{E}[\epsilon^{2}]=1$, $0<\sigma<\infty$. Let $\left\{Y_{1},...,Y_{n} \right\}$ be the response observations from the iid sample $\left\{\left(Y_{i},\textbf{X}_{i} \right) \right\}_{i=1}^{n}$ and suppose $||f_{A}||_{\infty}\leq L$.

\medskip
Then, for $\beta_{n}=4\sigma\sqrt{\log(n)}$ it follows:

\begin{equation}\label{eqProoflemma3:1}
\mathbb{P}\left\{\max\left\{Y_{1},...,Y_{n} \right\}>\beta_{n} \right\}=\mathcal{O}\left(\frac{1}{n}\right)\,.
\end{equation}
\subsubsection*{Proof}

Denote $Y_{(n)}=\max\left\{Y_{1},...,Y_{n} \right\}$. For some $\delta>0$ it holds:

\begin{eqnarray}
\nonumber
\mathbb{P}\left\{Y_{(n)}>\beta_{n} \right\} &\leq & \mathbb{P}\left\{\cup_{i=1}^{n}Y_{i}>\beta_{n} \right\}\\
\nonumber
&\leq & \sum_{i=1}^{n}\mathbb{P}\left\{Y_{i}>\beta_{n} \right\} \\
\nonumber
&\leq & n\,\int_{[0,1]^{p}}\mathbb{P}\left\{f_{A}(\textbf{u})+\sigma\,\epsilon>\beta_{n} \mid \textbf{X}=\textbf{u}  \right\}h(\textbf{u})d\textbf{u} \\
\nonumber
&\leq & n\,\int_{[0,1]^{p}}\mathbb{P}\left\{\left|\epsilon\right|>\frac{\beta_{n}-L}{\sigma} \mid \textbf{X}=\textbf{u}  \right\}h(\textbf{u})d\textbf{u} \\
\nonumber
&\leq & n\,\mathbb{P}\left\{\left|\epsilon\right|>\frac{\beta_{n}-L}{\sigma}\right\}\,.
\end{eqnarray}
Since $\epsilon$ is assumed to be sub-gaussian ($\mathbb{E}[\epsilon]=0$, $\mathbb{E}[\epsilon^{2}]=1$, $0<\sigma<\infty$), we have that $\forall s\in\mathbb{R}$, $\mathbb{E}\left[e^{s\,\epsilon} \right]\leq e^{\frac{s^{2}}{2}}$. Consequently, it is possible to show that $\mathbb{P}\left\{|\epsilon|>t \right\}\leq 2\,e^{-\frac{t^{2}}{2}}$. Using this result in the last equation, it follows:
\begin{eqnarray}
\nonumber
\mathbb{P}\left\{Y_{(n)}>\beta_{n} \right\} &\leq & 2\,n\,e^{-\frac{\left( \beta_{n}-L\right)^{2}}{2\sigma^{2}}}\,.
\end{eqnarray}
Suppose it is possible to choose $\beta_{n}$ in such a way that $2\,n\,e^{-\frac{\left( \beta_{n}-L\right)^{2}}{2\sigma^{2}}}\leq\frac{1}{n}$. This implies that $Y_{(n)}$ it's bounded in probability. Under this setting, assuming that for $n$ large enough $\sqrt{2}\,\sigma\,\sqrt{\log(n)}>L$, it follows:

\begin{equation}
\nonumber
\mathbb{P}\left\{\max\left\{Y_{1},...,Y_{n} \right\}>\sqrt{2}\,\sigma\,\sqrt{\log(n)} \right\}=\mathcal{O}\left(\frac{1}{n}\right)\,,
\end{equation}
which shows (\ref{eqProoflemma3:1}) holds.

\end{document}